\documentclass[a4paper,10pt]{article}
\usepackage[english]{babel}%
\usepackage[T1]{fontenc}
\usepackage[utf8]{inputenc}
\RequirePackage{fix-cm}
\usepackage{geometry}
\usepackage{graphicx}%
\usepackage{multirow}%
\usepackage{amsmath,amssymb,amsfonts}%
\usepackage{amsthm}%
\usepackage{mathrsfs}%
\usepackage{doi}
\usepackage{xcolor}%
\usepackage{textcomp}%
\usepackage{algorithm}%
\usepackage{algorithmicx}%
\usepackage{algpseudocode}%
\usepackage{listings}%
\usepackage[normalem]{ulem}
\usepackage{natbib}%
\usepackage{soul}%
\usepackage{environ}
\usepackage{mathtools}
\usepackage{bbold}
\usepackage{bm}%
\usepackage{bbm}%
\usepackage[rgb]{xcolor}%
\usepackage{dsfont}
\usepackage{url}
\usepackage{verbatim}%
\usepackage[all]{xy}
\usepackage{hyperref}
\hypersetup{
    colorlinks,
    linkcolor={gray!50!black},
    citecolor={blue!50!black},
    urlcolor={blue!80!black}
}
\usepackage{orcidlink}

\newcommand{\Mn}{\ensuremath{{M_{-} - n}}}
\renewcommand{\abstractname}{\vspace{-\baselineskip}}
\usepackage[position=top,font={sf,md,up}]{subfig}
\usepackage{hyperref}
\usepackage[right]{lineno}%
\newcommand{\mengi}[1]{\ensuremath{\left\{ #1 \right \}}}
\newcommand{\tolugildi}[1]{\ensuremath{\left| #1 \right|}}
\newenvironment{dsplit}[1]{%
\begin{displaymath}%
\begin{split}%
{#1}%
\end{split}%
\end{displaymath}%
}

\begin{document}%
\renewcommand{\abstractname}{\vspace{-\baselineskip-\baselineskip}}


\definecolor{OliveGreen}{cmyk}{0.64,0,0.95,0.40}
\definecolor{testcolor}{cmyk}{0.96,0,0.96,0.86}
\definecolor{beige}{rgb}{0.96,0.96,0.86}%
\definecolor{aw}{rgb}{0.98,0.92,0.84}%
\definecolor{aliceblue}{rgb}{0.94, 0.97, 1.0}
\definecolor{floralwhite}{rgb}{1.0, 0.98, 0.94}
\definecolor{cosmiclatte}{rgb}{1.0, 0.97, 0.91}



\raggedbottom

\newtheorem{thm}{Theorem}[section]
\newtheorem{propn}[thm]{Proposition}
\newtheorem{lemma}[thm]{Lemma}
\newtheorem{eg}[thm]{Example}
\newtheorem{defn}[thm]{Definition}
\newtheorem{remark}[thm]{Remark}
\newtheorem{notn}[thm]{Notation}
\newtheorem{corollary}[thm]{Corollary}
\newtheorem{conjecture}[thm]{Conjecture}
\newtheorem{assumption}[thm]{Assumption}
\newtheorem{condn}[thm]{Condition}
\newtheorem{example}[thm]{Example}
\newcommand{\braces}[1]{\ensuremath{\left( {#1} \right)}}
\newcommand{\curly}[1]{\ensuremath{\left\{ {#1} \right\}}}
\newcommand\numberthis{\addtocounter{equation}{1}\tag{\theequation}}
\newcommand{\norm}[1]{\left\lVert#1\right\rVert}
\newcommand{\ak}{\ensuremath{\left(\frac{1}{k^\alpha} - \frac{1}{(1+k)^\alpha}\right)}}
\newcommand{\Mm}{\ensuremath{{M_{-}}}}
\newcommand{\Mp}{\ensuremath{{M_{+}}}}
\newcommand{\C}{{\ensuremath{\mathcal{C}_N}}}%
\newcommand{\IN}{\ensuremath{\mathds N}}%
\newcommand{\vIN}{\ensuremath{\mathds N}}%
\newcommand{\IT}{\ensuremath{\mathbb T}}%
\newcommand{\I}{\ensuremath{\mathbb{I}} }
\newcommand{\one}[1]{\ensuremath{\mathds{1}_{ \left \{ #1 \right \} }}}%
\newcommand{\prob}[1]{\ensuremath{\mathds{P}\left[ #1 \right] }}%
\newcommand{\pc}[2]{\ensuremath{\mathbb{P}^{\left( #1 \right)}\left( #2 \right)}}%
\newcommand{\EE}[1]{\ensuremath{\mathds{E}\left[ #1 \right]}}%
\newcommand{\ES}[2]{\ensuremath{\mathbb{E}^{\left( #1 \right)} \left[#2\right]}}%
\newcommand{\vEE}[1]{\ensuremath{\varmathbb{E}\left[ #1 \right]}}%
\newcommand{\cO}{\mathcal{O}}
\newcommand{\co}{{o}}
\newcommand{\dx}{\ensuremath{{\mathrm{d}}}}%
\newcommand{\mutation}{\scalebox{1.2}{$\circ$}}
\newcommand{\set}[1]{\ensuremath{\left\{ #1 \right\}}}
\newcommand{\comp}[1]{\ensuremath{{#1}^{c}}}
\newcommand{\OO}[1]{\ensuremath{\mathcal{O}\left( #1 \right)}}  
\newcommand{\m}{\ensuremath{\mathbb m}}%
\newcommand{\mo}{\ensuremath{\mathfrak o}}%
\newcommand{\oo}[1]{\ensuremath{\mathbb{o}\left( #1 \right) }}%
\newcommand{\BO}[1]{\ensuremath{\mathcal{O}\left( #1 \right) }}%
\newcommand{\uk}{\ensuremath{{\underline{k}}}}%
\newcommand{\uX}{\ensuremath{{\underline{X}}}}%
\newcommand{\id}{\ensuremath{{\mathrm{d}}}}
\newcommand{\vIP}{\ensuremath{\varmathbb R}}%
\newcommand{\IP}{\mathds P}%
\newcommand{\vIQ}{\ensuremath{\varmathbb Q}}%
\newcommand{\IQ}{\mathbb Q}%
\newcommand{\IE}{\mathds E}%
\newcommand{\IR}{\mathds R}%
\newcommand{\IZ}{\mathbb Z}%

\newcommand{\rmP}{\mathrm{P}}
\newcommand{\rmE}{\mathrm{E}}
\newcommand{\rmT}{\mathrm{T}}
\newcommand{\rmt}{\mathrm{t}}

\newcommand{\tX}{\tilde{X}}

\newcommand{\hX}{\hat{X}}
\newcommand{\hS}{\hat{S}}
\newcommand{\hq}{\hat{q}}
\newcommand{\hQ}{\hat{Q}}
\newcommand{\cP}{\mathcal{P}}
\newcommand{\cA}{\mathcal{A}}
\newcommand{\cL}{\mathcal{L}}
\newcommand{\cG}{\mathcal{G}}
\newcommand{\cC}{\mathcal{C}}
\newcommand{\cV}{\mathcal{V}}
\newcommand{\cE}{\mathcal{E}}
\newcommand{\cT}{\mathcal{T}}
\newcommand{\cD}{\mathcal{D}}
\newcommand{\cZ}{\mathcal{Z}}
\newcommand{\cB}{\mathcal{B}}
\newcommand{\cK}{\mathcal{K}}
\newcommand{\cN}{\mathcal{N}}
\newcommand{\cX}{\mathcal{X}}
\newcommand{\cF}{\mathcal{F}}
\newcommand{\dd}{\; \mathrm{d}}
\newcommand{\ind}{\ensuremath{\mathds{1}}}
\newcommand{\ldual}{\langle}
\newcommand{\rdual}{\rangle}
\newcommand{\lquad}{\langle}
\newcommand{\rquad}{\rangle}
\newcommand{\ballg}[1]{B_{#1}}
\newcommand{\ball}[2]{B_{#1}(#2)}
\newcommand{\stopped}[1]{\widehat{#1}}
\newcommand{\conditional}{|}
\newcommand{\compliment}[1]{#1^c}
\newcommand{\cinftyc}{C^\infty_c}
\newcommand{\abs}{|}
\newcommand{\limit}[2]{\lim_{#1 \to #2}}
\newcommand{\Id}{\text{Id}}
\newcommand{\strikayfir}[1]{\text{\sout{\ensuremath{\textcolor{red}{#1}}}}}
\newcommand{\krassayfir}[1]{\text{\xout{\ensuremath{\textcolor{red}{#1}}}}}
\newcommand{\komma}{\ensuremath{,\, }}
\newcommand{\svigi}[1]{\ensuremath{\left( #1 \right)}}
\newcommand{\bjarki}[1]{\textcolor{blue}{\tt #1\newline}}%

\numberwithin{equation}{section}

\renewcommand {\theequation}{\arabic{section}.\arabic{equation}}
\def \non{{\nonumber}}
\def \hat{\widehat}
\def \tilde{\widetilde}
\def \bar{\overline}


\title{\flushleft\textsf{\textbf{Beta-coalescents when sample size is large}}}
\author{Jonathan A.
Chetwynd-Diggle\footnote{The Mathematical Institute, University of
Oxford,
\href{mailto:jonathan.chetwynd-diggle@maths.ox.ac.uk}{jonathan.chetwynd-diggle@maths.ox.ac.uk}}
 \quad  and  Bjarki Eldon\footnote{Corresponding
 author:\href{mailto:beldon11@gmail.com}{beldon11@gmail.com}; \today;
 comments and questions  welcome! }\orcidlink{0000-0001-9354-2391}}
 \date{}

\color{black}

\maketitle


\abstract{Individual recruitment success,  the offspring number
distribution in natural populations, is a fundamental element in
ecology and evolution.  Sweepstakes reproduction refers to mechanisms
generating  highly right-skewed distribution of   recruitment success  without involving natural
selection; it  may apply (for example)  to broadcast spawning
populations characterised by Type III survivorship. We consider an
extension of the~\cite{schweinsberg03} model of sweepstakes reproduction 
 for a haploid panmictic population of constant size $N$;
the extension also works as an alternative to the Wright-Fisher model.
Our model applies an upper bound to the random number of
potential offspring  produced by a given individual.
Depending on how the bound behaves relative to $N$, 
 we obtain the ~\cite{K82,K82b,K82c} coalescent, an incomplete Beta-coalescent,
or the  (complete) Beta-coalescent of
~\cite{schweinsberg03}. We believe that  such a  bound is
biologically reasonable. Moreover, we estimate the error of the
coalescent approximation.
The error estimates reveal that convergence can be
 slow, and  small sample size can be sufficient to invalidate
convergence, for example  if the  stated  bound is of the form
$N/\log N$.  
 Simulation results indicate that increasing sample size will not
 affect the  site-frequency spectrum when the limit is a  Beta-coalescent,
even though full  gene genealogies may  deviate from the coalescent trees.
 When in the domain of attraction of the Kingman coalescent the
effect of increasing sample size depends on the effective population
size as has been noted in the case of the Wright-Fisher
model. Conditioning on the population ancestry (the
random ancestral relations of the entire population at all times) 
may have an  effect on the site-frequency spectrum for the
models considered here (as evidenced by simulation results).
}


\maketitle



\setcounter{equation}{0}


\section{Introduction}\label{intro}

The convergence of gene genealogies describing the random ancestral
relations of sampled gene copies  to the  Kingman 
coalescent~\citep{K82,K82b,K82c} is based on two key assumptions.  First, the sample
size is a negligible fraction of the effective population size
($N_{e}$). Second, any given individual in the corresponding
population model has a negligibly small number of offspring relative
to $N_e$. These assumptions can sometimes be called into question.

Regarding the first assumption, in some  cases $N_e$ has
been estimated to be much smaller than census population size
\citep{A04,Tenesa2007}. In addition,  progress in DNA sequencing
technology and algorithms  for  simulating   gene genealogies
enables population genetic  samples of ever   increasing size  
 to  be efficiently processed and analysed
\citep{Bycroft2018,Kelleher2016,Baumdicker2021}.

Regarding the second assumption, observations from ecology and
empirical population genetics suggest that in highly fecund species
(ranging from fungi to gadids) family sizes may be highly variable in
any given generation, such that a small number of individuals may
produce numbers of offspring  proportional to the population size
(sweepstakes reproduction)
\citep{B94,may67,hedgecock_etal07_marinebiology,A88,H94,Li1998,HP11,birkner12,AH2015,CHRISTIE2010,
williams75_sex_evolut,Niwa2016,arnason22:_sweep,15315674,vendrami21_sweep}.
In such populations sample gene genealogies are in the domain of
attraction of \emph{multiple merger} coalescents where a random number
of ancestral lineages (ancestors of sampled gene copies) merges
whenever mergers occur~\citep{S99,P99,DK99,S00,MS01,MS03,SW08}.
Beta-coalescents, a subfamily of multiple-merger coalescents,  and a 
topic of our study,   arise
naturally from a haploid population model of sweepstakes reproduction  that
we shall refer to as the \emph{Schweinsberg model}
\citep{schweinsberg03}. The Schweinsberg model is a model of discrete (non-overlapping)
generations, in which all current individuals have a chance of
producing  \emph{potential offspring}  according to a given  non-trivial law.  
 A given number of the potential offspring  is then  sampled without
replacement (conditional on there being enough of them) to survive to
maturity.
It is assumed that, with $X$ denoting  the random number of potential
offspring 
produced by a given individual ($\alpha, C > 0$ fixed),  
\begin{equation}
\label{eq:19}
\lim_{x\to \infty}x^{\alpha} \prob{X\ge x} = C 
\end{equation}
\citep[Equation~11]{schweinsberg03}.

The key assumption in~\eqref{eq:19} is that the number of potential
offspring produced by any given individual can be arbitrarily large,
a big ask even for broadcast spawners. 
It is plausible that the number of potential offspring of any given
parent pair may, in general, be at most a fraction of the population
size, even among highly fecund populations.

Here, we shall consider an extension of the Schweinsberg model, and
incorporate an upper bound on the random number of potential offspring
that can be produced by a single individual.  The bound will depend on
the total population size.  We then identify the corresponding
coalescents while keeping track of the remainder terms.  Depending on
the bound, we obtain  \textit{(i)} the complete Beta coalescent of
~\cite{schweinsberg03},
\textit{(ii)} an \emph{incomplete} Beta coalescent, or \textit{(iii)}
the Kingman coalescent.  Our extension has given us a better
understanding of what 
~\eqref{eq:19} means in terms of the limiting coalescents.  
We will also use informal arguments   and simulations to investigate the  coalescent
approximation as  sample size increases. Moreover, we consider the
site-frequency spectrum in the context of quenched gene genealogies. 

The layout of the paper is as follows.
%
In Section~\ref{results} we
introduce our extension of the Schweinsberg model (see 
~\eqref{eq:PXiJ} in Section~\ref{model}), and give a precise statement
of our main results, i.e.\ the rescaling of time required for
convergence (as the population size $N\to \infty$) to a coalescent
(i.e.\ the unit of time used to measure the time between events
involving ancestors of the sample) 
(Proposition~\ref{pr:CNversion0}), and convergence to the coalescents 
(Propositions~\ref{errorkc} and~\ref{pn:Lambda-convergence 1<a<2 full
version}). In Sections~\ref{mschm} and~\ref{nLt} we
consider the impact of increasing sample size; in Section~\ref{mschm} in connection to the Kingman coalescent; and in
Section~\ref{nLt} we give arguments for the sample size required to
see its effect on the site-frequency spectrum (the number  of mutations
of given frequency in a sample)  in the case of the
Beta-coalescent of~\cite{schweinsberg03}.  Section~\ref{numerics} is
devoted to numerical experiments, and Section~\ref{concl} summarizes
and briefly discusses our main findings.
Key lemmas are reviewed in Section~\ref{key lemmas}, followed by
proofs in Sections~\ref{sec:convergencealpha12} to
~\ref{convergence_to_lambda_proofs}.  Many of the ideas follow
~\cite{schweinsberg03}, although we have taken care to keep track of
the error bounds.
In Section~\ref{sec:convergencealpha12} we give a proof of
the timescaling for our model, in Section~\ref{prerrorkc} we give a
proof of convergence to the Kingman coalescent from our model, in
Section~\ref{convergence_to_lambda_proofs} we give proofs of
convergence to Beta-coalescents.  In Appendices~\ref{sec:code} and
~\ref{sec:estimatequenched} we briefly describe algorithms used for
the numerical examples in Section~\ref{numerics} and in
Appendix~\ref{sec:quenched}; Appendix~\ref{sec:furth-numer-exampl}
contains examples of site-frequency spectra predicted by complete and
incomplete  Beta coalescents.

\section{Background}\label{backgroundsection}%

In this section we collect together notation and key definitions
used throughout.
\begin{defn}[Notation]\label{def:standardnotation}
For each $i\in \set{0,1,2,\ldots}$ write  $\IN_{i} \equiv
\set{i,i+1,\ldots}$, $\IN\equiv \IN_{1}$.    Let   $N\in \IN_{2}$  denote the population
size.   Asymptotics will be understood to hold as $N\to \infty$, unless
otherwise noted. 

Write $[n] := \{1,2,\ldots, n\}$ for any natural number $n \in
\IN_{2}$; unless stated otherwise  the sample size will be represented by $n$.

Given two positive sequences  ${(x_{i})}_{i\in \IN}$ and  ${(y_{i})}_{i\in
\IN}$ with $(y_{i})$ bounded away from zero   write
\begin{displaymath}
x_{i} \sim y_{i}
\end{displaymath}
if $x_{i}/y_{i} \to 1$ as $i \to \infty$.  When  $x_{i}/y_{i} \to c$
as $i\to \infty$  where $0 < c < \infty$ is fixed  we write
\begin{equation}
\label{eq:simc}
x_{i} \overset{c}{\sim} y_{i}
\end{equation}
In~\eqref{eq:simc}  $c$ is not specified,  and will change
depending on what $(x_{i})$ and $(y_{i})$ are each time; when we use
~\eqref{eq:simc} we are empasizing the conditions under which
~\eqref{eq:simc} holds with $(x_{i})$ and $(y_{i})$ as given. 
 %
For positive functions $f,g$ such that
$\limsup_{x\to \infty} f(x)/g(x) < \infty$ we write
$f=  \mathcal{O}(g)$.

For any real number $a$,   and  any  $i \in \IN_{0}$, 
we will write  ${(a)}_0 := 1$,  and 
\begin{equation}
\label{eq:10}
{(a)}_i  := a(a-1)\cdots (a-i + 1).
\end{equation}
For a  given condition/event $E$ we define 
\begin{equation}
\label{eq:27}
\one{E} := 1
\end{equation}
if  $E$ holds, and take $\one{E} = 0$
otherwise.

Let $0<p\le 1$ and $a,b > 0$  all be  fixed.  We will write
\begin{equation}
\label{eq:28}
  B(p;a,b) := 
\int_{0}^{1}\one{0<t\le p}t^{a-1}{(1-t)}^{b-1}{\rm d}t 
\end{equation}
It then holds that  $B(1;a,b) = B(a,b) :=
\int_{0}^{1}t^{a-1}{(1-t)}^{b-1}{\rm d}t = \Gamma(a)\Gamma(b)/\Gamma(a+b)$.

A $\Lambda$-coalescent is a continuous-time Markov chain on the
partitions of $[n]$ such that a given group of  $j \in
\{2,3,\ldots,k\}$  blocks for all $k\ge 2$  merges at
the rate ($a \ge  0$ fixed)
\begin{equation}\label{lambdacoalescentrates}
b_{k,j} =    a\one{j=2} +    \int_0^1 x^{j-2}{(1-x)}^{k-j}\Lambda_{+}(\mathrm{d}x), 
\end{equation}
where $\Lambda_{+}$ is a finite measure on $(0,1]$, and no other
transitions are possible. 
\end{defn}

 We formally define the (complete)   Beta-coalescent~\citep{schweinsberg03}.
\begin{defn}[The (complete) Beta coalescent]\label{betacoal}
The Beta-coalescent with parameter $\alpha$ is a 
$\Lambda$-coalescent in which, if there are currently $k\ge 2$ lineages
ancestral to the sample, the rate at which a particular subset of
$j\in \set{2,3,\ldots,k}$
of them coalesces into a single lineage is obtained from
~\eqref{lambdacoalescentrates}  by setting $a = 0$,  and,   for $0< x
\le 1$, 
\begin{equation}
\label{betadens}
\Lambda_{+}(\mathrm{d}x)=\frac{1}{B(2-\alpha,\alpha)}x^{1-\alpha}(1-x)^{\alpha-1}\mathrm{d}x,
\quad 1 < \alpha < 2.
\end{equation}
\end{defn}%
We will restrict to $1 < \alpha < 2$ (see~\citep{Eldon2026} for
Beta-coalescents with $0 < \alpha \le  1$).   In what follows, we shall
restrict our attention to a population as defined in
Definition~\ref{Schwm}. 
Beta-coalescents describe
the random ancestral trees of samples drawn from a population evolving
according to following model  in which the law of the
random family sizes has a specific tail behaviour.


\begin{defn}[The Schweinsberg model \citep{schweinsberg03}]\label{Schwm}%
Consider a haploid panmictic population of fixed size $N$ which
evolves in discrete (non-overlapping) generations.  In each
generation, each individual, independently, produces a random number
of potential offspring according to a given law.  From the pool of
potential offspring $N$ of them are then sampled uniformly at random
and without replacement to survive to maturity and replace the
parents.  If there are fewer than $N$ potential offspring from which
to sample, we suppose that the population is unchanged over the
generation (all the potential offspring perish before reaching
maturity).
\end{defn}%

In Definition~\ref{Schwm} we do not specify a law for the
potential offspring.  The key idea is that the reproducing individuals
independently contribute random  numbers of  potential offspring  according to the same nontrivial
law (i.e.\ the offspring numbers are independent and 
identically distributed; abbreviated i.i.d.).
Since we are only interested in models in which the total number of
potential offspring produced at the same time exceeds $N$
 with high probability (the mean number of
potential offspring produced by any given individual exceeds 1) the
specification of what happens when there are fewer than $N$ potential
offspring will be completely irrelevant.

\begin{defn}[The coalescence probability $c_{N}$]%
\label{cN}%
Define $c_N$ as the probability of the event that two individuals,
chosen uniformly at random without replacement and at the same time
from a given population derive from the same individual present in the
previous generation.
\end{defn}
In Definition~\ref{cN} the subscript $N$ stands for population size,
and it will be assumed that $c_{N} > 0$ for all $N \ge 2$.   The
quantity $c_{N}$ determines the timescale on which one views gene
genealogies in a large population. Morever, $c_{N}$ is related to the
effective size.
\begin{defn}[Effective population size]\label{def:effectivesize}
  The  effective population  size  $N_{e}$ is 
\begin{displaymath}
N_{e} := \frac{1}{c_{N}}
\end{displaymath}
\end{defn}
  
When the population evolves according to Definition~\ref{Schwm} with law as
in~\eqref{eq:19} for the number of potential offspring
$N_{e} \overset{c}{\sim} N^{\alpha - 1}$ as $N\to \infty$  when $1 < \alpha < 2$
\citep{schweinsberg03}.   An ancestral process
$\set{\xi^{n,N}(r); r \in \mathds N_{0}}$  is a Markov chain on the
partitions of  $[n]$
tracking the random ancestral relations of sampled gene copies from  a
finite population~\citep[][Section~1.2]{schweinsberg03}.  
\begin{thm}[\cite{schweinsberg03}, Theorem~4]\label{schweinsberg theorem}
Suppose that a haploid population of size $N$ evolves according to
Definition~\ref{Schwm}. Let $X$ be the random number of 
potential offspring  
produced by an arbitrary  individual so that~\eqref{eq:19} holds
for  $\alpha>0$  and a normalising constant $C>0$.
  Then
$\set{\xi^{n,N}(\lfloor t/c_N\rfloor), t\geq 0}$ converges, in the sense of
convergence of 
finite-dimensional distributions,  as $N\to\infty$
to a process whose law if $\alpha\geq 2$ is that of 
the Kingman  coalescent 
restricted to $[n]$ and if $\alpha\in [1,2)$ is
that of the Beta-coalescent (Definition~\ref{betacoal})  
 restricted to $[n]$.
\end{thm}%
Under the assumptions of Theorem~\ref{schweinsberg theorem}, if
$0 < \alpha<1$ the ancestral process converges to a discrete time
$(c_{N} \overset c \sim 1)$
analogue of a multiple-merger  coalescent \citep[Theorem~4d]{schweinsberg03}.   We do not
consider this scenario here.

\section{Mathematical results}\label{results}%

In this section we present the main mathematical results,
Propositions~\ref{pr:CNversion0},~\ref{errorkc},
~\ref{pn:Lambda-convergence 1<a<2 full version}, and
Lemma~\ref{lemma:ksamplesize}.     We start by
introducing an extension of~\eqref{eq:19}  behind the
main results.

\subsection{The extended Schweinsberg  model}\label{model}%

We will be concerned with a population  as described  in
Definition~\ref{Schwm}. 
We will consider an extension  of~\eqref{eq:19} for the law of the
number of potential offspring where there is an upper bound 
on  the number of potential  offspring produced by each 
individual.  
 Write   $X_i^N$ for the
random number of potential offspring produced 
in a given generation  by the $i$th 
individual (arbitrarily labelled).  Let 
 $X^N$ be  the random number of potential offspring of an  arbitrary
individual. Suppose  that $1<\alpha<2$ (see Remark~\ref{kingmag2}),   $\psi(N)$ is  a positive function of $N$, and 
there exist non-negative bounded functions  $f$ and $g$  on $\IN$,
both independent of $N$, such  that 
for all  $k \in  [\psi(N)]$  and $N \in \IN_{2}$ 
\begin{equation}
\label{eq:PXiJ}
 \left( \frac{1}{k^\alpha} - \frac{1}{ {(k+1)}^\alpha} \right) g(k)
\le \prob{X^N = k} \le  \left( \frac{1}{k^\alpha} - \frac{1}{ {(k+1)}^\alpha} \right) f(k).
\end{equation}
Any remaining mass is assigned  to $\left\{X^N = 0\right\}$.  
The 
quantity $\psi(N)$ in~\eqref{eq:PXiJ} is  such that 
$\prob{X^N \le \psi(N)} = 1$ for each $N\in \IN$.  
Since   $\prob{X^{N} = k} = \prob{X^{N} \ge k} - \prob{X^{N}
\ge k+1}$ we see that   $\prob{X^{N} = \psi(N)} =
\prob{X^{N}\ge \psi(N)}$ since $\prob{X^{N} \ge \psi(N) + 1} = 0$ by assumption.
  We will identify conditions on $f$ and $g$ in~\eqref{eq:PXiJ}  such that the ancestral process
will converge to a non-trivial limit.    Equivalently, we
can define a random variable   $X^{(\infty)} \equiv X^{N}\one{X^{N} \le \psi(N)}
+ \one{X^{N} > \psi(N)}$,  and take $X_{1}^{N},\ldots, X_{N}^{N}$  as independent
copies of $X^{(\infty)}$ for all $N\ge 2$.   Working with the mass function
of the form $\IP [X^{N} = k] = (k^{-\alpha} - {(1+k)}^{-\alpha})h(k)$
for a suitable  $h$ should lead to similar  results.


Let  $g_\infty$,
$f_\infty$, and $m_\infty$ be  positive constants such that 
\begin{equation}%
\label{gfinf}%
\begin{split}
      \lim _{k\to \infty}g(k) & = g_\infty >0
,\\
\lim_{k\to \infty}f(k) & = f_\infty >0
,
\end{split}\end{equation}%
 and, as $N \to \infty$, 
\begin{equation}
\label{eq:4}
 m_N := \EE{X^N}   \to m_\infty > 1.
\end{equation}
We take   $f$ and $g$ in~\eqref{eq:PXiJ}  such that $\EE{X^{N}} > 1$ for all $N$.   
We  define, for $k \in  \IN$, 
\begin{equation}%
\label{supinff}%
\begin{split}%
\overline{f}(k) & := \sup_{i \geq k} f(i),
\\
\underline{f}(k) & := \inf_{i \geq k} f(i),
\end{split}%
\end{equation}%
with $\overline{g}(k)$ and $\underline{g}(k)$ similarly defined.
Then, for all   $b \in \{1,\ldots, \psi(N)\}$ and $N\ge 2$,   
\begin{equation}
\label{eq:23}
\left( \frac{1}{b^\alpha} -   \frac{1}{{(\psi(N)+1)}^\alpha}  \right)\underline{g}(b) \le  \prob{X^N \ge b } \le  \left( \frac{1}{b^\alpha} -   \frac{1}{(\psi(N)+1)^\alpha}  \right)\overline{f}(b)
\end{equation}

As $g_\infty >0$ by assumption (see~\eqref{gfinf}), $g$ can be chosen to ensure $\underline{g}(i)
>0$ for all $i$ and we will assume this throughout.  From 
~\eqref{eq:PXiJ}  we see that, with $1 < \alpha < 2$,  for all $N\ge 2$, 
\begin{equation}
\label{mNbound}
 \underline{g}(1)\left( 1 + \frac{2^{1-\alpha}} {\alpha - 1} + \OO{
 \psi{(N)}^{1-\alpha} } \right)  \le m_N \le
  \overline{f}(1) \left( 1 + \frac{1 } {\alpha - 1} + \OO{
 \psi{(N)}^{1-\alpha} } \right).
\end{equation}
It is clear from~\eqref{mNbound} that if $\underline{g}(1)$ is big
enough then $m_N$ will be greater than 1.  In Section~\ref{tmcp} we
discuss the units of time associated with the  coalescents  derived from
our model.

\subsection{Timescale and coalescents}%
\label{tmcp}

An important quantity for us will be, for each $N\ge 2$, 
\begin{equation}
\label{fullSN} 
S_N :=\sum_{i=1}^{N}X_i^N
\end{equation}
the random total number of potential offspring  produced in any given generation.
  For stating the timescaling
associated with our model in~\eqref{eq:PXiJ} it will be
convenient to define (cf.\ Lemma~6 in~\cite{schweinsberg03}; recall
$\one{E}$ from~\eqref{eq:27} in
Definition~\ref{def:standardnotation}), for each $N\ge 2$, 
\begin{equation}
\label{curlyc}
{\mathcal C}_N =N\IE\left[\frac{X_1^N \left(X_1^N - 1\right)}{S_N^2}\one{S_N \ge N}\right].
\end{equation}%
In Lemma~\ref{lemma:Schw6} we prove that ($c_N$ is defined in
Definition~\ref{cN}), for all $N\ge 2$, 
\begin{equation}
\label{cncn}
c_N = \C \left(1 + \cO\left(\frac{1}{N}\right)\right).
\end{equation}


\begin{remark}\label{kingmag2}%
For $\alpha \ge 2$ we recover the Kingman coalescent from 
~\eqref{eq:PXiJ} regardless of $\psi(N)$.   The effect of large
sample size on the approximation of the Kingman coalescent to gene
genealogies of samples from a population evolving according to the
Wright-Fisher model has been investigated to some extent 
\citep{WT2003,Melfi2018b,Melfi2018,BCS2014,Fu2006}.  Our focus is on
sweepstakes reproduction  and $\Lambda$-coalescents. Unless otherwise
stated, Definition~\ref{Schwm} is in force  with
numbers of potential offspring produced according to~\eqref{eq:PXiJ}
restricted to $1 < \alpha < 2$.
\end{remark}%

Proposition~\ref{pr:CNversion0} identifies ${\mathcal C}_N$
(see~\eqref{curlyc}) with error estimates given that
Definition~\ref{Schwm} and~\eqref{eq:PXiJ} are in force
(Remark~\ref{kingmag2}).  Section~\ref{sec:convergencealpha12}
contains a proof of Proposition~\ref{pr:CNversion0}.

\begin{propn}%
\label{pr:CNversion0}%
Suppose a haploid population evolves according to
Definition~\ref{Schwm} and~\eqref{eq:PXiJ}   with $1
< \alpha < 2$.   Let $K > 0$ be a constant,  recalling~\eqref{gfinf} suppose   $f_{\infty} = 
g_{\infty}$,  recall  $\overline{f}, \underline{g}$ from
~\eqref{supinff},  $m_{N}$ from~\eqref{eq:4},   and $\mathcal{C}_{N}$ from~\eqref{curlyc}.   
In each of  the   statements    below, 
$L$ is a function of $N$, over which one can optimise
for any specific choice of
model (see Section~\ref{sec:convergencealpha12}, e.g.\
Lemma~\ref{lm:newexp1}); the quantities $ {\sf E}_{i}^{N}$ denote
the errors in our estimates.
\begin{enumerate}
\item~\label{JCD1psi<N 0}
Suppose that  $\psi(N)/N \to 0$, and
$L(N)/\psi(N)\to 0$.  Then,  
\begin{displaymath}
\frac{N}{\psi(N)^{2-\alpha}}\mathcal{C}_N
  = \frac{\alpha f_\infty}{(2 - \alpha)m_N^2}
 +  {\sf E}_{1}^{N}
\end{displaymath}
where (recall~\eqref{eq:6} and Remark~\ref{rm:eta}   with $\eta = (\alpha + 1)/(2\alpha)$)
\dsplit{
 {\sf E}_{1}^{N}
&
=    \mathcal O\left(   \left(  \frac{L }{\psi(N) }
\right)^{2-\alpha}  \right) 
+  \mathcal O \left(  \frac{\psi(N)}{N}  \right)
  +
    \mathcal O\left( N^{ \frac{1 - \alpha }{2\alpha}}\right)   +
    \mathcal O\left( N^{\frac{1-\alpha }{2}} \right)
.\\ }%
\item~\label{JCD1psi=N 0}
Suppose that  $\psi(N)/N \to K$, and
$L(N)/N\to 0$.  Then,  
\begin{displaymath}
N^{\alpha-1} \mathcal{C}_N  =   f_\infty \frac{\alpha }{m_N^\alpha} \int^1_{\frac{m_N}{K+m_N}} y^\alpha (1- y)^{1- \alpha} \mathrm{d}y +   {\sf E}_{2}^{N}
\end{displaymath}
where  (recall~\eqref{eq:6} and Remark~\ref{rm:eta}  with $\eta = (4-\alpha)/3$)
\begin{displaymath}
  {\sf E}_{2}^{N} =  
  \mathcal O \left(  \int_{\frac{m_{N} }{m_{N} + L/N  } } ^{1}  y^{\alpha - 1}(1-y)^{1-\alpha} dy  \right)
+  \mathcal O\left( \frac{L^{1-\alpha} - 1 }{N^{2 - \alpha}}\right)
+  \mathcal O\left( {N^{\frac{1-\alpha}{3}}}\right) +\mathcal O\left(  {N^{ \frac{(\alpha - 1)(\alpha - 3) }{3} }}\right)
.
\end{displaymath}
\item\label{JCD1psi>N 0}
Suppose that  $\psi(N)/N \to \infty$ and
$L(N)/N\to 0$.  Then, 
\begin{displaymath}
N^{\alpha-1} \mathcal{C}_N =   f_\infty \alpha m_N^{-\alpha}  B(2-\alpha,\alpha)  +   {\sf E}_{3}^{N} 
\end{displaymath}
where (recall \eqref{eq:6} and Remark~\ref{rm:eta}  with $\eta = 2/(1+\alpha)$)
\begin{displaymath}
 {\sf E}_{3}^{N}  {=}
   \mathcal O\left( \int_{\frac{m_{N} }{m_{N} + L/N  } } ^{1}  y^{\alpha - 1}(1-y)^{1-\alpha} dy \right)
+  \mathcal O\left( \frac{1 + L^{1-\alpha}}{N^{2-\alpha}} \right)
+  \mathcal O\left(  { N^{\frac{1 - \alpha }{1+\alpha}} } \right)
. 
\end{displaymath}
\end{enumerate}
\end{propn}
The constant term in Case~\ref{JCD1psi>N 0} in Proposition
~\ref{pr:CNversion0} (that is the limit as $N\to\infty$ of
$N^{\alpha-1}\mathcal C_N$) is the same as that obtained in Lemma~13 of
~\cite{schweinsberg03} where there is no restriction on the number of
potential  offspring. Further,
$\int_{p}^{1}y^{\alpha} {(1-y)}^{1-\alpha} {d}y =
\int_{0}^{1-p}u^{1-\alpha} {(1-u)}^{\alpha}  d u$ for any $0 \le p \le 1$.


Before discussing the implications of Proposition~\ref{pr:CNversion0},
we state two more propositions, which detail the limiting coalescents
for the three cases in Proposition~\ref{pr:CNversion0}.  We identify
the limiting coalescent for fixed sample sizes and maintain control
over the errors. Section~\ref{prerrorkc} contains a proof of
Proposition~\ref{errorkc}.

\begin{propn}\label{errorkc}
Under the conditions of  Proposition~\ref{pr:CNversion0} Case (\ref{JCD1psi<N 0}), the
scaled ancestral process
$\set{\xi^{n,N}(\lfloor t/c_N\rfloor), t\geq 0}$ for  a
sample of size $n$ converges 
  to the  Kingman
coalescent restricted to $\{1,\ldots,n\}$.  Moreover (recall
~\eqref{eq:10} in 
Definition~\ref{def:standardnotation}),  in a large population
\begin{equation}
\label{eq:Kingman1}
 \frac{N}{c_N} \IE\left[\frac{ {(X_1^N)}_3}{S_N^3} \one{S_N \geq N}
 \right] = \mathcal O\left(  \frac{\psi(N)}{N}\right)
, 
\end{equation}
\begin{equation}
\label{eq:Kingman2}
 \frac{N^2}{c_N} \IE\left[\frac{ {(X_1^N)}_2 {(X_2^N)}_2}{S_N^4} \one{S_N
 \geq N} \right]  = \mathcal O\left( \frac{ {\psi(N)}^{2-\alpha}}{N} \right)
. 
\end{equation}
\end{propn}%
We now specify the limiting coalescent   for the cases in which
multiple-merger coalescents are
obtained. Section~\ref{convergence_to_lambda_proofs}
 contains a proof of Proposition
~\ref{pn:Lambda-convergence 1<a<2 full version}.  
\begin{propn}\label{pn:Lambda-convergence 1<a<2 full version}
 Under the conditions of  Proposition~\ref{pr:CNversion0} Case~\ref{JCD1psi=N 0}, 
the scaled ancestral process $\set{\xi^{n,N}(\lfloor t/c_N\rfloor), t\geq 0}$
corresponding to a sample of size $n$ converges 
to a process whose law is given by a $\Lambda$-coalescent (see~\eqref{lambdacoalescentrates})  restricted
to $\{1,\ldots ,n\}$ and without an atom at zero (corresponding to
$a=0$ in~\eqref{lambdacoalescentrates}), with  $m_\infty$  defined
in~\eqref{eq:4} and $K > 0$ fixed. The $\Lambda_{+}$ measure in
~\eqref{lambdacoalescentrates} is given by 
\begin{equation}
\label{incbeta}
\Lambda_{+}(\mathrm{d}x)= \one{0 < x\leq \frac{K}{m_\infty+K}}  \frac{1}{B\left(\frac{K}{m_\infty+K}; 2-\alpha,\alpha \right)}
x^{1-\alpha}
{(1-x)}^{\alpha-1}\mathrm{d}x,
\end{equation}
where we have used the notation  (recall $B(p;a,b)$ from~\eqref{eq:28})
\begin{displaymath}
B\left(\tfrac{K}{m_\infty+K}; 2-\alpha,\alpha\right) :=
\int_0^{1} \one{0 < x \le  \frac{K}{m_\infty+K}   }   x^{1-\alpha}
{(1-x)}^{\alpha-1}\mathrm{d}x
\end{displaymath}

{Under the conditions of}  Proposition~\ref{pr:CNversion0} Case~\ref{JCD1psi>N 0},
the scaled 
ancestral process $\set{\xi^{n,N}(\lfloor t/c_N\rfloor), t\geq 0}$
corresponding to a sample of size $n$ converges  
to a process whose law is given by a $\Lambda$-coalescent  restricted
to $\{1,\ldots ,n\}$, with the $\Lambda_{+}$-measure given in~\eqref{betadens}.
Moreover, in Cases~\ref{JCD1psi=N 0} and~\ref{JCD1psi>N 0}, as $N\to \infty$, 
\begin{equation}%
\label{eq:Lambdaeq1}%
\frac{N^2}{c_N} \IE\left[\frac{\braces{X_1^N}_2 \braces{X_2^N}_2}{S_N^4} \one{S_N \geq N}
\right] = \mathcal O\left(  N^{1-\alpha} \right)
.
\end{equation}%
\end{propn}%
\begin{remark}\label{randomtrunc}%
Whenever $\alpha$ is estimated to be close to  1 it may be difficult to
recover the  observed amount of  genetic variation used to estimate
$\alpha$ for the  given sample  with the timescale  of the
multiple-merger coalescents we are considering  (recall $N^{\alpha-1}c_{N} \overset c\sim
1$,  Case~\ref{JCD1psi=N 0} of Proposition~\ref{pr:CNversion0}).  One
way  to overcome  this is 
to  take $\psi(N)$ as  random.  Suppose ${(\psi_{1,N})}_{N\in \IN}$ and
${(\psi_{2,N})}_{N\in \IN}$ are sequences of positive numbers, where
 $\psi_{1,N} / N \to 0$, and $\psi_{2,N} / N \gneqq 0 $. 
It is  understood that if $x_{n}/y_{n} \gneqq 0$ for some given positive
sequenches  $(x_{n})$  and    $(y_{n})$,     then either  $x_{n}/y_{n}$
converges to a strictly positive  constant, or   $x_{n}/y_{n}$
diverges.     Take, for all
$N\ge 2$, 
\begin{displaymath}
\begin{split}
  \prob{\psi(N) = \psi_{1,N} } = & 1 -{p}_{N}, \\
 \prob{\psi(N) = \psi_{2,N} } = & p_N, \\
\end{split}
\end{displaymath}
independently in each generation. To clarify, in  every generation the
$X_1^N, \ldots, X_N^N$ are i.i.d., so $\psi(N)$ takes the same value
for all $X_i^N$ in each generation, but $\psi(N)$ may vary between
generations as just described.  If $p_N = {(\psi_{1,N}/N)}^{2-\alpha} $
then $c_N \overset c \sim  \psi_{1,N}^{2-\alpha} / N$ and the
scaled ancestral process converges to a $\Lambda$-coalescent which is
a mixture between Kingman and Beta-coalescent (cf.\
~\eqref{lambdacoalescentrates}).  This is an easy consequence of the
calculations in Sections~\ref{prerrorkc} and
~\ref{convergence_to_lambda_proofs} and Proposition~\ref{prop:lambdaRule}.
         (recall Proposition~\ref{pn:Lambda-convergence 1<a<2 full
        version}).  The new timescaling (proportional to
        $N$ generations whenever  $\psi_{1,N} = O(1)$, 
         taking $\psi_{1,N} = N/\log N$ leads to $N^{\alpha
         - 1} {(\log N)}^{\alpha-2} c_{N} \overset c \sim 1$) means that lower mutation rates are required
        to explain  observed genetic diversity.
\end{remark}%

Propositions~\ref{pr:CNversion0}--\ref{pn:Lambda-convergence 1<a<2 full version}
give us an insight into the sources of the errors  in the
coalescent approximation.
An  uncertainty in the estimate of the effective size  can come
through the scaling constant  by either  assuming a wrong model for the
upper bound  $(\psi(N))$, or if  $\psi(N)$ varies considerably  over the
ancestral history of our sample.  A further source of error can come from
ignoring the error terms.

Case~\ref{JCD1psi<N 0} in Proposition~\ref{pr:CNversion0} reveals that the
timescale and some of the error terms  depend  on $\psi(N)$, even though the limiting process is
the Kingman coalescent.  Suppose $\psi(N)= N/\log N $.  Then
$\psi(N)/N \to 0$ so we would be in the domain of attraction of the
Kingman coalescent, but time would be measured in units proportional to 
$N^{\alpha - 1}{(\log N)}^{2-\alpha}$ generations, so we would be
measuring time in essentially the same way as for the Beta-coalescents.  
Thus, even when we are in the domain of attraction of the
Kingman coalescent, the effective size can be small relative to $N$.
The effective size can range from $cN$ to
$c^{\prime} N^{\alpha-1} {(\log N)}^{2-\alpha}$ for some
$c,c^{\prime} > 0$ fixed and with $1 < \alpha < 2$.  Considering the
error terms, two of them are of order ${N^{1-\alpha}}$, and
this can be arbitrarily close to $\OO{1}$.  Further, with
$\psi(N) = N/\log N$ we see that the error term  of order
$\psi(N)/N$ decreases only as $1/\log N$.

For both  Case~\ref{JCD1psi=N 0}  and Case~\ref{JCD1psi>N 0},  the leading
term decreases as $\alpha \to 1$.  The error terms of the form $N^{c(1-\alpha)}$ for some  $c > 0$ 
  can be arbitrarily  close to $\OO{1}$.     The uncertainty
in estimating  effective size can come from two directions; in Case~\ref{JCD1psi<N 0}  from uncertainty
in  determining the order of  $\mathcal{C}_N$, and  in all three cases from  the (at
least potentially) large errors.    Since  $  \int^1_{r} y^\alpha {(1- y)}^{1- \alpha}
\mathrm{d}y  \to  (1/(2-\alpha)){( 1-r
)}^{2-\alpha}$ as $r\to 1$, we see   $1/\mathcal{C}_N$ for
Case~\ref{JCD1psi=N 0} can be much larger than  $1/\mathcal{C}_N$ for
Case~\ref{JCD1psi>N 0} if $K$ is much smaller than  $m_N$.  




Propositions~\ref{errorkc} and~\ref{pn:Lambda-convergence 1<a<2 full
version} tell us that when $1 < \alpha < 2$ (see Remark~\ref{kingmag2})  it is the behaviour of
$\psi(N)/N$ (as $N\to \infty$) that determines the limiting process.
We obtain a Kingman coalescent in case $\psi(N)/N \to 0$, and a
multiple merger coalescent when $\psi(N)/N \gneqq  0$ (recall Remark~\ref{randomtrunc}), i.e.\ the
limiting measure $\Lambda$  is of the form
\begin{equation}
\label{eq:22}
\Lambda  =   \one{\tfrac{\psi(N)}{N} \to 0  }\delta_{0}    + \one{\tfrac{\psi(N)}{N} \gneqq 0} \Lambda_{+}
 \end{equation}
 where $\Lambda_{+}$ is a finite measure on $(0,1]$. 
\subsection{Increasing sample size and the Kingman coalescent}\label{mschm}

In this section we consider increasing  sample size in connection with the
extended  Schweinsberg model as given in Definition~\ref{Schwm} and
~\eqref{eq:PXiJ}  assuming $\psi(N)/N \to 0$
 (recall Proposition~\ref{errorkc}).  We will argue (informally) 
 that a sample size proportional to  $\sqrt{N/\psi(N)}$  can
be sufficient for the Kingman coalescent to be a poor approximation of
the genealogy. For the extended  Schweinsberg model $\sqrt{N/\psi(N)}$
can be much smaller than $N_{e}^{1/3}$ (recall
Definition~\ref{def:effectivesize}), the reference point for the
Wakeley-Takahashi model  \citep{WT2003,Melfi2018}.  For an example, 
compare $\sqrt{N/\psi(N)}$ and  ${(N/\psi{(N)}^{2-\alpha})}^{1/3}$ when $\psi(N) =
N/\log N$.

Recall $S_{N}$ from~\eqref{fullSN}, and~\eqref{eq:simc} in 
Definition~\ref{def:standardnotation}. 
Under the conditions of 
Proposition~\ref{errorkc},
when there are  $k$ lineages, we see simultaneous mergers at rate
 (relative to the rate of pairwise mergers; recall~\eqref{eq:Kingman2}),
\begin{displaymath}
\frac{ \tbinom{N}{2} \IE\left[\frac{\svigi{X_1^N}_2 \svigi{X_2^N}_2}{S_N^4} 
\one{S_N \geq N} \right] \tbinom{k}{4}\tbinom{4}{2} }
{N\IE\left[ \frac{\svigi{X_1^N}_2}{S_N^2} \one{S_N \geq N} \right]\binom{k}{2}}
\overset{c}{\sim} 
\frac{k^4 \psi{(N)}^{2-\alpha} \frac 1N }{k^2} =  \frac{k^{2} \psi{(N)}^{2-\alpha}  }{N }.
\end{displaymath}
  This is of order  $k^2/N_e$ (cf.\ Case~\ref{JCD1psi<N 0} of Proposition~\ref{pr:CNversion0}). 
Summing over $k = 2, \ldots, n$ gives 
 that we can expect to see  simultaneous mergers in the genealogy of a
sample of size $n$ when  $n$ is at least proportional to  $N_e^{1/3}$.

On the other hand, we see $3$-mergers at rate (relative to the rate of
pairwise mergers; recall~\eqref{eq:Kingman1})
\begin{displaymath}
\frac{N\IE\left[\frac{ \svigi{X_1^N}_3}{S_N^3} \one{S_N \geq N} \right]\binom{k}{3}}
{N\IE\left[ \frac{\svigi{X_1^N}_2}{S_N^2} \one{S_N \geq N} \right]\binom{k}{2}}
\overset{c}{\sim} 
\frac{k^3 \psi(N) \frac 1N }{k^2}=k\frac{\psi(N)}{N}.
\end{displaymath}
Summing as before  gives, that we can expect to see  3-mergers in the
genealogy when the sample size  (where $c > 0$ is fixed)
\begin{equation}
\label{eq:11}
n \ge  c\sqrt{ \frac{ N }{ \psi(N)}}
\end{equation}
Suppose   $ N^{1/(2\alpha - 1)}/\psi(N) \to 0$ and 
that  $\psi(N)/N \to 0$ (and recall $1 < \alpha < 2$). Then
$\sqrt{N/\psi(N)}/N_{e}^{1/3}\to 0$ (recall Case~\ref{JCD1psi<N 0} of  Proposition~\ref{pr:CNversion0})  so we expect to see 3-mergers at
a smaller sample size than simultanous pairwise mergers, so the number of lineages is going down one at a time
and the  summation over  $k$ from $2$ to $n$  is justified.  When
$N^{1/(2\alpha - 1)}/\psi(N) \gg 0$  then $c\sqrt{N/\psi(N)}$  is
an upper bound for the sample size at which we expect to see
3-mergers.  

If $\psi(N)$ is unbounded as $N\to \infty$   (but with $\psi(N)/N \to 0$), 
~\eqref{eq:11}   suggests that the Kingman approximation for this model will be
poor at sample sizes closer to $\sqrt{N/\psi(N)}$, which could be
substantially smaller than $(N/\psi(N)^{2-\alpha})^{1/3}$ (see Proposition
~\ref{pr:CNversion0}, Case~\ref{JCD1psi<N 0}).  As an example, take
$\psi(N) = N/\log N$  to see that a sample size
proportional to  $\sqrt{\log N }$ would then be  sufficient for a poor approximation
of the Kingman coalescent.

Suppose $\psi(N) = N$.  Then we are in the domain of
attraction of a  Beta-coalescent and  $N_{e} \overset{c}{\sim} 
N^{\alpha - 1}$ (recall Case~\ref{JCD1psi=N 0} of
Proposition~\ref{pr:CNversion0}); arguing as before then $N^{(\alpha -
1)/3}$ is an upper bound for the sample size  at which we expect to
see  simultaneous mergers.

\subsection{Increasing  sample size and  Beta-coalescents}%
\label{nLt}%

In this section we focus on increasing sample size when 
the extended Schweinsberg model (Definition~\ref{Schwm}
and~\eqref{eq:PXiJ}) is  in the domain of attraction of the Beta-coalescents 
({recall}~\eqref{betadens} and~\eqref{incbeta}).
In particular, we show that as a result of approximating sampling
without replacement by sampling with replacement we consistently
overestimate the rate of $k$-mergers (a merger of $k$ lineages) and
that the error is on the same order as the rate of mergers when $k$
 is proportional to  $\sqrt{N}$ ($k\propto \sqrt N$).  
In Section~\ref{sec:heur-argum-sample} we  provide a heuristic 
argument, based on known results for the number of blocks in the limiting
coalescent, that suggests that this should  significantly distort our
prediction of statistics based on the site frequency spectrum 
for sample sizes of order $N^{\alpha/2}$.

The result $k\propto \sqrt N $ follows from
Lemma~\ref{lemma:ksamplesize} which is similar to
Proposition~\ref{pr:CN}.  
   Equation~\eqref{eq:lemmapsi=N} in   Lemma~\ref{lemma:ksamplesize}  
shows that   one would  choose $L$ (see   Section~\ref{sec:proofksamplesize} for the role of
$L$)  so that the maximum of ${(L/N)}^{k-\alpha}$
and $k^2/L$ is as small as possible. In a similar way one would 
optimise over $\beta_{N}$.     Section~\ref{sec:proofksamplesize}  contains a
proof of Lemma~\ref{lemma:ksamplesize}.

\begin{lemma}\label{lemma:ksamplesize}
Consider a sample of $n$ individuals from a haploid panmictic
population of constant size  evolving
according to Definition~\ref{Schwm} and  
~\eqref{eq:PXiJ}  with $1 < \alpha < 2$.   Let ${(\beta_{N})}_{N\in \IN}$ be an 
increasing positive sequence.  Suppose that $n/N \to 0$,   $f_{\infty} = g_{\infty}$ with   $f_{\infty}$  and  $g_{\infty}$ from
~\eqref{gfinf}, $m_{N}$ from~\eqref{eq:4},  
$S_{N}$ from~\eqref{fullSN},  $K >0$ a constant,   and
suppose  $L$ is a function of $N$ with $L/N \to 0$.
\begin{enumerate}
\item
If    $\psi(N)/N \to K$, then for any $k = 2,3,\ldots, n$, 
\begin{equation}
\begin{split}
\label{eq:lemmapsi=N}
&   N \IE\left[
\frac{\binom{X_1^N}{k} \binom{S_N-X_1^N}{n-k}}{\binom{S_N}{n}}\one{S_N \geq N}
\right] \\
= &   N^{1-\alpha} f_\infty  \binom{n}{k} \left( 
\frac k{m_{N}^{\alpha}} \int_0^{\frac{K}{K+ m_N}} u^{k-1-\alpha} (1-u)^{n-k+\alpha} \mathrm{d}u  
   - \frac k {K^{\alpha}} \int_{0}^{\frac K{K+m_{N}} }u^{k-1}(1-u)^{n-k}{\rm d}u
   \right. 
\\
& \left. 
- \frac{n-k}{m_{N}^{\alpha}} \int_0^{\frac{K}{K+ m_N}} u^{k-\alpha}(1-u)^{n-k+\alpha-1} \mathrm{d}u
+ \frac{n-k}{K^{\alpha}}\int_{0}^{\frac K{m_{N} + K}}u^{k}(1-u)^{n-k-1}{\rm d}u  \right )\\
& \times  \svigi{
1   + \cO\left( \frac{\beta_N}{N} \right)
+ \cO\left( \left( \frac  LN \right)^{k-\alpha}\right) + \cO\left( \frac{k^{2}}{L}  \right)
 + \cO\left( \frac{ N }{ \beta_N^\alpha} \right)
}
.
\end{split}
\end{equation}
\item
If    $\psi(N)/N \to \infty$,  then for any $k = 2,3,\ldots, n$, 
\begin{equation}
\label{eq:lemmapsi>N}
\begin{split}
&  N \IE\left[
\frac{\binom{X_1^N}{k} \binom{S_N-X_1^N}{n-k}}{\binom{S_N}{n}}\one{S_N \geq N}
\right] \\
& =
N^{1-\alpha} f_\infty m_N^{-\alpha} \binom{n}{k} \svigi{
k \int_0^1 u^{k-1-\alpha} (1-u)^{n-k+\alpha} \mathrm{d}u
- (n-k) \int_0^1 u^{k-\alpha}(1-u)^{n-k+\alpha-1} \mathrm{d}u} \\
& 
\times \svigi{
1  + \cO\left( \frac{\beta_N}{N} \right)
+ \mathcal O\svigi{ \svigi{\frac N{\psi(N)}}^{\alpha} }
+ \cO\left( \left( \frac{L}{N} \right)^{k-\alpha}\right) + \cO\left( \frac{k^{2}}{L} \right)
+ \cO\left( \frac{ N}{ \beta_N^\alpha} \right)
}
.
\end{split}
\end{equation}
\end{enumerate}
\end{lemma}
Equation~(\ref{eq:lemmapsi>N}) arises naturally in our proof 
and so we have left it in this form.    However, observe that (recalling $\Gamma(x + 1) = x\Gamma(x)$)
\begin{displaymath}
\begin{split}
k\int_0^1u^{k-1-\alpha}(1-u)^{n-k \textcolor{black}{+} \alpha}\mathrm{d}u
-(n-k)\int_0^1u^{k-\alpha}(1-u)^{n-k+\alpha-1}\mathrm{d}u
\\=\frac{k(n-k+\alpha)\Gamma(k-\alpha)\Gamma(n-k+\alpha)}{\Gamma(n+1)}
-\frac{(k-\alpha)(n-k)\Gamma(k-\alpha)\Gamma(n-k \textcolor{black}{+} \alpha)}{\Gamma(n+1)}
\\
=\alpha n \frac{\Gamma(k-\alpha)\Gamma(n-k+\alpha)}{\Gamma(n+1)}
=\alpha\int_0^1u^{k-\alpha-1}(1-u)^{n-k \textcolor{black}{+} \alpha-1}\mathrm{d}u,
\end{split}
\end{displaymath}
as expected from Proposition~\ref{pn:Lambda-convergence 1<a<2 full version}.

\begin{corollary}
\label{cor:largemerger}
When there are at least $\sqrt{N}$ lineages, then for large
$N$   we will not see convergence to the 
limiting rates suggested by 
Proposition~\ref{pn:Lambda-convergence 1<a<2 full version}.
\end{corollary}
In particular, we see from the proof of Lemma~\ref{lemma:ksamplesize}  (see Section~\ref{sec:proofksamplesize})
 that the $\cO(k^2/L)$ error 
term is always  negative. Therefore, when the limiting $\Lambda$-coalescent 
has mergers of at least $\sqrt{N}$  lineages we will be seeing a 
reduced number of such mergers, and so a different topology 
in the coalescent tree corresponding to the gene genealogy of the
ancestral process.
In Section~\ref{sec:heur-argum-sample} we present  heuristic calculations to establish 
that a  sample size proportional to  $N^{\alpha/2}$ is sufficient
for there to be an $\cO(1)$
probability of seeing a merger of at least
$\sqrt{N}$ lineages  in the limiting
complete  Beta$(2-\alpha,\alpha)$ coalescent tree  (Definition~\ref{betacoal}).
However,   the site-frequency
spectrum is not distorted   by increasing sample size when the limiting
tree is that predicted by the  Beta coalescents considered here (see
Section~\ref{numerics}).

\section{Increasing sample size and relative branch lengths}\label{numerics}%

In this section we use simulations to investigate if and how
increasing sample size may distort the site-frequency spectrum (the
count of mutations of a given frequency in a sample)
relative to the predictions of a given coalescent.

\begin{figure}[htp]
\centering
\subfloat[$\alpha = 1.05$, $\psi(N) = N$]{\label{fig:resbetapsiN}\includegraphics[angle=0,scale=.7]{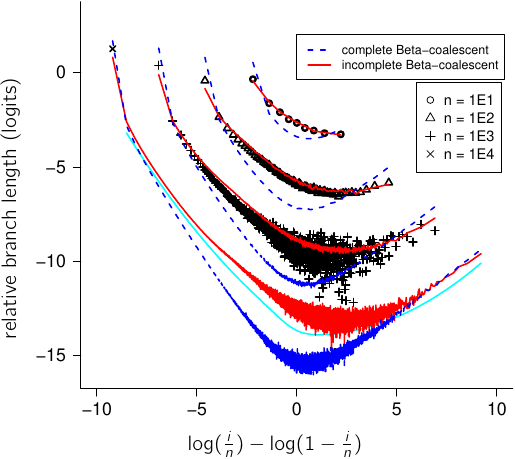}}
\subfloat[$\alpha=1.05$, $\psi(N) = \infty$]{\label{fig:resbetapsiunbounded}\includegraphics[angle=0,scale=.7]{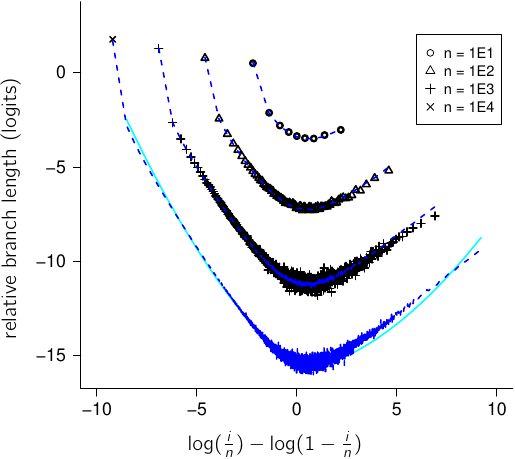}}\\
\caption{Comparing relative branch lengths - Beta-coalescents.
Comparison of $\overline \varrho_i^N(n)$ (
symbols) and of $\overline\varrho_i(n)$ (lines) shown as
$\log(e_{i}(n)) - \log(1 -e_{i}(n))$ as a function of
$\log(i/n) - \log(1-i/n)$, where $e_{i}(n)$ denotes the corresponding
mean relative branch length estimate for sample size $n$ as shown from
a haploid population of constant size $N = 10^{4}$ evolving according
to Definition~\ref{Schwm} with number of potential offspring 
distributed according to ~\eqref{eq:haploid_pxi} with
$\alpha=1.05$ and $\psi(N) = N$ (a), and
according to ~\eqref{eq:26} with $\alpha = 1.05$ (b).  The blue
dashed lines show $\overline \varrho_i(n)$ for the complete
Beta-coalescent, and the red solid lines in (a) show $\overline
\varrho_{i}(n)$ for 
the incomplete Beta-coalescent with $K=1$.  The cyan lines for the
case $n = N$ is a loess regression (using the function {\tt loess} in
{\tt R} \citep{rsystem}) through $\overline\varrho_{i}^{N}(n)$
for $i\in \{2,3, \ldots, n-1\}$; the estimates of $\EE{R_{i}^{N}(n)}$
are the results from $10^{4}$ experiments. Appendix~\ref{sec:code}
contains a brief description of an algorithm for estimating
$\EE{R_{i}^{N}(n)}$ }\label{fig:relativebranchesbetapsiN}
\end{figure}

For the numerical examples we apply a special case of the model in
~\eqref{eq:PXiJ}.  Consider a haploid population of fixed size $N$ and
evolving according to Definition~\ref{Schwm}.  
Recalling  $[n] = \set{1,2,\ldots, n}$ for all  $n\in \IN$, suppose
(recall $X_{1}^{N},\ldots,X_{N}^{N}$ are i.i.d.\ for all $N$)
\begin{equation}
\label{eq:haploid_pxi}
\prob{X_1^N = k} =  \one{k\in [\psi(N)]} \left( \frac{1}{k^\alpha} - \frac{1}{(1+k)^\alpha} \right)\frac{1}{1 - (\psi(N) + 1)^{-\alpha} }.
\end{equation}
Then the mass function in~\ref{eq:haploid_pxi} is monotone 
decreasing on $[\psi(N)]$, i.e.\
$\prob{X_1^N = k + 1} \le \prob{X_1^N = k}$.  Suppose $K > 0$ is a
constant, $\psi(N)/N \gneqq 0$ (recall Remark~\ref{randomtrunc}),  and write
\begin{equation}
\label{eq:3}
 \phi(\psi(N))  :=  \one{ \frac{\psi(N)}{N} \to K } \frac{K}{m_\infty + K}   +
   \one{ \frac{\psi(N)}{N} \to \infty},
\end{equation}
{We will}  approximate $m_{\infty}$   (see~\eqref{eq:4} and
~\eqref{mNbound}) with 
\begin{equation}
\label{eq:1}
m_{\infty} \approx  1 +  \frac{1 + 2^{1-\alpha}}{2(\alpha - 1)}
\end{equation}
Our calculations then show
that one obtains, with  $1 < \alpha < 2$ and $\phi$ as in
~\eqref{eq:3}, a Beta coalescent with $\Lambda_{+}$-measure given by
~\eqref{incbeta} with $K/(m_{\infty} + K)$ replaced with $\phi$.
We also consider an unbounded distribution of number of potential
offspring  (see
Figure~\ref{fig:resbetapsiunbounded}),
where $X$ denotes the number of potential offspring  produced by an arbitrary
individual, and
\begin{equation}
\label{eq:26}
\prob{X \ge k} = k^{-\alpha}, \quad k \in \IN
\end{equation}



\begin{figure}[htp]
\centering
\includegraphics[scale=1]{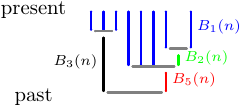}
\caption{An illustration of the functionals  $B_{i}(n)$ resp.\
$B_{i}^{N}(n)$ for an example gene genealogy; $B_{1}(n)$ is the sum of
the length of the  blue branches, and  $B_{2}(n)$ (green branch),
$B_{3}(n)$ (black branch), and
$B_{5}(n)$ (red branch) as shown. }\label{fig:illB}
\end{figure}

To describe the simulation results we need to define notation.  Let
$B_{i}(n)$ denote the random total length of branches supporting
$i \in [n-1]$ leaves when the sample size is $n$ and the gene
genealogy of the sample is described by a given coalescent
$\set{\xi^{n}}$; let $B_{i}^{N}(n)$ denote the corresponding quantity
for a given ancestral process $\set{\xi^{n,N}}$.  With $\#A$ denoting
the number of elements in a given (finite) set $A$, for all
$i\in [n-1]$,
\begin{subequations}
\begin{align}
\tau(n) & \equiv \inf\set{t \ge 0 : \#\xi^{n}(t) = 1}, \quad \tau^{N}(n) \equiv   \inf\set{t \ge 0 : \#\xi^{n,N}(t) = 1}, \label{eq:32} \\
B_{i}{(n)} &  \equiv \int_{0}^{\tau(n)} \#\set{\xi\in \xi^{n}(t) : \#\xi = i} dt, \quad B(n) \equiv \int_{0}^{\tau(n)}\#\xi^{n}(t) dt \label{eq:33} \\
B_{i}^{N}(n) & \equiv  \sum _{r=0}^{\tau^{N}(n)} \#\set{\xi \in \xi^{n,N}(r) : \#\xi = i }, \quad B^{N}(n) \equiv \sum_{r=0}^{\tau^{N}(n)}\#\xi^{n,N}(r) \label{eq:34}
\end{align}
\end{subequations}
An example gene genealogy  illustrating  $B_{i}(n)$ resp.\ $B_{i}^{N}(n)$ is given in
Figure~\ref{fig:illB}.  

The $B_{i}^{N}(n)$ are measured in generations. 
Write, for $i \in [n-1]$,   
 \begin{equation}
 \label{eq:8}
 \begin{split}
 R_i(n)  :=
 \frac{ B_i(n) }{B_1(n) + \cdots + B_{n-1}(n)}, \quad 
 R_i^N(n)  :=
 \frac{B_i^N(n)}{B_1^N(n) + \cdots + B_{n-1}^N(n)} 
\end{split}
\end{equation}
The quantities $R_i(n)$ and $R_i^N(n)$ are well defined, since $B_1^N(n) \ge
n$, and  $B_1(n) > 0$ a.s.  Write $\overline \varrho_{i}(n)$
resp.\ $\overline \varrho_{i}^{N}(n)$ for the  means of
independent realisations of 
$R_{i}(n)$ resp.\ $R_{i}^{N}(n)$.

We will compare corresponding  $\overline\varrho_{i}^{N}(n)$ and
$\overline \varrho_{i}(n)$.  The $\overline\varrho_i^N(n)$ are
obtained by sampling $\set{\xi^{n,N}}$ of a  population evolving
according to Definition~\ref{Schwm} with numbers of potential
offspring distributed according to~\eqref{eq:haploid_pxi}
or~\eqref{eq:26}.  The empirical means $\overline \varrho_{i}^{N}(n)$
(and $\overline \varrho_{i}(n)$)  are obtained from a finite number of
simulated gene genealogies; see Appendix~\ref{sec:code} for  a
brief description of the  algorithm for computing
$\overline\varrho_i^N(n)$.

\begin{figure}[htp]
\centering
\subfloat[$\alpha = 3$, $\psi(N) = N$]{\label{fig:reskingNe4a3psiN}\includegraphics[angle=0,scale=.7]{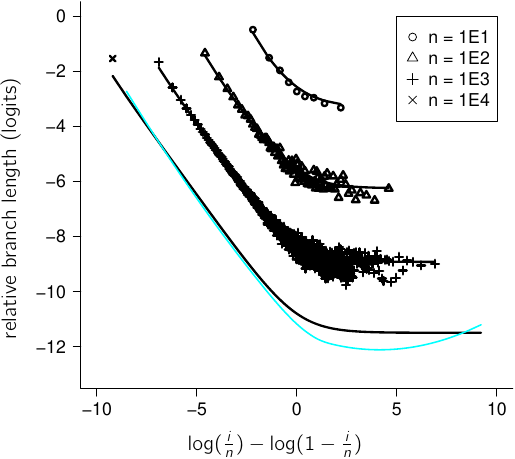}}
\subfloat[$\alpha = 1.05$, $\psi(N) = N/\log N$]{\label{fig:reskingNe4psiNoverlogN}\includegraphics[angle=0,scale=.7]{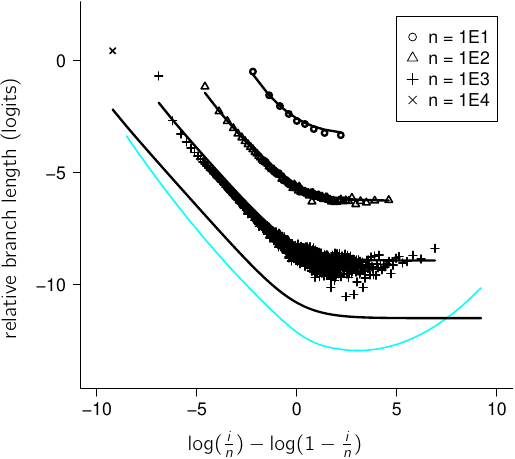}}
\caption[Large sample size does not distort the site-frequency
spectrum]{Comparing relative branch lengths --  the Kingman coalescent.    Comparison of
$\overline \varrho_{i}^{N}(n)$ (symbols) and 
 of  $\overline \varrho_{i}(n)$ (black lines)  shown as  $\log(e_{i}(n)) - \log(1 -e_{i}(n))$ as
 a function of   $\log(i/n) - \log(1-i/n)$  where $e_{i}(n)$ denotes
 the corresponding mean relative branch length  estimate       for sample size $n$ as
 shown from a haploid panmictic population of constant size 
$N = 10^{4}$   evolving according to 
Definition~\ref{Schwm}  with number of potential offspring  distributed according to
~\eqref{eq:haploid_pxi} with  $\alpha$ and $\psi(N)$ as shown.
The black lines are the approximation
$\EE{B_{i}(n)}/\EE{B(n)} = i^{-1}/\sum_{j=1}^{n-1}j^{-1} $ of
$\EE{R_{i}(n)}$ predicted by the Kingman coalescent \citep{F95}. The
cyan lines for the case $n = N$ is a loess regression (using the
function {\tt loess} in {\tt R} \citep{rsystem}) through
 $\overline \varrho_{i}^{N}(n)$ for $i\in \{2,3, \ldots, n-1\}$; the estimates
of $\EE{R_{i}^{N}(n)}$ are the results from $10^{4}$ experiments.
Appendix~\ref{sec:code} contains a brief description of an algorithm
for computing $\overline \varrho_{i}^{N}(n)$  estimating  $\EE{R_{i}^{N}(n)}$ }\label{fig:indomainkingman}
\end{figure}

In Figure~\ref{fig:indomainkingman} we compare
$\overline\varrho_{i}^{N}(n)$  (symbols)   and  $\overline
\varrho_{i}(n)$  (black lines) when  $\set{\xi^{n,N}}$ 
 is in the domain of attraction of the
Kingman-coalescent.  In Figure~\ref{fig:reskingNe4a3psiN} we have
taken $\psi(N) = N$ and $\alpha = 3$ so the effective size (recall
Definition~\ref{def:effectivesize}) is proportional to $N$.  In
Figure~\ref{fig:reskingNe4psiNoverlogN} we have taken
$\psi(N) = N/\log N$ and $\alpha = 1.05$, so the effective size is
proportional to $N^{\alpha - 1}(\log N)^{2-\alpha}$ (recall
Case~\ref{JCD1psi<N 0} of Proposition~\ref{pr:CNversion0}), and
$N^{\alpha - 1}(\log N)^{2-\alpha} \approx 13$ for $N=10^{4}$ and
$\alpha = 1.05$.  The agreement between
{$\overline \varrho_{i}^{N}(n)$ and $\overline \varrho_{i}(n)$
} depends on the relation between the sample size and the effective
size when the ancestral process is in the domain of attraction of the
Kingman-coalescent.  This is in line with results based on the
Wright-Fisher model \citep{WT2003,Melfi2018}.

In Figure~\ref{fig:relativebranchesbetapsiN} we compare 
$\overline \varrho_i^N(n)$ (symbols)  and $\overline \varrho_i(n)$ (blue and
red lines)  when { $\set{\xi^{n,N}}$}  is in the domain of attraction of  the
incomplete Beta-coalescent (\eqref{eq:haploid_pxi}; Figure~\ref{fig:resbetapsiN})  and  the
complete Beta-coalescent (\eqref{eq:26}; Figure~\ref{fig:resbetapsiunbounded}).  In
Figure~\ref{fig:relativebranchesbetapsiN}  the
blue dashed  lines are for the complete Beta-coalescent,  and the red
solid  lines in
Figure~\ref{fig:resbetapsiN} for the incomplete Beta-coalescent with
$K=1$.  Overall  the agreement between {$\overline \varrho_i^N(n)$ and
$\overline \varrho_i(n)$}
 as predicted by the corresponding Beta-coalescent   is  good,  indicating that the site-frequency spectrum 
predicted by   our  model    is not distorted  by  increasing sample
size when  in the domain of attraction of a Beta coalescent.
Figure~\ref{fig:resbetapsiN} further shows that $\overline \varrho_{i}(n)$
 for   the complete
Beta-coalescent (blue dashed lines) do not match  $\overline \varrho_{i}^{N}(n)$
 when  $\psi(N) = N$, in contrast to the incomplete
Beta-coalescent (red solid lines).  

  In
Appendix~\ref{sec:furth-numer-exampl} we compare $\overline \varrho_{i}(n)$
 predicted by the Beta-coalescents (see Figure~\ref{figK}). The
red line in   Figure~\ref{figK} is for the complete
Beta$(2-\alpha,\alpha)$ coalescent, and the remaining ones (except the
black one is for the Kingman coalescent) are for the incomplete
Beta$(\gamma; 2-\alpha,\alpha)$
coalescent with $\gamma = K/(K+m_{\infty})$ with $K$ as shown, and
$m_{\infty}$ as in~\eqref{eq:1}.  
Figure~\ref{figK} shows  that an upper bound on
the number of potential offspring  can  markedly affect the
predicted site-frequency  spectrum;    the Beta-coalescent lines
are all for the same  $\alpha$ ($\alpha = 1.01$). 

\clearpage
\pagebreak
\newpage

\section{Conclusion}%
\label{concl}%

Our main results are \textit{(i)} the extension~\eqref{eq:PXiJ} of
~\eqref{eq:19} that should be applicable to a broad class
of study systems (see also Remark~\ref{randomtrunc}).  In particular, when
$\alpha \ge 2$ or $\psi(N)/N\to 0$ the ancestral process  is in the domain of
attraction of the Kingman coalescent.  \textit{(ii)} Restricting to
$1 < \alpha < 2$, then depending on the upper bound $\psi(N)$ (see~\eqref{eq:22}) the
limiting coalescent is a Kingman coalescent (Proposition~\ref{errorkc}), an
incomplete Beta$(2-\alpha,\alpha)$ coalescent, or the complete
Beta$(2-\alpha,\alpha)$ coalescent (Proposition~\ref{pn:Lambda-convergence
1<a<2 full version}).  \textit{(iii)} The error in the coalescent
approximation can be quite large, especially if $\psi(N)$ has a
$\log N $ term (e.g.\ \eqref{eq:Kingman1} and~\eqref{eq:Kingman2}). 
\textit{(iv)} The effective size (recall Definition~\ref{def:effectivesize}) can
be much smaller than the population size with the ancestral process  nevertheless
in the domain of attraction of the Kingman coalescent
(Case~\ref{JCD1psi<N 0} of Proposition~\ref{pr:CNversion0}), for example if
$\psi(N)$ is of the form $N/\log N$. \textit{(v)} The sample size at
which the gene genealogies of the ancestral process  start  to deviate from the
Kingman trees  can be much smaller than for the Wright-Fisher
model, especially if the upper bound is of a specific form (see
~\eqref{eq:11}). \textit{(vi)} When the number of potential offspring   are not
bounded and the ancestral process  is in the domain of attraction of
the complete Beta-coalescent one can expect deviations in the shape 
of gene genealogies from the one predicted by the complete
Beta$(2-\alpha,\alpha)$-coalescent when sample size is at least
$cN^{\alpha/2}$ (some $c> 0$ fixed;  see~\eqref{intmk}).
\textit{(vii)} The effect of increasing sample size on the
site-frequency spectrum depends on the relation between the sample
size and the effective population size when the model is in the domain
of attraction of the Kingman coalescent
(Figure~\ref{fig:indomainkingman}). Increasing sample size does not
distort the site-frequency spectrum when the model is in the domain of
attraction of a Beta-coalescent
(Figure~\ref{fig:relativebranchesbetapsiN}).  Thus, when the sample
size is large enough that the gene genealogy  should deviate from the
limiting coalescent  tree (e.g.\ Corollary~\ref{cor:largemerger}, and the
$cN^{\alpha/2}$ result from
Section~\ref{sec:heur-argum-sample}) the  site-frequency spectrum
appears not to be affected.  \textit{(viii)} The site-frequency
spectrum is sensitive to the upper bound 
(Figure~\ref{figK} in Appendix~\ref{sec:furth-numer-exampl}).  \textit{
(ix)} Conditioning on the population ancestry appears to  have an  effect on
the site-frequency spectrum for the models considered here
(Figure~\ref{fig:quenchedannealedsfsA} in
Appendix~\ref{sec:quenched}).

In our model, a potential offspring is seen as a fertilised egg.  For
  many populations it would seem
plausible that family sizes would, on average, be at most a fraction
of the population size.  By extending~\eqref{eq:19}
to include an upper bound on the number of potential offspring we
incorporate an important ecological reality, especially of broadcast
spawners. Restricting  $\alpha$ to $(1,2)$ leaves  it to  the upper
bound to 
determine  the limiting  coalescent (recall~\eqref{eq:22}).

The coalescents we have considered are obtained by ignoring the
   ancestral relations of the gene copies in the population.  {At the
time of sampling (and all previous times)  the ancestral relations of
the gene copies in a  
given population 
is fixed,}  but we derive our coalescents
by (implicitly) averaging over the relations.       Gene genealogies when the {pedigree
(ancestral relations between diploid individuals)}  is fixed
(``quenched'' or conditional gene trees) have been considered in
diploid biparental organisms evolving according to the Wright-Fisher
model, where the predictions of the quenched trees are shown to be
similar to the ones of the classical (or ``annealed'') Kingman
coalescent \citep{wakeley2012gene}.  However, when the underlying
(diploid)  population evolves according to a particular model of sweepstakes
reproduction,~\cite{Diamantidis2024} show that the coalescent time of
two gene copies, when conditioning on the {population pedigree},    does not converge to the time predicted by the
corresponding annealed  multiple-merger coalescent.  It is, therefore, of
interest to investigate conditional gene genealogies in populations
with sweepstakes reproduction.  Here we consider  haploid populations. 
  In 
    Appendix~\ref{sec:quenched} we use simulations to  investigate  if   predictions of
    genetic variation will change   when  conditioning on
    {the population ancestry, the ancestral relations between the  gene
    copies in the population at all times}.    
    Figure~\ref{fig:quenchedannealedsfsA} records  an example  comparing
   $\overline \varrho_{i}^{N}(n)$ to estimates  of 
$\EE{ \widetilde R_{i}^{N}(n) }$, where we use 
$\EE{ \widetilde R_{i}^{N}(n)}$ to denote  the mean relative
branch lengths read  off quenched gene genealogies 
(Appendix~\ref{sec:estimatequenched} contains a  brief description of the
algorithm for computing the empirical mean  $\overline
\rho_{i}^{N}(n)$  estimating 
 $\EE{\widetilde R_{i}^{N}(n)}$).
 Figure~\ref{fig:quenchedannealedsfsA} shows   there is a noticeable difference
 between $\overline \varrho_{i}^{N}(n)$ and $\overline \rho_{i}^{N}(n)$
for the models considered here.      A   quenched multiple-merger coalescent would be required to
properly investigate the effect of increasing sample size on the
predictions of conditional gene genealogies, and this we leave to
future work, as well as the task
\citep{arnason22:_sweep,Eldon2020,Freund2022.04.12.488084} of
extending our model to include diploidy \citep{MS03,BLS15,BBE13}, many
chromosomes, and complex demography along the lines of  
~\cite{Koskela2019} and~\cite{freund2020cannings}.

\section{Key Lemmas}%
\label{key lemmas}%

In this section  we briefly  recall  key lemmas we will need.   Throughout we
will be considering a haploid {panmictic}  population of
constant size $N$ evolving as in Definition~\ref{Schwm} and
~\eqref{eq:PXiJ} {(recall Remark~\ref{kingmag2})}.    Our model is easily recast as a
Cannings model \citep{C74,Cannings1975}, enabling us to make use of
~\cite{MS01}.
We fix notation by describing the Cannings model for reproduction in a
haploid panmictic  population of constant size $N$  evolving in
discrete generations. 
{Write}  $\nu_k$ for the random  number of
\emph{surviving} offspring of the $k$th individual {in an arbitrary}
generation, $\underline{\nu}\equiv (\nu_1,\nu_2,\ldots ,\nu_N)$ is an
exchangeable random vector with $\sum_{k=1}^N\nu_k=N$ for all $N\ge 2$.  The vectors
$\underline{\nu}$ are assumed to be i.i.d.\  across generations.

Recall  $c_N$ from  Definition~\ref{cN}. 
For the Cannings model, and for all $N\ge 2$, it holds 
\begin{equation}
\label{def of cN}
c_N  =   \frac{\IE[\nu_1(\nu_1 - 1)]}{N - 1}.
\end{equation}

For ease of reference we record general criteria for the convergence of the
corresponding rescaled ancestral processes.  Throughout, the
convergence (as $N\to \infty$) of the pre-limiting (time rescaled)
ancestral process $\set{\xi^{N,n}(\lfloor t/c_{N} \rfloor ); t \ge 0}$
to a coalescent $\set{\xi^{n}}$ is in the sense of convergence of
finite-dimensional distributions.  The first criterion,  due to
~\cite{Mhle2000}, guarantees convergence to the  Kingman coalescent.
Recall~\eqref{eq:10} in Definition~\ref{def:standardnotation}.
\begin{propn}[~\cite{Mhle2000}, Section~4]\label{prop:kingmanRule}
Recall $c_{N}$ from~\eqref{def of cN}.  Suppose  
\begin{equation}%
\label{no triples}%
\lim_{N\to\infty}\frac{\IE [{(\nu_1)}_3]}{N^2c_N}=0.
\end{equation}
Then $c_{N}\to 0$,  and for each  finite sample size  $n\in\set{2,3,\ldots}$ 
 the rescaled ancestral processes
$\set{ \xi^{n,N}(\lfloor t/c_N\rfloor), t\ge 0}$ converge 
 to the Kingman coalescent restricted to $\{1,\ldots,n\}$.
\end{propn}
If the family-size distribution,  on the timescale determined by
$c_N$, admits `large families' (comprising a non-trivial proportion of the total population) then we recover 
a coalescent with multiple mergers. Proposition~\ref{prop:lambdaRule}   follows from Theorem~3.1 of~\cite{S99}.
\begin{propn}[\cite{S99}]\label{prop:lambdaRule}
Suppose $c_{N}\to 0$ as $N\to \infty$  and that 
\begin{equation}
\label{no simultaneous}
\lim_{N\to\infty}\frac{\IE [{(\nu_1)}_2 {(\nu_2)}_2]}{N^2c_N} =0
\end{equation}
holds.    If there exists a probability measure $\Lambda_{+}$ on $(0, 1]$
such that for all $x \in (0,1]$
\begin{equation}
\label{eq:20}
\lim_{N\to\infty} \frac{N}{c_N}\IP[\nu_1 > Nx] = \int_x^1 y^{-2} \Lambda_{+}(dy)
\end{equation}
 then, for each finite sample size $n\ge 2$,   the time-rescaled ancestral processes 
$\set{\xi^{n,N}(\lfloor t/c_N\rfloor), t\geq 0} $ 
converge as $N\to\infty$ to a $\Lambda$-coalescent (recall~\eqref{lambdacoalescentrates})  restricted to
$\{1,\ldots, n\}$.
\end{propn}



\section{Asymptotic rate of the coalescent: calculating $c_N$}
\label{sec:convergencealpha12}%
In this section we give a proof of Proposition~\ref{pr:CNversion0}. First we set notation.
\begin{notn}\label{defn:ANandM}
Recall   $S_{N}$ from~\eqref{fullSN}. 
As we are interested in the behaviour of $(N \ge 2)$
\begin{displaymath}
N\IE\left[\frac{X_{1}^N \braces{ X_{1}^N - 1}}{S_N^2} \one{S_N \geq N}\right]
\end{displaymath}
we fix a notation for $X_{2}^{N} + \cdots + X_{N}^{N}$; we  define,
for all $N\ge 2$, 
\begin{equation}
\label{st}%
\tilde{S}_N := \sum_{i=2}^N X_i^N.
\end{equation}
{Throughout}   $(\beta_{N})_{N\in \IN}$ will denote an
increasing positive  sequence. 
    We will be partitioning over  events (recall $m_{N}$
    from~\eqref{eq:4}), for all $N\ge 2$,  
\begin{align} \label{eq:betaN}
A_N := \curly{ \tolugildi{\tilde{S}_N - (N-1) m_N } \leq \beta_N}
\end{align}
and 
\begin{equation}
\label{eq:5}
A_N^c :=  \curly{ \tolugildi{\tilde{S}_N - (N-1) m_N } > \beta_N}
\end{equation}
(recall  $\tilde S_{N}$ from~\eqref{st}), $A_{N}^{c}$ being  
 the complement of $A_N$ in~\eqref{eq:betaN}.  
Write, for all $N\ge 2$, 
\begin{equation}
\label{Mpm}
\begin{split}
M_- := & (N-1)m_N - \beta_N
,
\\
M_+ := & (N-1)m_N + \beta_N.
\end{split}
\end{equation}

\end{notn}

\subsection{Preliminary lemmas}%
\label{sec:prelimlemmas}%

\begin{lemma}\label{sec:usefulapproxs}
Suppose $G$ and $H$ are positive  functions on $[1,\infty)$, $G$
be monotone decreasing and $H$ monotone increasing, and that
$\int HG^{\prime}$ exists.  Then, for all $\ell,m\in \IN$, $\ell \le
m$, 
\begin{displaymath}
-\int_{\ell}^{m+1} H(x-1) G^{\prime}(x) {d} x \le  \sum_{k=\ell}^{m} H(k)(G(k) - G(k+1)) \le  - \int_{\ell}^{m+1} H(x)G^{\prime}(x)  dx
\end{displaymath}
\end{lemma}
\begin{proof}[Proof of  Lemma~\ref{sec:usefulapproxs}]
  {It holds that}
\begin{displaymath}
	G(k)-G(k+1)=-\int_k^{k+1}G'(x)  dx.
\end{displaymath}
 {Then,  }
\begin{displaymath}
	-\int_k^{k+1} H(x-1)G'(x)   d x\leq H(k)\big(G(k)-G(k+1)\big)
	\leq-\int_k^{k+1}H(x)G'(x)    d x
\end{displaymath}
and all these quantities are non-negative. It remains to sum over $k\in \{\ell,\ell + 1, \ldots, m\}$.
\end{proof}

\begin{lemma}[Bounding $\sum_{k}{(k)_{2}(k + M)^{-2}}\left(k^{-\alpha} - (k+1)^{-\alpha}\right)$]\label{lm:boundonER}
For all  $k \in \IN$,  with  $M>1$ fixed,  and $1 < \alpha < 2$,
{it holds that }
\begin{displaymath}
\begin{split}
& \frac{\alpha}{ {(M-1)}^{\alpha}}\int_{\frac{M-1}{M+m}}^{\frac{M-1}{M-1+\ell} }y^{\alpha-1}{(1-y)}^{1-\alpha}   dy + \mathcal{O}\left( \frac{1}{M^{2}}  \right) \\
& \le  \sum_{k=\ell}^{m }\frac{k(k-1)}{{(k+M)}^2}\left( \frac{1}{k^\alpha}-
\frac{1}{{(k+1)}^\alpha}\right)   \\
\le &  \frac{\alpha}{M^{\alpha}}\int_{\frac{M}{M+m+1 }   } ^{ \frac{M}{M+\ell }   } y^{\alpha-1 }{(1-y)}^{1-\alpha}   dy +  \mathcal{O}\left( \frac{1}{M^{2}}  \right)   \\
\end{split}
\end{displaymath}
\end{lemma}
\begin{proof}[Proof of Lemma~\bjarki{\ref{lm:boundonER}}]
Substituting  $k(k-1) {(k+M)}^{-2}$  for   $H$ and  $k^{-\alpha}$ for  $G$ in Lemma~\ref{sec:usefulapproxs}  gives
\begin{equation}
\label{eq:13}
	\int_{\ell }^{m+1}\frac{(x-1)(x-2)}{{(x-1+M)}^2}
	\frac{\alpha}{x^{\alpha +1}} dx
\leq 
\sum_{k=\ell }^{m }\frac{k(k-1)}{ {(k+M)}^2} \left( \frac{1}{k^{\alpha}} - \frac{1}{{(1+k)}^{\alpha}}  \right)
\leq 
	\int_{\ell }^{m +1}\frac{x(x-1)}{{(x+M)}^2}
	\frac{\alpha}{x^{\alpha +1}}  dx.
\end{equation}
We  evaluate the integrals appearing in~\eqref{eq:13}. For
the integral on the left in~\eqref{eq:13} we obtain, using the
substitution  $y = (M-1)/(x + M-1)$,  
\begin{displaymath}
\alpha\int_{\ell}^{m+1} \frac{x^{1-\alpha}}{(x+M - 1)^{2} }dx =   	\frac{\alpha }{(M-1)^\alpha}\int_{\frac{M-1}{ M+m} }^{ \frac{M-1}{M-1+\ell} }y^{\alpha -1}(1-y)^{1-\alpha}dy.
\end{displaymath}
 Integration  by parts and the same substitution give us 
\begin{displaymath}
\begin{split}
& 3\alpha \int_{\ell }^{m+1} \frac{x^{-\alpha}}{ (x+M -1)^{2}  }dx  =  \frac{3\alpha}{1-\alpha}\left[ \frac{x^{1-\alpha}}{(x+M-1)^{2} }  \right]_{\ell}^{m+1} + \frac{6\alpha}{1-\alpha}\int_{\ell}^{m+1} \frac{x^{1-\alpha}}{(x+M-1 )^{3} }dx  \\
 =&  \frac{3\alpha}{\alpha-1}\left(  \frac{\ell^{1-\alpha}}{(\ell +M-1)^{2}} -  \frac{(m+1)^{1 - \alpha} }{(m+M)^{2}} \right) -  \frac{6\alpha}{(\alpha-1)M^{\alpha + 1}  }\int_{\frac{M-1}{M+m}  }^{      \frac{M-1 }{M-1+\ell}  }(1-y)^{1-\alpha}y^{\alpha}dy   \\
\end{split}
\end{displaymath}
and  using integration by parts twice and then the same substitution
gives
\begin{displaymath}
\begin{split}
& 2\alpha\int_{\ell}^{m+1}\frac{x^{-\alpha-1}}{(x+M-1 )^{2} }dx   =  - 2 \left[ \frac{x^{-\alpha}}{(x+M-1)^{2} } \right]_{\ell}^{m+1}  - 4 \int_{\ell}^{m+1} \frac{x^{-\alpha}}{(x+M-1)^{3} }dx \\
= & 2\left( \frac{\ell^{-\alpha}}{(\ell+M-1)^{2} }  - \frac{(m+1)^{-\alpha}}{(m+M )^{2} } \right)  - \frac{4}{1-\alpha}\left[ \frac{x^{1-\alpha}}{(x+M-1 )^{3} }  \right]_{\ell}^{m+1} +  \frac{12}{1-\alpha}\int_{\ell}^{m+1}\frac{x^{1-\alpha}}{(x+M-1)^{4} }dx \\ 
= &  2\left( \frac{\ell^{-\alpha}}{(\ell+M-1)^{2} }  - \frac{(m+1)^{-\alpha}}{(m+M )^{2} } \right) + \frac{4}{\alpha - 1} \left( \frac{(m+1)^{1-\alpha} }{(m+M)^{3} } -           \frac{\ell^{1-\alpha} }{(\ell + M -1)^{3} }  \right) \\
& - \frac{12}{(\alpha - 1)M^{2+\alpha}}\int_{\frac{M-1}{m+M}}^{ \frac{M-1}{M-1+\ell}} (1-y)^{1-\alpha}y^{\alpha + 1} dy  \\ 
\end{split}
\end{displaymath}
Similar calculations for the integral on the right in~\eqref{eq:13} give
\begin{displaymath}
\begin{split}
 &  \int_{\ell}^{m+1}\frac{x(x-1)}{(x+M)^{2} } \frac{\alpha}{x^{\alpha+1}} =  \frac{\alpha}{M^{\alpha}}\int_{\frac{M}{M+1+m}}^{\frac{M}{M+\ell}} y^{\alpha-1}(1-y)^{1-\alpha} dy \\
+ &  \frac{\alpha}{\alpha - 1}\left( \frac{(m+1)^{1-\alpha} }{(M+1+m )^{2} } -  \frac{\ell^{1-\alpha} }{(\ell+M)^{2} }  \right) + \frac{2\alpha}{(\alpha-1)M^{\alpha+1}}\int_{     \frac{M}{M+1+m}  }^{\frac{M}{M+\ell} }(1-y)^{1-\alpha}y^{\alpha}dy  \\
\end{split}
\end{displaymath}
\end{proof}

\begin{lemma}\label{lm:newexp1}
For $1 < \alpha < 2$, any constant   $M > 1$,  with  $\overline{f}$ and
$\underline{f}$ defined in~\eqref{supinff},  and  $X_{1}^{N}$ distributed as in
~\eqref{eq:PXiJ}, {and for }  any positive function $L \equiv L(N) \le \psi(N)$,
\begin{displaymath}
\begin{split}
\IE\left[  \frac{X_{1}^N\svigi{X_{1}^N - 1} }{\svigi{X_1^N + M}^2} \right] \le  &  \overline{f}(1) \frac{ \alpha}{M^{\alpha}}\int_{\frac{M}{M+1+L-1 } }^{ \frac{M }{M+1}  }y^{\alpha-1}{(1-y)^{1-\alpha}}dy   \\
& +    \overline{f}(L) \frac{ \alpha}{M^{\alpha}}\int_{\frac{M}{M+1+ \psi(N) } }^{ \frac{M }{M+L}  }y^{\alpha-1}{(1-y)^{1-\alpha}}dy    \\
& +  \overline{f}(1)\frac{\alpha}{\alpha - 1}\left( \frac{L^{1-\alpha} }{(M+1+L-1 )^{2} } -  \frac{1 }{(1+M)^{2} }  \right) \\
& +  \overline{f}(L)\frac{\alpha}{\alpha - 1}\left( \frac{ (\psi(N) + 1) ^{1-\alpha} }{(M+1+\psi(N) )^{2} } -  \frac{L^{1-\alpha} }{(L +M)^{2} }  \right) \\
& +   \overline{f}(1)\frac{2\alpha}{(\alpha-1)M^{\alpha+1}}\int_{     \frac{M}{M+1+L-1 }  }^{\frac{M}{M+ 1 } }(1-y)^{1-\alpha}y^{\alpha}dy \\
& +   \overline{f}(L)\frac{2\alpha}{(\alpha-1)M^{\alpha+1}}\int_{     \frac{M}{M+1+\psi(N) }  }^{\frac{M}{M+ L } }(1-y)^{1-\alpha}y^{\alpha}dy \\
\end{split}
\end{displaymath}
and
\begin{displaymath}
\begin{split}
& \IE\left[  \frac{X_{1}^N \svigi{ X_{1}^N - 1}}{ \svigi{X_1^N + M}^2} \right] \\
& \ge     \underline{g}(1) 	\frac{\alpha }{(M-1)^\alpha}\int_{\frac{M-1}{ M+L-1  } }^{ \frac{M-1}{M-1+1 } }y^{\alpha -1}(1-y)^{1-\alpha}dy   +    \underline{g}(L) 	\frac{\alpha }{(M-1)^\alpha}\int_{\frac{M-1}{ M+\psi(N) } }^{ \frac{M-1}{M-1+ L  } }y^{\alpha -1}(1-y)^{1-\alpha}dy   \\   
&  -  \underline{g}(1) \frac{3\alpha}{\alpha-1}\left(  \frac{1 }{M^{2}} -  \frac{(L-1 +1)^{1 - \alpha} }{(L-1 +M)^{2}} \right) +  \frac{6\alpha \underline{g}(1)  }{(\alpha-1)M^{\alpha + 1}  }\int_{\frac{M-1}{M+ L-1 }  }^{      \frac{M-1 }{M }  }(1-y)^{1-\alpha}y^{\alpha}dy  \\
&  -  \underline{g}(L) \frac{3\alpha}{\alpha-1}\left(  \frac{L^{1-\alpha} }{(L + M - 1)^{2} } -  \frac{(\psi(N) +1)^{1 - \alpha} }{(\psi(N) +M)^{2}} \right) +  \frac{6\alpha \underline{g}(L)  }{(\alpha-1)M^{\alpha + 1}  }\int_{\frac{M-1}{M+ \psi(N) }  }^{      \frac{M-1 }{M - 1 + L }  }(1-y)^{1-\alpha}y^{\alpha}dy \\
&  +  2 \underline{g}(1)  \left( \frac{1 }{M^{2} }  - \frac{( L-1 +1)^{-\alpha}}{( L-1  +M )^{2} } \right) + \frac{4\underline{g}(1) }{\alpha - 1} \left( \frac{( L-1 +1)^{1-\alpha} }{( L-1   +M)^{3} } -           \frac{1 }{M^{3} }  \right) \\
& - \frac{12\underline{g}(1) }{(\alpha - 1)M^{2+\alpha}}\int_{\frac{M-1}{ L-1 +M}}^{ \frac{M-1}{M  }} (1-y)^{1-\alpha}y^{\alpha + 1} dy \\
&  +  2 \underline{g}(L)  \left( \frac{L^{-\alpha} }{(L+M - 1)^{2} }  - \frac{( \psi(N) +1)^{-\alpha}}{( \psi(N)  +M )^{2} } \right) + \frac{4\underline{g}(L) }{\alpha - 1} \left( \frac{(\psi(N) +1)^{1-\alpha} }{( \psi(N)   +M)^{3} } -           \frac{L^{1-\alpha} }{(L+ M-1)^{3} }  \right) \\
& - \frac{12\underline{g}(L) }{(\alpha - 1)M^{2+\alpha}}\int_{\frac{M-1}{ \psi(N) +M}}^{ \frac{M-1}{M-1+L  }} (1-y)^{1-\alpha}y^{\alpha + 1} dy \\
\end{split}
\end{displaymath}
\end{lemma}
\begin{proof}[Proof of Lemma~\ref{lm:newexp1}]
The lemma follows from  ~\eqref{eq:PXiJ} and Lemma~\ref{lm:boundonER} and we  recall $\overline{f}$ and $\underline{f}$ from 
 \eqref{supinff};  by  ~\eqref{eq:PXiJ}
\begin{displaymath}
\begin{split}
& \underline{g}(1)\sum_{k=1}^{L-1}\frac{k(k-1)}{(k+M)^{2} }\left( \frac{1}{k^{\alpha}} -  \frac{1}{(k+1)^{\alpha}}  \right) +  \underline{g}(L)\sum_{k=L}^{\psi(N)}\frac{k(k-1)}{(k+M)^{2} }\left( \frac{1}{k^{\alpha}} -  \frac{1}{(k+1)^{\alpha}}  \right) \\
\le &  \IE\left[  \frac{X_{1}^N(X_{1}^N - 1)}{(X_1^N + M)^2} \right] =   \sum_{k=1}^{\psi(N)} \frac{k(k-1)}{(k+M)^{2} } \prob{X_{1}^{N} = k}     \\
\le&  \overline{f} (1) \sum_{k=1}^{L-1}\frac{k(k-1)}{(k+M)^{2} }\left( \frac{1}{k^{\alpha}} -  \frac{1}{(k+1)^{\alpha}}  \right)   +  \overline{f}(L)\sum_{k=L}^{\psi(N)}\frac{k(k-1)}{(k+M)^{2} }\left( \frac{1}{k^{\alpha}} -  \frac{1}{(k+1)^{\alpha}}  \right) \\
\end{split}
\end{displaymath}
\end{proof}

\begin{remark}
The integrals in Lemma~\ref{lm:newexp1} will all  tend to zero unless $\psi(N)/N \gneqq 0$.  
\end{remark}

Proposition~\ref{pr:CNversion0} will follow from Proposition \ref{pr:CN} as 
an easy corollary. Section \ref{proofprCN} contains a  proof of
Proposition 
\ref{pr:CN}.  
\begin{propn}%
\label{pr:CN}%
Suppose that the population evolves according to 
Definition \ref{Schwm} and \eqref{eq:PXiJ}   with $1 < \alpha < 2$. 
Let   $L$ be  a function of $N$ as
specified in each case,   recall $m_{N}$ from ~\eqref{eq:4},
$\overline{f}$ and $\underline{f}$ from ~\eqref{supinff} and suppose
$f_{\infty} = g_{\infty}$ with 
$f_{\infty}, g_{\infty} $ from ~\eqref{gfinf}. Recall
$\mathcal{C}_{N}$ from ~\eqref{curlyc}.   
\begin{enumerate} 
\item \label{JCD1psi<N}
If    $\psi(N)/N \to 0$ and $L(N)/\psi(N)\to 0$  then
\begin{displaymath}
\begin{split}
\frac{N}{\psi(N)^{2-\alpha}}\mathcal{C}_N
& =
 \frac{\alpha f_\infty}{(2 - \alpha)m_N^2}
  + \OO{  \frac{L^{2-\alpha}}{\psi(N)^{2-\alpha}}}
  + \cO\left(\frac{\beta_N}{N} \right) + \cO\left(\frac{\psi(N)}{N} \right)
+ \OO{\frac{N}{\beta_{N}^{\alpha}} }
\\
\end{split}
\end{displaymath}
\item \label{JCD1psi=N}
If  $\psi(N)/N \to K$ where $K > 0$ a
constant, and $L(N)/N \to 0$   then
\begin{displaymath}
\begin{split}
N^{\alpha-1} \mathcal{C}_N  & =  f_{\infty}\frac{\alpha}{m_{N}^{\alpha}} \int_{\frac{m_{N} }{m_{N} + K } }^{\frac{m_{N} }{m_{N} + L/N } }y^{\alpha - 1}(1-y)^{1-\alpha}dy 
\\
& + \OO{\int_{ \frac{m_{N} }{m_{N} + L/N }  }^{1}y^{\alpha - 1}(1-y)^{1-\alpha}dy } + \OO{\frac{\beta_{N} }{N}} +  \OO{ \frac{L^{1-\alpha} - 1}{N^{2-\alpha}}} + \OO{ \frac {N}{\beta_{N}^{\alpha}} }    \\ 
\end{split}
\end{displaymath}
\item \label{JCD1psi>N}
{If}  $\psi(N)/N \to \infty$ and $L(N)/N \to 0$ then 
\begin{displaymath}
\begin{split}
N^{\alpha-1} \mathcal{C}_N
& =  f_{\infty} \frac{\alpha}{m_{N}^{\alpha}} \int_{ \frac{m_{N}}{m_{N} + \psi(N)/N } }^{\frac{m_{N} }{ m_{N} + L/N  } }y^{\alpha - 1}(1-y)^{1-\alpha}dy
\\
& +   \OO{ \int_{\frac{m_{N}}{ m_{N} + L/N  } }^{1} y^{\alpha - 1}(1-y)^{1-\alpha}  dy} + \OO{\frac{\beta_{N}}{N}} +  \OO{\frac{1 + L^{1-\alpha}}{N^{2-\alpha}} } + \OO{ \frac{N}{\beta_{N}^{\alpha}} } \\
\end{split}
\end{displaymath}
\end{enumerate}
\end{propn}
\begin{remark}
The constant order term in Case~\ref{JCD1psi>N}
reduces to a  familiar form, since 
\begin{displaymath}
 \alpha \int_0^1y^{\alpha - 1}(1-y)^{1-\alpha} {d}y= \alpha \int_{0}^{1}u^{1-\alpha}(1-u)^{\alpha -1}{d}u = 
\alpha\Gamma(2-\alpha)\Gamma(\alpha).
\end{displaymath}
\end{remark}
\begin{remark}[The integral remainder term in Proposition~\ref{pr:CN}]
Since $L/N\to 0$ it holds   
\begin{displaymath}
\begin{split}
\int_{\frac{m_{N}}{m_{N} + L/N} }^{1} y^{\alpha - 1}(1-y)^{1-\alpha}dy \le  \int_{\frac{m_{N}}{m_{N} + L/N} }^{1}(1-y)^{1-\alpha}dy = \frac{1}{2-\alpha}\svigi{\frac{L/N}{m_{N} + L/N}}^{2-\alpha}
\end{split}
\end{displaymath} 
and this is of order  $\mathcal O \svigi{ \svigi { L / N}
}^{2-\alpha}  $    as $N\to \infty$. Moreover,  as $N\to \infty$, 
\begin{displaymath}
\begin{split}
\int_{\frac{m_{N}}{m_{N} + L/N} }^{1} y^{\alpha - 1}(1-y)^{1-\alpha}dy &   \ge  \svigi{\frac{m_{N}}{m_{N} +  L/N }}^{\alpha - 1}\int_{\frac{m_{N}}{m_{N} + L/N} }^{1} (1-y)^{1-\alpha}dy \\
& =  \frac{m_{N}^{\alpha - 1}}{m_{N} + L/N }\frac{1}{2-\alpha}\svigi{ \frac LN}^{2-\alpha} 
 =   \mathcal  O\svigi{ \svigi{\frac LN }^{2-\alpha} }
\end{split}
\end{displaymath}
\end{remark}

\begin{proof}[Proof of Proposition~\ref{pr:CNversion0}]

This now follows immediately 
from Proposition~\ref{pr:CN}
on setting 
\begin{equation}
\label{eq:6}
{ \beta_N  =  N^{\eta}, \quad \text{with } \frac{1}{\alpha} < \eta < 1}
\end{equation}
\end{proof}

\begin{remark}[$\eta$ in \eqref{eq:6}]
\label{rm:eta}
{Examples of $\eta$ in  \eqref{eq:6} such that $1/\alpha < \eta < 1$   include 
\begin{displaymath}
\eta = \frac{\alpha + 1}{2\alpha}, \quad \eta = \frac{4-\alpha}{3}, \quad \eta = \frac{ 2}{1 + \alpha}
\end{displaymath}}
\end{remark}


To prove Proposition~\ref{pr:CN} we will partition over the event
$A_N$ (defined in Notation~\ref{defn:ANandM}), and its complement $\comp{A}_N$. 
The random variable  $\tilde{S}_{N}$ is defined in  ~\eqref{st} in 
Notation~\ref{defn:ANandM}; and  $S_{N}$ in
~\eqref{fullSN}.  
Our main focus will be the
value, for large $N$, of the term 
\begin{displaymath}
{\IE\left[ \frac{X_{1}^N \svigi{ X_{1}^N - 1}}{S_N^2} \one{S_N \geq N} \one{ A_N} \right]
\to
\IE\left[ \frac{X_{1}^N \svigi{ X_{1}^N - 1} }{ \svigi{ X_1^N + \tilde{S}_N}^2} \one{ A_N} \right]}
,
\end{displaymath}
under the assumption that {(recall $m_{N}$ from \eqref{eq:4})}  $m_{N} > 1$ (so
that $\prob{S_{N} < N} \to 0$ exponentially fast
\cite[Lemma~5]{schweinsberg03}) and $\beta_N/N \to 0$.
Proposition~\ref{pr:CN}
will then follow if we prove that, recalling  $A_{N}^{c}$ from
~\eqref{eq:5}, 
\begin{align*}
N\IE\left[  \frac{X_{1}^N \svigi{X_{1}^N - 1} }{S_N^2} \one{{S_N} \geq N} \one{ \comp{A}_N} \right]
\end{align*}
is  {asymptotically}   sufficiently small.

\begin{lemma}
\label{lemma:largepsibound}
In the notation of Proposition~\ref{pr:CN}, 
 \eqref{eq:5} in    Notation~\ref{defn:ANandM},   as $N\to \infty$ 
\begin{displaymath}
\begin{split}
 \prob{A_{N}^{c} }  = \OO{ \frac{N}{\beta_N^\alpha}}  \\
\end{split}
\end{displaymath}
\end{lemma}
\begin{remark}\label{rmk:stableLaw}
{Taking} $\psi(N) = \infty$ the rate in 
Lemma~\ref{lemma:largepsibound} is optimal
as $X_{1}^{N}, \ldots, X_{N}^{N}$
will then lie in the 
domain of attraction of a stable law of index $\alpha$ for $0 < \alpha
< 2$  and so 
we expect
\begin{displaymath}
\prob{ |S_N - N m_N | \geq \beta_N}
=
\prob{ \left|\frac{S_N - N m_N}{N^{1/\alpha}} \right| \geq \frac{\beta_N}{N^{1/\alpha}} } = \OO{ \frac{N}{\beta_N^\alpha}}
,
\end{displaymath}
for large $N$~\citep{hall1981rate,Phillips1985}.
\end{remark}

\begin{proof}[Proof of Lemma~\ref{lemma:largepsibound}] 
Define, for all $N\ge 2$, with  ${(\beta_{N})}$ an
increasing positive sequence, 
\begin{equation}
\label{BN}
B_N := \left\{ X_i^N \leq \beta_N \; \text{for all $i = 1, \ldots, N$}\right\}.
\end{equation}
Observe that, recalling $\widetilde{S}_N$ from~\eqref{st} in 
Notation~\ref{defn:ANandM}, and with $B_N^c$ the complement of $B_N$,
\begin{displaymath}
\begin{split}
& \prob{ \tolugildi{\tilde{S}_N - (N-1) m_N } \geq \beta_N } \\
= & 
\prob{  \left(\one{B_N} \tolugildi{\tilde{S}_N - (N-1) m_N } 
+
\one{\comp{B}_N} \tolugildi{ \tilde{S}_N - (N-1) m_N } \right) \geq \beta_N }
\\
\leq &
\prob{  \one{B_N} \tolugildi{ \tilde{S}_N - (N-1) m_N } \geq
\frac{\beta_N}{2} }
+
\prob{ \one{\comp{B}_N} \tolugildi{ \tilde{S}_N - (N-1) m_N } \geq
\frac{\beta_N}{2}  }
\\
\leq &
\prob{ \one{B_N} \tolugildi{ \tilde{S}_N - (N-1) m_N } \geq \frac{\beta_N}{2}}
+
\prob{ \comp{B}_N }
.
\end{split}
\end{displaymath}
We can see from~\eqref{eq:PXiJ} and~\eqref{eq:23},
\begin{displaymath}
\IP\left[\comp{B}_N\right] \leq N \IP\left[X_1^N > \beta_N\right] \leq  \frac{N \overline{f}(\beta_N + 1)}{(\beta_N + 1)^\alpha} -   \frac{N \overline{f}(\beta_N + 1)}{(\psi(N) + 1)^\alpha}
\leq
\frac{N \overline{f}(1)}{\beta_N^\alpha}
\end{displaymath}
        
Define
$\tilde{m}_N := \IE\left[X_1^N \one{X_1^N \leq \beta_N} \right ]$.
{Then,}   with  $m_N$ from~\eqref{eq:4} and $\tilde{S}_N$ from~\eqref{st}
        \begin{displaymath}
        \begin{split}
& \IE \left[ \one{B_N}\left(  \tilde{S}_N - (N-1) m_N \right)^2  \right] \\
= &
\IE \left[ \one{B_N}\left(  \tilde{S}_N - (N-1) \tilde{m}_N+  {(N-1)\left( \tilde{m}_N -  m_N\right)} \right)^2  \right] 
\\
= &
\IE\left[ \one{B_N}\left(  \tilde{S}_N - (N-1) \tilde{m}_N \right)^2 \right] +  2(N-1)\svigi{\tilde{m}_N - m_N }\EE{\one{B_N}\left(\tilde{S}_N - (N-1)\tilde{m}_N \right)   } \\
& + (N-1)^2(\tilde{m}_N -m_N)^2 
= \OO{ \max\set { N\beta_{N}^{2-\alpha}\komma   N^{2}\beta_{N}^{2-2\alpha}}}    
.
\end{split}
\end{displaymath}
To explain the last line above we have by independence of the    $X_{1}^{N},\ldots, X_{N}^{N}$
\begin{displaymath}
\begin{split}
\EE{\one{B_{N} }\svigi{\tilde S_{N} - (N-1)\tilde m_{N}}^{2}  } & =  (N-1)\EE{ \svigi{X_{1}^{N}}^{2}\one{B_{N}} } - (N-1)\tilde{m}_{N}^{2}
\end{split}
\end{displaymath}
Using \eqref{eq:PXiJ} we have (recall $B_{N}$ from \eqref{BN}, and
that $\svigi{\beta_{N}}$ is an increasing positive sequence) 
\begin{displaymath}
\begin{split}
\EE{ \svigi{X_{1}^{N}}^{2}\one{B_{N}} } &  \le  \overline{f}(1)\sum_{k=1}^{\beta_{N}} k^{2}\left( \frac {1}{k^{\alpha}} - \frac{1}{(k+1)^{\alpha}} \right) \\
= & \overline{f}(1)\sum_{k=1}^{\beta_{N}}k^{2-\alpha} -  \overline{f}(1)\sum_{k=2}^{\beta_{N} + 1} k^{2-\alpha} +  2\overline{f}(1)\sum_{k=2}^{\beta_{N} + 1} k^{1-\alpha} -  \overline{f}(1)\sum_{k=2}^{\beta_{N} + 1}k^{-\alpha} \\
\le  & \overline{f}(1) - \overline{f}(1)(\beta_{N} + 1)^{2-\alpha} + \frac{2 \overline{f}(1)}{2-\alpha} \beta_{N}^{2-\alpha} =   \OO{ \beta_{N}^{2-\alpha} }
\end{split}
\end{displaymath}
Similarly we have
$\EE{ \svigi{X_{1}^{N}}^{2}\one{B_{N}} } \ge
\underline{g}(1)\sum_{k}k^{2}\svigi{ k^{-\alpha} - (k+1)^{-\alpha}}
= \OO{ \beta_{N}^{2-\alpha}} $ so that
$N\EE{ \svigi{ X_{1}^{N}} ^{2} \one{B_{N}} }  = \OO{
N\beta_{N}^{2-\alpha} }$.  Since $\alpha > 1$ we have
$ \limsup_{N\to \infty} \tilde m_{N} < \infty$  so that
$(N-1) \tilde{m}_{N}^{2} = \OO{  N^{1}}$.  It follows that
$\EE{\one{B_{N}} \left( \tilde S_{N}  - (N-1)\tilde m_{N} \right)^{2}  }
 =  \OO{  N\beta_{N}^{2-\alpha}}$. 
Since 
$\tilde{m}_{N}\le  m_{N}$  the term $ 2(N-1)(\tilde{m}_N - m_N )\EE{\one{B_N}\left(\tilde{S}_N - (N-1)\tilde{m}_N \right)   }$
can be discarded; and  $(\tilde{m}_{N} - m_{N})^{1} $ can be bounded
by a constant times  $\one{\beta_{N} <
\psi(N)}\sum_{k=\beta_{N}+1}^{\psi(N)}k(k^{-\alpha} -
(k+1)^{-\alpha})$, which is of order  $\beta_{N}^{1-\alpha }$ (use
Lemma~\ref{sec:usefulapproxs} with $H(x) = x$ and $G(x) = x^{-\alpha}$).

Chebyshev's inequality then implies
\begin{displaymath}
\begin{split}
\IP\left[\one{B_N} \left|\tilde{S}_N - (N-1) m_N \right| \geq \frac{\beta_N}{2}\right] 
\leq &
4\frac{\IE \left[ \one{B_N} \left(  \tilde{S}_N - (N-1) m_N \right)^2  \right]}{\beta_N^2}
\\
= & \cO\left( \frac{N \beta_N^{2- \alpha}}{\beta_N^2}  +  \frac{N^2
}{\beta_N^{2\alpha}} \right) \\  
=&  \cO\left( \frac{N}{\beta_N^\alpha} \right)
\end{split}
\end{displaymath}
for ${N}/{\beta_N^\alpha}\to 0$. { This concludes the proof of Lemma~\ref{lemma:largepsibound}.}
\end{proof}

\subsection{Proof of Proposition~\ref{pr:CN}}\label{proofprCN}%

We are now in a position to prove Proposition~\ref{pr:CN}.

\begin{proof}[Proof of { Case~\ref{JCD1psi<N} of}  Proposition~\ref{pr:CN}]

We see using~\eqref{eq:PXiJ} 
\begin{displaymath}
\begin{split}
\EE{X_{1}^{N}\svigi{ X_{1}^{N}-1} }  &  \le  \overline{f}(2) \sum_{k=2}^{L(N)} k(k-1)(k^{-\alpha} - (k+1)^{-\alpha}) +   \overline{f}(L) \sum_{k=L+1}^{\psi(N)} k(k-1)(k^{-\alpha} - (k+1)^{-\alpha})  \\
\end{split}
\end{displaymath}
For  $\ell, m \in \IN$ with $\ell \le m$ 
\begin{displaymath}
\begin{split} 
&  \sum_{k=\ell}^{m}k(k-1)(k^{-\alpha} - (1+k)^{-\alpha}) \\
& = \sum_{k=\ell}^{m} k(k-1)k^{-\alpha} -  \sum_{k=\ell + 1}^{m+1} (k-1)(k-2)k^{-\alpha} \\
& =  \sum_{k=\ell}^{m }k^{2-\alpha} -  \sum_{k=\ell}^{m}k^{1-\alpha} -  \sum_{k=\ell+1}^{m+1}k^{2-\alpha}  + 3\sum_{k=\ell+1}^{m+1}k^{1-\alpha} - 2 \sum_{k=\ell + 1}^{m+1}k^{-\alpha} \\
& = \ell^{2-\alpha} - (m+1)^{2-\alpha} - \ell^{1-\alpha} + 3(m+1)^{1-\alpha} + 2\sum_{k=\ell + 1}^{m+1}k^{1-\alpha} -  2 \sum_{k=\ell+1}^{m+1}k^{-\alpha}  \\
\end{split}
\end{displaymath}
Since  $ \sum_{k=\ell+1}^{m+1}k^{1-\alpha} \le  \int_{k=\ell}^{m+1}x^{1-\alpha}dx  = \left( (m+1)^{2-\alpha} - \ell^{2-\alpha} \right) /(2-\alpha) $
we have upon substituting for $\ell$ and $m$ and writing $L$ for $L(N)$
\begin{displaymath}
\begin{split}
\EE{X_{1}^{N} \svigi{X_{1}^{N}-1}} & \le \overline{f}(2)\left( \frac{2}{2-\alpha}(L+1)^{2-\alpha} - (L+1)^{2-\alpha} +\OO{1}  \right) \\
& +   \overline{f}(L) \left(  \frac{ 2  }{2-\alpha}(\psi(N) + 1)^{2-\alpha} - (\psi(N) + 1)^{2-\alpha} + \OO{L^{2-\alpha}}   \right) \\
& =  \psi(N)^{2-\alpha} \left(  \overline{f}(L)\frac{\alpha}{2-\alpha} + \OO{ \left(  \frac{L(N)}{\psi(N)} \right)^{2-\alpha} } \right) \\
\end{split}
\end{displaymath}
Similar calculations give
\begin{displaymath}
\begin{split}
\EE{X_{1}^{N} \svigi{ X_{1}^{N}-1} } &   \ge  \psi(N)^{2-\alpha}\left( \underline{g}(L) \frac{\alpha}{2-\alpha}  + \OO{ \left(  \frac{L(N)}{\psi(N)} \right)^{2-\alpha} } \right) \\
\end{split}
\end{displaymath}

Recall $A_N$ from  \eqref{eq:betaN} and $m_{N}$ from
~\eqref{eq:4} and $M_{-}$ from  \eqref{Mpm} and $\tilde S_{N}$
from  \eqref{st}.  On $A_{N}$ we have
$M_{-} \le M_{-} + X_{1}^{N} \le \widetilde{S}_{N} + X_{1}^{N} $.
Choose  $N$ large enough that
  $S_N\geq N$ on $A_{N}$. Then 
\begin{displaymath}
\begin{split}
& \EE{ \frac{X_{1}^N \svigi{X_{1}^N - 1}}{ \svigi{ X_1^N + \tilde{S}_N } ^2} \one{S_N \ge N} {\one{  A_N}} }   \le  M_{-}^{-2} \EE{ X_{1}^{N} \svigi{ X_{1}^{N} - 1} } \\
& \le  \frac{\psi(N)^{2-\alpha}}{ M_{-}^{2}} \left( \frac{\alpha \overline{f}(L) }{2-\alpha}  + \OO{ \left( \frac{L(N)}{\psi(N)} \right)^{2-\alpha} } \right) \\
& =   \frac{\psi(N)^{2-\alpha}}{ N^{2}} \left( \frac{\alpha \overline{f}(L) }{(2-\alpha)m_{N}^{2} }  + \OO{ \left( \frac{L(N)}{\psi(N)} \right)^{2-\alpha}} + \OO{\frac{\beta_{N}}{N}} \right) \\
\end{split}
\end{displaymath}
With $M_{+}$ from  \eqref{Mpm} we have
$X_{1}^{N} + \tilde S_{N} \le X_{1}^{N} + M_{+} \le \psi(N) + M_{+}$  on $A_{N}$.
Then choosing $N$ as before we have 
\begin{displaymath}
\begin{split}
& \EE{ \frac{X_{1}^{N} \svigi{ X_{1}^{N} - 1} }{ \svigi{X_{1}^{N} + \tilde S_{N}}^{2}}\one{S_{N}\ge N} { \one{ A_{N}}}   }   \ge \frac{ \EE{X_{1}^{N} \svigi{X_{1}^{N} - 1} }}{         \svigi{ \psi(N) + M_{+}}^{2} }  \\
& =  \frac{\psi(N)^{2-\alpha}}{N^{2}}\left( \frac{ \alpha \underline{g}(L) }{ (2-\alpha)m_{N}^{2} }  +   \OO{ \left(\frac{L(N)}{\psi(N)}\right)^{2-\alpha}}  +   \mathcal{O}\left( \frac{\beta_{N}}{N} \right) +  \OO{ \frac{\psi(N)}{N}} \right)
\end{split} 
\end{displaymath}
Recalling  $A_{N}^{c}$ from ~\eqref{eq:5},  and that
$\EE{ \svigi{ X_{1}^{N}}_{2}} = { \OO{ \psi_{N}^{2-\alpha}}}$ by our calculations above, 
\begin{displaymath}
\begin{split}
\frac{N^2}{\psi(N)^{2-\alpha}}\IE\left[  \frac{X_{1}^{N} \svigi{ X_{1}^{N} - 1}}{S_N^2} \one{S_N \geq N} \one{\comp{A}_N} \right]
& \leq \frac{N^2}{\psi(N)^{2-\alpha}}
\frac{\IE \left[X_{1}^{N} \svigi{ X_{1}^{N} - 1} \right]\IP\left[\comp{A}_N \right]}{N^{2}} \\
& =  \mathcal{O}\svigi{ \IP \left[A_{N}^{c}\right]} 
 = \mathcal{O}\left(  {N}{\beta_{N}^{-\alpha}} \right)
\end{split}
\end{displaymath}
by  Lemma~\ref{lemma:largepsibound}.
{  The  proof  of Case~\ref{JCD1psi<N} of  Proposition~\ref{pr:CN}   is complete.}
\end{proof}

\begin{proof}[Proof of  {Case~\ref{JCD1psi=N}}  of  Proposition~\ref{pr:CN}]

Recall the assumption $\psi/N \to K$    with $K > 0$ constant,  and 
the notation $M_{-}$
from ~\eqref{Mpm} in  Notation~\ref{defn:ANandM}, and   $f_{\infty}$ from ~\eqref{gfinf}.  
We use Lemma~\ref{lm:newexp1} to see that if $L(N)/N$ is bounded, and
that $\overline{f}(L) > 0$ for any $L$ by assumption and recalling
that by
\eqref{eq:6} we have $\beta_{N}/N \to 0$ (and
assuming $N$ is large enough that $N-1 \approx N$)
\begin{displaymath}
\begin{split}
&   N^{\alpha} \overline{f}(L)  \frac{\alpha}{M_{-}^{\alpha} } \int_{\frac{M_{-} }{M_{-} + 1 + \psi(N) } }^{\frac{M_{-}}{M_{-} + L } } y^{\alpha - 1}(1-y)^{1-\alpha}dy  =      \frac{\alpha \overline{f}(L) }{  m_{N}^{\alpha}  } \int_{\frac{m_{N}  }{K+m_{N} } }^{ \frac{m_{N}  }{m_{N} + L/N  } }   y^{\alpha - 1}(1-y)^{1-\alpha}dy  +    \OO{\frac{\beta_{N}}{N} }  \\
\end{split}
\end{displaymath}
 We then see
 \begin{displaymath}
 \begin{split}
N^\alpha \IE\left[  \frac{X_{1}^N \svigi{ X_{1}^N - 1} }{ \svigi{X_1^N +
\tilde{S}_N}^2} \one{S_N \ge N} {\one{ A_N}} \right]  & \le  \overline{f}(L) \frac{\alpha}{m_{N}^{\alpha}}\int_{\frac{m_{N} }{m_{N} + K } }^{ \frac{m_{N} }{m_{N} +  L/N}  }y^{\alpha - 1}(1-y)^{1-\alpha }dy \\
 & + \overline{f}(1) \frac{\alpha}{m_{N}^{\alpha}}  \int_{ \frac{m_{N}}{ m_{N} + L/N  }}^{ 1  }y^{\alpha - 1}(1-y)^{1-\alpha}dy   + \OO{\frac{\beta_{N}}{N} } \\
&  +    \OO{ \frac{L^{1-\alpha}  - 1 }{N^{2-\alpha} }   } + \OO{ \frac{1}{N} }  \\
\end{split}
\end{displaymath}
We also see, recalling $M_{+}$ from ~\eqref{Mpm},
\begin{displaymath}
\begin{split}
N^\alpha \IE\left[  \frac{X_{1}^{N}\svigi{ X_{1}^{N} - 1} }{\svigi{ X_1^N + \tilde{S}_N} ^2} \one{S_{N} \ge N} {\one{ A_N}} \right]  & \ge \underline{g}(L) \frac{\alpha}{m_{N}^{\alpha}} \int_{\frac{m_{N} }{m_{N} +  K  } }^{ \frac{m_{N}}{m_{N} + L/N }  }y^{\alpha - 1}(1-y)^{1-\alpha}dy \\
 & + \underline{g}(1)\frac{\alpha}{m_{N}^{\alpha}} \int_{ \frac{m_{N}}{ m_{N} + L/N  } }^{1}y^{\alpha-1}(1-y)^{1-\alpha}dy  + \OO{\frac{\beta_{N} }{N} } \\
&  +    \OO{ \frac{1 - L^{1-\alpha}}{N^{2-\alpha}}  } + \OO{\frac{1}{N} } \\ 
\end{split}
\end{displaymath}
It remains to check that 
\begin{equation}
\label{eq:psi k prob}
N^\alpha\IE\left[  \frac{X_{1}^N \svigi{ X_{1}^N - 1} }{S_N^2} \one{S_N \geq N}  \one{\comp{A}_N} \right]
\leq \frac{\IE \left[X_{1}^N \svigi{ X_{1}^N - 1} \right]\IP[\comp{A}_N ]}{N^{2- \alpha}}
=
\cO\left(
\frac{N}{\beta_N^\alpha}
\right)
\end{equation}
by Lemma~\ref{lemma:largepsibound} and the assumption
$\psi(N)/N\to K$.
{ The proof of  Case~\ref{JCD1psi=N} of  Proposition~\ref{pr:CN} is complete.}
\end{proof}

\begin{proof}[Proof of Proposition~\ref{pr:CN}, Case~\ref{JCD1psi>N}]
Now assuming that $\psi(N)/N\to\infty$ and $L(N)/N$ bounded, 
we again turn to Lemma~\ref{lm:newexp1} to see
\begin{displaymath}
\begin{split}
& N^\alpha \IE\left[  \frac{X_{1}^N \svigi{ X_{1}^N - 1} }{ \svigi{X_1^N + \tilde{S}_N}^2} \one{S_{N} \ge N} {\one{ A_N}} \right]
\leq   \overline{f}(L)  \frac{\alpha}{m_N^\alpha} \int_{ \frac{m_{N}}{ m_{N} +  \psi(N)/N  } }^{ \frac{m_{N}}{m_{N} + L/N }  } y^{\alpha - 1} (1- y)^{1- \alpha} \mathrm{d}y    \\
 &  +  \OO{ \int_{\frac{m_{N}}{m_{N} + L/N } }^{1 } y^{\alpha - 1}(1-y)^{1-\alpha} dy}   +  \OO{\frac{\beta_{N}}{N}} +  \OO{ \frac{L^{1-\alpha} - 1 }{N^{2-\alpha} }  } + \OO{\frac 1N }   \\
\end{split}
\end{displaymath}
Similarly, 
\begin{displaymath}
\begin{split}
& N^\alpha \IE\left[  \frac{X_{1}^{N} \svigi{ X_{1}^{N} - 1} }{ \svigi{X_1^N + \tilde{S}_N}^2} { \one{ A_N}} \right]
\geq  \underline{g}(L) \frac{\alpha}{m_{N}^{\alpha}} \int_{\frac{m_{N}}{m_{N} + \psi(N)/N }  }^{ \frac{m_{N}}{m_{N} + L/N }  } y^{\alpha -1}(1-y)^{1-\alpha}dy     \\
 &   +  \OO{ \int_{ \frac{m_{N}}{m_{N} + L/N }  }^{1}y^{\alpha - 1}(1-y)^{1-\alpha} dy} + \OO{\frac{\beta_{N}}{N}}+  \OO{ \frac{1 + L^{1-\alpha}}{N^{2-\alpha} } } + \OO{\frac 1N}  \\
\end{split}
\end{displaymath}
When $S_{N} \ge N$ then $S_{N}^{2} \ge \max\mengi{
\svigi{X_{1}^{N}}^{2}\komma  N^{2}}$,  and  $\svigi{X_{1}^{N} +
N}^{2} \le 4\max\mengi{ \svigi{ X_{1}^{N}}^{2}, N^{2}} $ always holds.  
With this in mind we see 
\begin{equation}
\label{eq:psi infty prob}
\begin{split}
 N^\alpha\IE\left[  \frac{X_{1}^N \svigi{ X_{1}^N - 1} }{S_N^2} \one{S_N \geq N}  \one{\comp{A}_N} \right]
\leq N^\alpha\IE\left[\frac{X_{1}^N \svigi{ X_{1}^N - 1} }{\max\mengi{ \svigi{X_1^N}^2, N^2}} \one{\comp{A}_{N}}  \right] 
\\
 \leq
 4 N^\alpha \IE\left[\frac{X_{1}^N \svigi{ X_{1}^N - 1}}{ \svigi{X_1^N+N}^2}\right]\IP[\comp{A}_N ]
=
\cO \svigi{\IP\left[\comp{A}_N \right]}
=
\cO\left( \frac{N}{\beta_N^\alpha} \right)
,
\end{split}
\end{equation}
where we have used Lemma~\ref{lemma:largepsibound} to bound
$\IP[\comp{A}_N]$ and Lemma~\ref{lm:newexp1} to control all other
terms.
{ The proof of Case~\ref{JCD1psi>N} of   Proposition~\ref{pr:CN} is complete.} 
\end{proof}

We conclude this section with a slight modification of  Lemma~6 in  
  \citep{schweinsberg03},  and a proof of~\eqref{cncn}.
\begin{lemma}[\cite{schweinsberg03}, Lemma~6]\label{lemma:Schw6}
Suppose a haploid population of size $N$ evolves according to 
Definition~\ref{Schwm}.  Recall our
notation $\nu_{1,N}, \dots, \nu_{N,N}$ for the random number of
surviving offspring of the $N$ parents, in a given generation, where
$\nu_{1,N} + \cdots + \nu_{N,N} = N$.  For all $r \geq 1$, 
$k_1, \dots, k_r \geq 2$, and $N\ge 2$,  we have, 
\begin{equation}
\label{eq:schw6}
\begin{split}
\frac{\IE[(\nu_{1,N})_{k_1} \dots (\nu_{r,N})_{k_r}]}{N^{k_1 + \cdots + k_r - r} c_N}
& =
\frac{N^r}{c_N} \IE\left[
\frac{(X_1^N)_{k_1} \dots (X_r^N)_{k_r}}{S_N^{k_1+ \cdots + k_r}} \one{S_N \geq N}
\right]
\left(1 + \cO\left(\frac{(k_1+\cdots +k_r)^2}{N}\right)\right)
,\\
\end{split}
\end{equation}
where $c_{N}$ is defined in Definition~\ref{cN}, recall also~\eqref{def of cN}
and~\eqref{cncn},  $\C$ in~\eqref{curlyc}, and 
\begin{equation}
\label{eq:cNworks}
c_N = \C \left(1 + \cO\left(\frac{1}{N}\right)\right)
.
\end{equation}
\end{lemma}

Our proof follows that presented in~\cite{schweinsberg03}.  However,
we include it in order to emphasize that the source of the error term
in~\eqref{eq:schw6} is the distinction between sampling with and
without replacement from the pool of potential offspring.
\begin{proof}[Proof of Lemma~\ref{lemma:Schw6}]

Label individuals in the current generation by $\{1,2,\ldots
,N\}$. For each $1\leq r\leq N$, set $k_0=0$ and write
$B_{k_1,\ldots ,k_r}$ for the event that for each $i=1,\ldots ,r$, the
individuals labelled
$\{k_0+\cdots +k_{i-1}+1,\ldots ,k_0+\cdots+k_{i-1}+k_i\}$ are all
descended from the individual with label $i$ in the previous
generation.

We then see that
\begin{displaymath}
\IP[B_{k_1,\dots,k_r}]
=
\IE\left[ \IP[B_{k_1,\dots,k_r} | \nu_{1,N} , \dots , \nu_{N,N} ] \right]
=
\IE\left[
\frac{(\nu_{1,N})_{k_1} \dots (\nu_{r,N})_{k_r}}{(N)_{k_1+\cdots + k_r}}
\right]
,
\end{displaymath}
and
\dsplit{
\IP[B_{k_1,\dots,k_r}]
= &
\IP[B_{k_1,\dots,k_r}\cap \{S_N \geq N\}]+\IP[B_{k_1,\dots,k_r}\cap \{S_N < N\}]
\\
= &
\IE\left[ \IP[B_{k_1,\dots,k_r}\cap \{S_N \geq N\} | X_1^N, \dots, X_N^N ] \right] + \IP[B_{k_1,\dots,k_r}\cap \{S_N < N\}]
\\
= &
\IE\left[
\frac{(X_1^N)_{k_1} \dots (X_r^N)_{k_r}}{(S_N)_{k_1+\cdots + k_r}} \one{S_N \geq N}
\right]
,
}
where in the case  $S_N < N$ we will keep  the previous 
population, and so  one individual in the given  generation will have 
descended from each individual in the previous generation.


The result follows on replacing ${(S_N)}_{k_1+\cdots +k_r}$ by
$S_N^{k_1+\cdots +k_r}$;  expanding  ${(n)}_m$ (see
~\eqref{eq:10}) gives {(recall $\sum_{j=1}^{m}j = m(m+1)/2$)}
\dsplit{
(n)_m = n^m\left(1 + \OO{\tfrac{\sum_{j=1}^m j }{n} }\right) = n^m\left(1 + \OO{\tfrac{m^2 }{n}} \right).
}
Recall  $\C$ from~\eqref{curlyc}. 
We may deduce~\eqref{eq:cNworks} from~\eqref{eq:schw6} with $r=1$ 
and $k_1 = 2$, by using that $c_N =\IE[(\nu_{1,N})_2]/(N-1)$ and that, 
in this case, the left hand side of~\eqref{eq:schw6} is equal to
$(N-1)/N$. { This concludes the proof of Lemma~\ref{lemma:Schw6}.}
\end{proof}

\section{Proof of Lemma~\ref{lemma:ksamplesize}}\label{sec:proofksamplesize}

Recall the notation in Definition~\ref{defn:ANandM} in
Section~\ref{sec:convergencealpha12}, $M_{-}$ and $M_{+}$ are defined
in~\eqref{Mpm}, $\tilde S_{N}$ in~\eqref{st} and $A_{N}$ in
~\eqref{eq:betaN}.  In Lemma~\ref{lemma:sumbound2} and
in~\eqref{useful integral} we use $M$ to denote an arbitrary
(positive) constant. We then intend to use the result with $M$
replaced by either $M_+$ or $M_-$.

For any natural numbers   $n \ge k  \ge 2$ let   $E$ be the event 
\begin{displaymath}
E = \{ \text{any   $k$ from $n$ lineages merge in one merger in one generation} \}
\end{displaymath}
Then, conditioning on $X_1^N, \ldots, X_N^N$, for each $N\in \IN_{2}$,  
 \begin{equation}
 \label{eq:29}
 \begin{split}
\prob{E}&  =
\EE{ \prob{ E | X_1^N,\dots, X_N^N }}
 =
N \EE{ 
\frac{\binom{X_1^N}{k} \binom{S_N-X_1^N}{n-k}}{\binom{S_N}{n}}
\one{S_N \geq N} }  \\
 & = N\binom nk  \EE{\frac{\svigi{X_{1}^{N}}_{k} \svigi{\widetilde S_{N}}_{n-k}}{\svigi{X_{1}^{N} + \widetilde S_{N}}_{n} }\one{S_{N}\ge N}  } 
\end{split}
\end{equation}
and we  have not prohibited simultaneous mergers.

On  $A_N$ (recall~\eqref{eq:betaN} in   Notation~\ref{defn:ANandM}), 
 for any $n$ with $n/S_N \to 0$,  
we approximate $(\widetilde S_{N})_{n-k}$ by $\tilde{S}_N^{n-k}$, and
$\svigi{X_{1}^{N} + \widetilde S_{N} }_{n} $ by 
 $\svigi{X_1^N+\tilde{S}_N}^n$.  Moreover, for any
$0 < a < b < \infty$, $M > 0$, and using the substitution
$y = M/(M + x)$ (so that $x = M/y - M$ and
$dx = -My^{-2}dy$), 
\begin{multline}
\label{useful integral}
\int_a^b \frac{x^p}{(x+M)^q} \mathrm{d}x
=
\int_{\frac{M}{b+M}}^{\frac{M}{a+M}} M^p \frac{(1-y)^p}{y^p} \frac{y^q}{M^q}  \frac{M}{y^2} \mathrm{d} y
\\
=
M^{1+p-q} \int_{\frac{M}{b+M}}^{\frac{M}{a+M}}
(1-y)^{p} y^{q-p-2}
\mathrm{d} y
=
M^{1+p-q} \int^{1-\frac{M}{b+M}}_{1-\frac{M}{a+M}}
u^{p} (1-u)^{q-p-2}
\mathrm{d} u
.
\end{multline}
using the substitution $u = 1-y$ for the last equality.  
We also recall   from the calculations in 
Section~\ref{convergence_to_lambda_proofs}
that, for $\psi(N)/N \to \infty$ and $0< x < 1$,
\begin{displaymath}
\lim_{N\to\infty}\frac{N}{c_N}\prob{ \nu_1>Nx}=
\int_x^1 y^{-2} \Lambda_{+}(\mathrm{d}y)
\end{displaymath}
with $\Lambda_{+}(\mathrm{d}y) =C(\alpha)
(1-y)^{\alpha-1}y^{1-\alpha} \mathrm{d}y$, where the constant
$C(\alpha)=1/(\Gamma(\alpha)\Gamma(2-\alpha))$ 
(recall ~\eqref{eq:20} in Proposition~\ref{prop:lambdaRule} in Section~\ref{key lemmas}).

\begin{lemma} \label{lemma:sumbound2}
For $M > 0$  and $0 < a < b < \infty$ constants  we have,  for
$p,q > 0$, and  $p \neq q$,
\begin{displaymath}
\begin{split}
M^{1+p-q} \int^{1-\frac{M}{b+M}}_{1-\frac{M}{a+M}}
u^{p} (1-u)^{q-p-2}
\mathrm{d} u
& \leq
\sum_{i=a}^{b} \frac{i^{p}}{(i+M)^q}
\\
& \leq M^{1+p-q} \int^{1-\frac{M}{b+1+M}}_{1-\frac{M}{a-1+M}}
u^{p} (1-u)^{q-p-2}
\mathrm{d} u
+ \cO(M^{p-q}) \\
\end{split}
\end{displaymath}
\end{lemma}
\begin{proof}[Proof of Lemma~\ref{lemma:sumbound2}]
We begin with, recalling \eqref{useful integral}, 
\begin{displaymath}
\begin{split}
\sum_{i=a}^{b} \frac{i^{p}}{(i+M)^q}
&
\leq \int_{a-1}^{b+1} \frac{x^p}{(x+M)^q} \mathrm{d} x + \cO(M^{p-q})
\\
&
=
M^{1+p-q} \int^{1-\frac{M}{b+1+M}}_{1-\frac{M}{a   -1   + M  } }
u^{p} (1-u)^{q-p-2}
\mathrm{d} u
+ \cO(M^{p-q})
\end{split}
\end{displaymath}
where the $\cO(M^{p-q})$ correction appears since for $p\not= q$,
$i^p/(i+M)^q$ has a 
maximum at $i=pM/(q-p)$.
A similar calculation provides the opposite inequality.
\end{proof}




\begin{proof}[Proof of Lemma~\ref{lemma:ksamplesize}]
We partition over $A_N$~\eqref{eq:betaN} and $A_N^c$
~\eqref{eq:5}. {On } $A_N$ it holds that  
\begin{displaymath}
M_- \leq \tilde{S}_N \leq M_+
\end{displaymath}
with $\tilde{S}_{N}$ as in~\eqref{st}.
Recall~\eqref{eq:29},  $M_+, M_-$ from 
~\eqref{Mpm},  $S_{N}$ from~\eqref{fullSN}, and   note that  
\begin{equation}\label{eq:AN for prob}
\begin{split}
(M_{-}-n+k+1)^{n-k} \EE{\frac{ \svigi{X_{1}^{N}}_{k} }{ \svigi{X_{1}^{N} + M_{+}}_{n} } } &  \le  \EE{ \frac{ \svigi{X_{1}^{N}}_{k}\svigi{ \tilde S_{N}}_{n-k}  } { \svigi{X_{1}^{N} + \tilde S_{N}}_{n}} \one{S_{N} \ge N} {\one{ A_{N}}} } \\
& \le  M_{+}^{n-k}  \EE{\frac{ \svigi{X_{1}^{N}}_{k} }{ \svigi{X_{1}^{N} + M_{-}}_{n}  }  }
\end{split}
\end{equation}
We will use   that  
$\EE{H(X_{1}^{N})} = \sum_{k\ge 1}(H(k) - H(k-1))\prob{X_{1}^{N} \ge
k}$ for a real-valued function $H$ with $H(0) = 0$ \citep[Lemma~9]{schweinsberg03}, and the falling
factorial (recall ~\eqref{eq:10})   identities $(i)_{k}(i-k) =
i(i-1)_{k}$,  $(i)_{k} = i(i-1)_{k-1}$, and  $(i-1)_{k} =
(i-1)_{k-1}(i-k)$.  We now look at $\EE{{\svigi{X_{1}^{N}}_{k} }/{
 \svigi{X_{1}^{N} + M_{-}}_{n}  }   } $ from \eqref{eq:AN for prob}. We see 
\begin{displaymath}
\begin{split}
  &    \EE{ \frac{ \svigi{X_{1}^{N}}_{k} }{ \svigi{X_{1}^{N} + M_{-}}_{n}  }   }  = \sum_{i=k}^{\psi(N)}\left( \frac{ (i)_{k}}{ (i+M_{-})_{n}} -    \frac{ (i-1)_{k}}{ (i+M_{-} - 1)_{n}}\right)\prob{ X_1^N \geq i}   \\
= &   \sum_{i=k}^{\psi(N)}\left( \frac{ (i)_{k}(i + M_{-} -n) }{  (i + M_{-})_{n}(i + M_{-} - n) } -   \frac{ (i-1)_{k} (i + M_{-}) }{  (i+ M_{-})(i+M_{-} -1)_{n}}  \right)\prob{ X_1^N \geq i}    \\
= &    \sum_{i=k}^{\psi(N)} \frac{ (i-1)_{k-1}\left( i(i + M_{-} -n) - (i-k)(i + M_{-})\right) }{  (i + M_{-})_{n}(i + M_{-} - n) }    \prob{ X_1^N \geq i}      \\
= &    \sum_{i=k}^{\psi(N)} \frac{  (i-1)_{k-1}\left( kM_{-} - i(n-k) \right) }{  (i + M_{-})_{n}(i + M_{-} - n)  }  \prob{ X_1^N \geq i}       \\
= &   \sum_{i=k}^{\psi(N)} \frac{ (i-1)_{k-1} kM_{-} }{  (i + M_{-})_{n}(i + M_{-} - n)  }  \prob{ X_1^N \geq i} -    \sum_{i=k}^{\psi(N)} \frac{ i(i-1)_{k-1}(n-k) }{  (i + M_{-})_{n}(i + M_{-} - n)  }   \prob{ X_1^N \geq i}    \\ 
\end{split}
\end{displaymath}
Our next task is to control the first term in this expression. For any
$L\leq\psi(N)$ with $L/N\to 0$, 
\begin{displaymath}
\begin{split}
&  \sum_{i=k}^{\psi(N)}\frac{(i-1)_{k-1}kM_{-}}{(i+M_{-})_{n}(i+M_{-}-n)}\prob{X_{1}^{N}\ge i} =   kM_{-}  \sum_{i=k}^{\psi(N)}  \frac{(i-1)\dots(i-k+1)}{(i+M_- ) \dots (i+M_- - n)}  \prob{ X_1^N \geq i } \\
& \le  k\Mm \sum_{i=k}^{L-1} \frac{i^{k-1}}{(i+ \Mm - n)^{n+1}}  \prob{ X_1^N \geq i } +   k\Mm \sum_{i=L}^{\psi(N)} \frac{i^{k-1}}{(i+ \Mm - n)^{n+1}} \prob{ X_1^N \geq i }  \\
\end{split}
\end{displaymath}
Lemma~\ref{lemma:sumbound2} and~\eqref{eq:23} now give ({with}
$\overline{f}$, $\underline{f}$ from~\eqref{supinff})
\begin{displaymath}
\begin{split}
&  k\Mm \sum_{i=k}^{L-1}\left( \frac{i^{k-1}}{(i+ \Mm - n)^{n+1}} \right) \prob{ X_1^N \geq i } \\
& \le kM_{-}\overline f(k) \sum_{i=k}^{L-1}\left( \frac{i^{k-1}}{(i+ \Mm - n)^{n+1}} \right) \svigi{ \frac 1{i^{\alpha}} - \frac 1{(\psi(N) + 1)^{\alpha} }  } \\
& \le  kM_{-}\svigi{M_{-}-n}^{k-n-\alpha-1} \overline f(k)  \int_{1 -  \frac{M_{-}-n}{k-1+M_{-}-n } }^{1 -  \frac{M_{-}-n}{L+M_{-}-n } }u^{k-1-\alpha}(1-u)^{n-k+\alpha}{\rm d}u \\
& + \mathcal O\svigi{kM_{-} \svigi{M_{-}-n}^{k-\alpha - n}} \\
& - \frac{kM_{-}  \svigi{M_{-}-n }^{k-n-1}}{\svigi{\psi(N)+1}^{\alpha}} \overline f(k) \int_{1 - \frac{M_{-}-n}{k+M_{-}-n }}^{1 - \frac{M_{-}-n}{L-1 + M_{-}-n } }u^{k-1}\svigi{1-u}^{n-k}{\rm d}u \\
& - \mathcal O\svigi{kM_{-} \psi(N)^{-\alpha}\svigi{M_{-}-n}^{k- n-2}} \\
& \le  \frac{\overline f(k)}{k-\alpha}  kM_{-}\svigi{M_{-}-n}^{k-n-\alpha-1}\svigi{ \svigi{ \frac{L-1}{M_{-}-n} }^{k-\alpha} - \svigi{ \frac k{M_{-}-n} }^{k-\alpha}} \\
& + \mathcal O\svigi{kM_{-} \svigi{M_{-}-n}^{k-\alpha - n}} \\
&  - \frac{M_{-}  \svigi{M_{-}-n }^{k-n-1}}{\svigi{\psi(N)+1}^{\alpha}} \overline f(k) \svigi{\frac{M_{-}-n }{L-1+M_{-}-n } }^{n-k}\svigi{ \svigi{ \frac{L-1 }{L-1+M_{-}-n} }^{k} - \svigi{\frac{k}{k+M_{-}-n}}^{k}} \\
&  - \mathcal O\svigi{kM_{-} \psi(N)^{-\alpha}\svigi{M_{-}-n}^{k- n-2}} \\
\end{split}
\end{displaymath}

Again combining Lemma~\ref{lemma:sumbound2} and   ~\eqref{eq:23} gives 
\begin{displaymath}
\begin{split}
& kM_{-} \sum_{i=L}^{\psi(N)} \frac{i^{k-1} }{\svigi{i+M_{-} -n}^{n+1} }\prob{X_{1}^{N}\ge i } \\
& \leq
\overline{f}(L) kM_{-} \sum_{i=L}^{\psi(N)}
\frac{i^{k-1}}{(i+ \Mm - n)^{n+1}} \left( \frac{1}{i^\alpha} -  \frac{1}{(\psi(N) + 1)^\alpha} \right)
\\
& = 
\overline{f}(L)k \Mm \sum_{i=L}^{\psi(N)}
\frac{i^{k-\alpha-1}}{(i+M_- - n)^{n+1}}
  -  
\frac{ \overline{f}(L)k\Mm }{  (\psi(N) + 1)^\alpha } \sum_{i=L}^{\psi(N)}
\frac{i^{k-1}}{(i+ \Mm - n)^{n+1}}
\\
& \leq
\overline{f}(L)
k \Mm ( \Mm -n )^{k-n-\alpha-1}
\int_{1-\frac{M_--n}{L+M_- -n - 1}}^{1 - \frac{M_- -n}{\psi(N)+M_- -n+1}}
u^{k-\alpha-1} (1-u)^{n-k+\alpha} \mathrm{d}u \\ 
& + \mathcal O\svigi{kM_{-} \svigi{M_{-}-n}^{k-\alpha-n-2} }
\\
& - 
\frac{ \overline{f}(L)
k M_- (M_- -n )^{k-n-1}}{ (\psi(N) +1)^\alpha  }
\int_{1-\frac{M_--n}{L+M_- -n  }}^{1 - \frac{M_- -n}{\psi(N)+M_- -n  }}
u^{k-1} (1-u)^{n-k} \mathrm{d}u  \\
& + \mathcal O \svigi{kM_{-} \psi(N)^{-\alpha} \svigi{M_{-}-n}^{k - n-2} }
\\
\end{split}
\end{displaymath}

To find a corresponding lower bound, first observe that
\begin{equation}
\label{eq:30}
(i-1)\cdots (i-k+1) = \svigi{1 - \frac 1i} \cdots \svigi{1 - \frac{k+1}i}i^{k-1} = i^{k-1}\svigi{1 + \mathcal O\svigi{k^{2}/i} }
\end{equation}
It then holds by 
~\eqref{eq:23} and   Lemma~\ref{lemma:sumbound2} 
\begin{displaymath}
\begin{split}
&k\Mm \sum_{i=k}^{\psi(N)} 
\frac{(i-1)\dots(i-k+1)}{(i+M_- ) \dots (i+M_- - n)}
\prob{ X_1^N \geq i}
\\
& \ge  k\Mm  \sum_{i=k}^{L-1} 
\frac{i^{k-1} }{ (i+M_- )^{n+1} }
 \prob{ X_1^N \geq i}\svigi{1  + \mathcal O(k)} \\
& +
k\Mm  \sum_{i=L}^{\psi(N)} 
\frac{ i^{k-1} }{(i+M_- )^{n+1} }
\prob{ X_1^N \geq i}\svigi{1 + \mathcal O \svigi{ \frac{k^{2}}L} } \\
& \ge  \underline g(k) \svigi{1  + \mathcal O(k)}  kM_{-} \sum_{i=k}^{L-1} 
\frac{i^{k-1} }{ (i+M_- )^{n+1} } \svigi{\frac{1}{i^{\alpha}} -  \frac{1}{(\psi(N)+1)^{\alpha}} } \\
& + \underline  g(L)  kM_{-}\svigi{1 + \mathcal O \svigi{ \frac{k^{2}}L} }\sum_{i=L}^{\psi(N)}  \frac{i^{k-1} }{(i+M_{-})^{n+1}} \svigi{\frac{1}{i^{\alpha}} -  \frac{1}{(\psi(N)+1)^{\alpha}} } \\
& \ge  \underline g(k)  \svigi{1  + \mathcal O(k)}  kM_{-}^{k-n-\alpha}\int_{1 - \frac{M_{-} }{k+M_{-} }}^{1 - \frac{M_{-}}{L-1+M_{-} } }u^{k-1-\alpha}(1-u)^{n-k+\alpha}{\rm d}u  + \mathcal O\svigi{kM_{-}^{k-n-1-\alpha} } \\
& - \frac{\underline g(k)(1+\mathcal O(k))k}{(\psi(N) +1)^{\alpha}}M_{-}^{k-n} \int_{1 - \frac{M_{-}}{k-1-M_{-}} }^{1 - \frac{M_{-}}{L+M_{-} } }u^{k-1}(1-u)^{n-k}{\rm d}u +  \mathcal O\svigi{kM_{-}^{k-n-1} } \\
& +  \underline g(L)k\svigi{1 + \mathcal O \svigi{ \frac{k^{2}}L} }M_{-}^{k-\alpha -n}\int_{1 - \frac{M_{-}}{L+M_{-} } }^{1 - \frac{M_{-}}{\psi(N) + M_{-} } }u^{k-1-\alpha}(1-u)^{n-k+\alpha}{\rm d}u + \mathcal O\svigi{kM_{-}^{k-\alpha-n-1} } \\
& - \underline g(L)k \svigi{1 + \mathcal O \svigi{ \frac{k^{2}}L} }\frac{M_{-}^{k-n}}{(\psi(N)+1)^{\alpha} } \int_{1 - \frac{M_{-}}{L+M_{-} }}^{1 - \frac{M_{-}}{\psi(N)+M_{-} }}
u^{k-1}(1-u)^{n-k}{\rm d}u  +  \mathcal O\svigi{kM_{-}^{k-n-1} }
\end{split}
\end{displaymath}

Analogous arguments  yield

\begin{displaymath}
\begin{split}
& \sum_{i=k}^{\psi(N)} 
\frac{(i-1)\dots(i-k+1)}{(i+ \Mm ) \dots (i+ \Mm - n)}
(n-k)i  \prob{ X_1^N \geq i}
\\
& \le  \sum_{i=k}^{L-1 } \frac{ (n-k) i^{k} }{\svigi{i+M_{-}-n}^{n+1} }  \prob{ X_1^N \geq i}  + \sum_{i=L}^{\psi(N) } \frac{(n-k) i^{k} }{\svigi{i+M_{-}-n}^{n+1} }  \prob{ X_1^N \geq i} \\
& \le \sum_{i=k}^{L-1 } \frac{\overline f(k) (n-k) i^{k} }{\svigi{i+M_{-}-n}^{n+1} }\svigi{\frac 1{i^{\alpha}} - \frac{1}{(\psi(N) + 1)^{\alpha}} }  +  \sum_{i=L}^{\psi(N) } \frac{\overline f(L) (n-k) i^{k} }{\svigi{i+M_{-}-n}^{n+1} }\svigi{\frac 1{i^{\alpha}} - \frac{1}{(\psi(N) + 1)^{\alpha}} }     \\
& \le \overline f(k) (n-k) \svigi{\Mn}^{k-\alpha -n} \int_{1 - \frac{\Mn }{k-1 + \Mn }}^{1 - \frac{\Mn}{L+\Mn }}u^{k-\alpha}(1-u)^{n-k+\alpha - 1}{\rm d}u  + \mathcal O\svigi{       \svigi{\Mn}^{k-\alpha -n-1} }   \\
& - \frac{\overline f(k)(n-k) }{(\psi(N)+1)^{\alpha} }\svigi{\Mn}^{k-n} \int_{1 - \frac{\Mn}{k+\Mn}}^{1 - \frac{\Mn}{L-1+\Mn}}u^{k}(1-u)^{n-k-1}{\rm d}u + \mathcal O\svigi{ \svigi{\Mn}^{k-n-1}  }  \\
& +  \overline f(L)(n-k)\svigi{\Mn}^{k-\alpha - n}\int_{1 - \frac{\Mn }{L-1 + \Mn }}^{1 - \frac{\Mn}{\psi(N)+1 + \Mn } }u^{k-\alpha}(1-u)^{n-k+\alpha-1}{\rm d}u + \mathcal O           \svigi{  \svigi{\Mn}^{k-\alpha - n - 1}} \\
& -  \frac{\overline f(L)(n-k)}{(\psi(N)+1)^{\alpha} }\svigi{\Mn}^{k-n}\int_{1 - \frac{\Mn}{L+\Mn} }^{1 - \frac{\Mn}{\psi(N) + \Mn } }u^{k}(1-u)^{n-k-1}{\rm d}u  + \mathcal O     \svigi{ \svigi{\Mn}^{k-n-1} }  \\
\end{split}
\end{displaymath}

Similar calculations give us the lower bound (recall \eqref{eq:30})
\begin{displaymath}
\begin{split}
& \sum_{i=k}^{\psi(N)} 
\frac{(i-1)\dots(i-k+1)}{(i+ \Mm ) \dots (i+ \Mm - n)}
(n-k)i  \prob{ X_1^N \geq i} \\
& \ge  \sum_{i=k}^{L-1} \frac{(n-k) i^{k}}{ \svigi{i+M_{-}}^{n+1} }  \prob{ X_1^N \geq i}\svigi{1 + \mathcal O(k)} +  \sum_{i=L}^{\psi(N)} \frac{(n-k) i^{k}}{ \svigi{i+M_{-}}^{n+1} }  \prob{ X_1^N \geq i}\svigi{1 + \mathcal O\svigi{\frac{k^{2}}L } }  \\
&  \ge  \underline g(k) \svigi{1 + \mathcal O(k)}(n-k)  \sum_{i=k}^{L-1} \frac{ i^{k}}{ \svigi{i+M_{-}}^{n+1} }\svigi{\frac 1{i^{\alpha}} - \frac{1}{(\psi(N) + 1)^{\alpha}}  }   \\
& +   \underline g(L) \svigi{1 + \mathcal O\svigi{\frac{k^{2}}L}}(n-k)  \sum_{i=L}^{\psi(N)} \frac{ i^{k}}{ \svigi{i+M_{-}}^{n+1} }\svigi{\frac 1{i^{\alpha}} - \frac{1}{(\psi(N) + 1)^{\alpha}}  } \\ 
\end{split}
\end{displaymath}
We can now apply   Lemma~\ref{lemma:sumbound2} to get 
\begin{displaymath}
\begin{split}
& \sum_{i=k}^{\psi(N)} 
\frac{(i-1)\dots(i-k+1)}{(i+ \Mm ) \dots (i+ \Mm - n)}
(n-k)i  \prob{ X_1^N \geq i} \\
& \ge  \underline g(k)  \svigi{1 + \mathcal O(k)} (n-k)M_{-}^{k-\alpha - n}\int_{1 - \frac{M_{-}}{k+M_{-} } }^{1 - \frac{M_{-}}{L-1+M_{-} } }u^{k-\alpha}(1-u)^{n-k+\alpha -1}{\rm d}u  +  \mathcal O\svigi {M_{-}^{k-\alpha -n-1} } \\
& - \frac{\underline g(k)  \svigi{1 + \mathcal O(k)}(n-k)}{(\psi(N)+1)^{\alpha} } M_{-}^{k-n}\int_{1 - \frac{M_{-}}{k-1+M_{-} } }^{1 - \frac{M_{-}}{L+M_{-} } }u^{k}(1-u)^{n-k-1}{\rm d}u + \mathcal O     \svigi{M_{-}^{k-n-1} }  \\
& +   \underline g(k)\svigi{1 + \mathcal O\svigi{\frac{k^{2}}L} }(n-k) M_{-}^{k-\alpha - n}\int_{1 - \frac{M_{-} }{L+M_{-} } }^{1 - \frac{M_{-} }{\psi(N) + M_{-} } }u^{k-\alpha}(1-u)^{n-k+\alpha - 1}{\rm d}u  +  \mathcal O\svigi{M_{-}^{k-\alpha -n-1}}   \\
& -  \frac{\underline g(L)  \svigi{1 + \mathcal O\svigi{\frac{k^{2}}L}}(n-k)}{(\psi(N) + 1)^{\alpha} } M_{-}^{k-n} \int_{1 - \frac{M_{-} }{L-1+M_{-} } }^{1 - \frac{M_{-}}{\psi(N) + 1+M_{-} } } u^{k}(1-u)^{n-k-1}{\rm d}u  + \mathcal O\svigi{M_{-}^{k-n-1} }  \\
\end{split}
\end{displaymath}

Bounds for $\EE{\svigi{X_{1}^{N}}_{k}/\svigi{X_{1}^{N} + M_+}_{n}}$  (recall \eqref{eq:AN for prob})  follow from  analogous arguments.

We must now estimate the terms {corresponding to  }  $\comp{A}_N$ \eqref{eq:5}.
When $n/N \to 0$, for sufficiently large $N$, using the identity
$(x)_{n} =  (x)_{n-k}(x-n+1)\cdots(x-n+k)$, 
\begin{multline*}
\EE{ \frac{ (X_{1}^{N})_{k}(\tilde S_{N})_{n-k}  }{ \svigi{X_{1}^{N} + \tilde S_{N}}_{n} }\one{S_{N} \ge N}   } \\
 =  \EE{ \frac{ (X_{1}^{N})_{k} (\tilde S_{N})_{n-k}  }{ \svigi{X_{1}^{N} + \tilde S_{N}}_{n-k} \svigi{X_{1}^{N} + \tilde S_{N} - n+k}\cdots  \svigi{ X_{1}^{N} + \tilde S_{N} - n+1}}  \one{S_{N} \ge N}  }
\\
\leq \IE\left[ \frac{X_1^N \svigi{X_1^N-1}\dots\svigi{X_1^N-k+1}}{ \svigi{X_1^N + \tilde{S}_N-n+k}\dots\svigi{ X_1^N+\tilde{S}_N - n+1}} \one{S_N \geq N} \right ] 
\\
\leq
\IE\left[  \frac{X_1^N(X_1^N-1)}{S_N^2} \one{S_N \geq N}
    \right] \left( 1 + \cO\left(\frac{(n-k)k}{N} \right) \right)
.
\end{multline*}
The error term is calculated by partitioning over $X_1^N\leq N$ and 
$X_1^N > N$. In the first case, each of the $k$ `spare terms' 
$(X_1^N-j)/(X_1^N+\tilde{S}_N-n+j)$ 
is bounded by $N/(N-(n-k))$; if $X_1^N>N$ each such term is bounded by
$X_1^N/(X_1^N+n-k)$.
Now using the 
elementary facts that $(i +b)^k/i^k = 1 + \cO(kb/i)$ and 
$(i-b)^k/i^k = 1 - \cO(kb/i)$  we recover the error term above.

We now  see that the error term $\mathcal O\svigi{N\beta_{N}^{-\alpha}}$
in \eqref{eq:lemmapsi=N} and \eqref{eq:lemmapsi>N} comes 
from \eqref{eq:psi k prob} in the case $\psi(N)/N \to K$ and 
from \eqref{eq:psi infty prob} when $\psi(N)/N \to
\infty$. {The proof of Lemma~\ref{lemma:ksamplesize} is complete.}
\end{proof}

\subsection{Heuristic arguments for sample size
$\mathcal{O} \svigi{ N^{\alpha/2}}   $}
\label{sec:heur-argum-sample}

Here we give heuristics for the sample size when we would expect  to
start seeing 
anomalies in the topology of the tree predicted by the complete
Beta$(2-\alpha,\alpha)$ coalescent {(Definition~\ref{betacoal})}.

Since
\begin{displaymath}
\lambda_{m,k} = \int_0^1 \frac{x^{k-\alpha-1} (1-x)^{m-k+\alpha-1}}{\Gamma(\alpha)\Gamma(2-\alpha)} \mathrm{d}x
=
\frac{\Gamma(k-\alpha) \Gamma(m-k+\alpha)}{\Gamma(m)\Gamma(\alpha)\Gamma(2-\alpha)}
,
\end{displaymath}
the total rate of a $k$-merger (with $k > 2$)   when  $m$ lineages is given by
\begin{displaymath}
\begin{split}
\binom{m}{k} \lambda_{m,k}
= &
\binom{m}{k}
\frac{\Gamma(k-\alpha) \Gamma(m-k+\alpha)}{\Gamma(m)\Gamma(\alpha)\Gamma(2-\alpha)}
\\
= &
\frac{m!}{k! (m-k)!(m-1)!\Gamma(\alpha)\Gamma(2-\alpha)}
{\Gamma(k-\alpha) \Gamma(m-k+\alpha)}
\\
> &
\frac{m!}{k! (m-k)!(m-1)!\Gamma(\alpha)\Gamma(2-\alpha)}
\frac{\Gamma(k-1) \Gamma(m-k+2) }{(k-1)^{\alpha-1}(m-k+2)^{2-\alpha}}
\\
= &
\frac{m!}{k! (m-k)! (m-1)!\Gamma(\alpha)\Gamma(2-\alpha)}
\frac{(k-2)! (m-k+1)! }{(k-1)^{\alpha-1}(m-k+2)^{2-\alpha}}
\\
= &
\frac{m(m-k+1)}{k(k-1) (k-1)^{\alpha-1}(m-k+2)^{2-\alpha}}\frac{1}{\Gamma(\alpha)\Gamma(2-\alpha)}
,
\end{split}
\end{displaymath}
obtained by writing  $\Gamma(k-\alpha) = \Gamma((k-2) + (2-\alpha))$ and 
$\Gamma(m-k+\alpha) =  \Gamma((m-k+1) + (\alpha-1))$ and 
using  Gautschi's inequality   \citep{Gautschi1959}
\begin{displaymath}
x^{1-s} < \frac{\Gamma(x+1)}{\Gamma(x+s)} < (x+1)^{1-s}
\end{displaymath}
(so that $\Gamma(x+s) > \Gamma(x+1)/(x+1)^{1-s}$)   for $x>0$ and $s \in (0,1)$.

{Suppose } $k = \sqrt{N} \to \infty$.  
 Approximating {  $\sum_{\ell=k}^{m} \binom{m}{\ell}\lambda_{m,\ell}$  }   by an integral,   the total rate of mergers 
involving {at least} $k$  {lineages}  can be approximated (up to a constant) by
\begin{displaymath}
\int_k^m \frac{m(m-x)^{\alpha-1}}{x^{\alpha+1}} \mathrm{d}x
=
\int_0^{\frac{m-k}{k}} u^{\alpha-1} \mathrm{d}u
=
\frac{1}{\alpha}
\left(
\frac{m-k}{k}
\right)^{\alpha}
,
\end{displaymath}
by  the substitution $u = {(m-x)}/{x}$ (so that  $x =
m/(1+u)$, and  $dx = -m/(1+u)^{2}du$). 

Now consider the number of ancestral lineages present in the
Beta$(2-\alpha,\alpha)$ coalescent at any given time. We recall the
following result   from   \cite{BBS08}. 

\begin{thm}[\cite{BBS08}, {Theorem~ 1.1}]
\label{thm:bere2008}
Let $\mathds{B}_t$ be the remaining number of ancestral lines  at time $t$ in
the complete Beta$(2-\alpha,\alpha)$ coalescent \eqref{betadens}
with $1 < \alpha < 2$.  Then 
\begin{displaymath}
\lim_{t \to 0} t^{\frac{1}{\alpha-1}}\mathds{B}_t = \left(\alpha \Gamma(\alpha)\right)^{\frac{1}{\alpha-1}}
\end{displaymath}
holds  almost surely.
\end{thm}
We now present some simple heuristic arguments.
We will be looking at small time results and so, 
with an abuse of notation, we take  (see Theorem~\ref{thm:bere2008})
\begin{displaymath}
\mathds{B}_t = \left(\frac{\alpha \Gamma(\alpha)}{t}\right)^{\frac{1}{\alpha-1}}
.
\end{displaymath}
We then see that the expected number of mergers involving more 
than $k$ individuals when started from $m$ individuals is approximated by
\begin{equation}
\label{intmk}
\int_{\frac{\alpha\Gamma(\alpha)}{m^{\alpha-1}}}^{\frac{\alpha\Gamma(\alpha)}{k^{\alpha-1}}}
\frac{1}{\alpha} \left(\frac{\left(\frac{\alpha\Gamma(\alpha)}{t}\right)^{\frac{1}{\alpha-1}} - k}{k} \right)^\alpha \mathrm{d}t
= \cO\left(\frac{m}{k^\alpha} \right)
.
\end{equation}
The integral is dominated by the contribution for times arbitrarily close
to the starting time, so as soon as $m=\cO(N^{\alpha/2})$ we expect
to see mergers of at least $\sqrt{N}$ lineages at arbitrarily small
times. 
Since the external branches of a Beta-coalescent extend into the `body' of the
tree, see e.g.\   \citep{DAHMER2014} (unlike in the Kingman case in which the ratio of external to internal
branches tends to zero),  such mergers can be expected to influence the 
site-frequency spectrum   for sample sizes of
$\cO(N^{\alpha/2})$, as claimed.   This holds even though $\tbinom{n}{k}\lambda_{n,k} \ge \tbinom{n}{k+1}\lambda_{n,k+1}$ and  
\begin{displaymath}
\frac{ \binom{n}{2}\lambda_{n,2} }{ n^\alpha } \to 1 
\end{displaymath}
as $n\to \infty$.   In words, the merger probabilities are decreasing
as a function of merger size, so a 2-merger  is always {the
most likely merger},  and
the probability of seeing a 2-merger tends to one as sample size increases.

\section{Convergence to the Kingman Coalescent}%
\label{prerrorkc}%
In this section we use Proposition~\ref{prop:kingmanRule}, where the
condition given in ~\eqref{no triples} requires us to control the
probability of three ancestral lines merging, to prove
Proposition~\ref{errorkc}.

\begin{proof}[Proof of Proposition~\ref{errorkc}.]

Recalling  $S_{N}$ from  ~\eqref{fullSN}
and setting $\beta_N=N$, we see
\begin{displaymath}
\frac{N}{c_N} \IE\left[\frac{ \svigi{X_1^N}_3}{ (S_N)_3} \one{S_N \geq N} \right]
=  \frac{N}{c_N} \IE\left[\frac{ \svigi{ X_1^N}_3}{S_N^3} \one{S_N \geq N} \right]\left(1 + \OO{\frac 1N } \right) 
\end{displaymath}
Then, according to  Lemma~\ref{lemma:Schw6}, it will suffice to replace
$c_N$ by $\mathcal{C}_N$, and from Proposition~\ref{pr:CNversion0},
$\mathcal{C}_N = \cO \svigi{\psi(N)^{2-\alpha} / N }$. Recall
 that we are assuming that $\psi(N)/N \to 0$. We then see
\begin{displaymath}
\frac{N}{c_N} \IE\left[\frac{ \svigi{ X_1^N}_3}{S_N^3} \one{S_N \geq N}
\right] \leq \frac{N}{c_N} \frac{\IE\left[ \svigi{ X_1^N}_3\right]}{N^3} = 
\cO\left(\frac{N^2}{\psi(N)^{2-\alpha}}
\frac{\psi(N)^{3-\alpha}}{N^3}\right) =
\cO\left(\frac{\psi(N)}{N}\right) .
\end{displaymath}
From this we can conclude convergence to Kingman coalescent by
Proposition \ref{prop:kingmanRule}. To establish \eqref{eq:Kingman1}
it remains to check the lower bound.
 Recalling $A_N$ from \eqref{eq:betaN} and $M_+$ from
 \eqref{Mpm} and  $\tilde S_{N}$ from ~\eqref{st},  {it holds that}
 \begin{displaymath}
 \begin{split}
\frac{N}{c_N} \IE\left[\frac{ \svigi{X_1^N}_3}{S_N^3} \one{S_N \geq N} \right]
\geq &
\frac{N}{c_N} \IE\left[\frac{ \svigi{ X_1^N}_3}{\svigi{\psi(N) + \tilde{S}_N}^3} \one{S_N \geq N} \one{A_N} \right]
\\
\geq & \frac{N}{c_N}
\frac{\IE\left[ \svigi{ X_1^N}_3\right]}{\svigi{\psi(N) + M_+}^3}
\IP[A_N]
\\
=   &
\cO\left(
\frac{N^2}{\psi(N)^{2-\alpha}}
\frac{\psi(N)^{3-\alpha}}{N^3}
\right)
= \cO\left(\frac{\psi(N)}{N} \right)
,
\end{split}
\end{displaymath}
where we used Lemma~\ref{lemma:largepsibound}
to deduce that $\IP[A_N]$ is $\mathcal{O}(1)$.

Considering \eqref{eq:Kingman2}, we obtain
\dsplit{
\frac{N^2}{c_N} \IE\left[\frac{ \svigi{ X_1^N}_2 \svigi{ X_2^N}_2}{S_N^4} \one{S_N \geq N} \right]
\leq &
\frac{N^2}{c_N} \frac{ \IE\left[ \svigi{ X_1^N}_2 \svigi{ X_2^N}_2 \right]}{N^4}
= 
\frac{\IE\left[ \svigi{ X_1^N}_2 \right]^2}{c_N N^2 }  
\\
=  & \cO\left(\frac{\psi(N)^{4-2\alpha}}{\psi(N)^{2-\alpha}N} \right)
= \cO\left(\frac{\psi(N)^{2-\alpha}}{N} \right)
.
}
For the opposite inequality, 
we define
\begin{displaymath}
\hat{A}_N := \left\{ \left|\sum_{i=3}^N X_i^N - (N-2) m_N \right| \leq \beta_N \right\}
.
\end{displaymath}
Proceeding  as before we  see
\dsplit{
\frac{N^2}{c_N} \IE\left[\frac{ \svigi{ X_1^N}_2 \svigi{ X_2^N}_2}{S_N^4} \one{S_N \geq N} \right]
\geq &
\frac{N^2}{c_N} \IE\left[\frac{ \svigi{ X_1^N}_2 \svigi{ X_2^N}_2}{ \svigi{ 2\psi(N) + \sum_{i=3}^N X_i^N}^4} \one{S_N \geq N} \one{\hat{A}_N} \right]
\\
\geq & \frac{N^2}{c_N}
\frac{\IE\left[ \svigi{ X_1^N}_2 \svigi{ X_2^N}_2 \right]}{ \svigi{ 2\psi(N) + (N-2)m_N + \beta_N}^4}
\IP \left[\hat{A}_N\right]
\\
=  &
\cO\left(
\frac{N^3}{\psi(N)^{2-\alpha}}
\frac{\psi(N)^{4-2\alpha}}{N^4}
\right)
= \cO\left(\frac{\psi(N)^{2-\alpha}}{N} \right)
,
}
where again we chose $\beta_N = N$, and used
Lemma~\ref{lemma:largepsibound}. {This concludes the proof of Proposition~\ref{errorkc}.}
\end{proof}

\section{Convergence to a Beta-coalescent}%
\label{convergence_to_lambda_proofs}%

In this section we use Proposition~\ref{prop:lambdaRule}
to prove Proposition~\ref{pn:Lambda-convergence 1<a<2 full version}.

We divide the proof of Proposition~{\ref{pn:Lambda-convergence 1<a<2
full version}} into two parts. First we control the probability of
simultaneous multiple mergers (recall the condition in ~\eqref{no
simultaneous} in Proposition~\ref{prop:lambdaRule}), and then we
identify the measure $\Lambda$ in the limiting coalescent, see
~\eqref{eq:20} in Proposition~\ref{prop:lambdaRule}.

\begin{proof}[Proof of \eqref{eq:Lambdaeq1}]

According to Lemma~\ref{lemma:Schw6} we can replace $c_N$ by
$\mathcal{C}_N$ (see \eqref{curlyc}), and according to
Proposition~\ref{pr:CNversion0}
$\mathcal{C}_N = \cO(N^{1-\alpha})$ (recall $1 < \alpha < 2$). Then, assuming that $N$ is
large enough that $m_N>1$ 
following   \cite{schweinsberg03} we see
\begin{equation}
\label{star}
\begin{split}
\frac{N^2}{c_N} \IE\left[\frac{ \svigi{ X_1^N}_2 \svigi{X_2^N}_2}{S_N^4} \one{S_N \geq N}
\right] \leq 
\frac{N^2}{c_N}
\IE\left[
\frac{ \svigi{ X_1^N}_2 \svigi{X_2^N}_2}{\max\mengi{ \svigi{X_1^N}^2, N^2} \max\mengi{ \svigi{X_2^N}^2, N^2}}
\right]
\\
=
\frac{N^2}{c_N}
\left(
\IE\left[
\frac{ \svigi{ X_1^N}_2}{\max\mengi{ \svigi{X_1^N}^2, N^2}}
\right]
\right)^2
\leq
16\frac{N^2}{c_N}
\left(
\IE\left[
\frac{ \svigi{ X_1^N}_2}{ \svigi{X_1^N + N}^2}
\right]
\right)^2
=  \cO\left( N^{ 1 - \alpha }\right)
,
\end{split}
\end{equation}
where we have used Lemma~\ref{lm:newexp1} for the final estimate.

We turn to the lower bound. 
Using the Notation~\ref{defn:ANandM}  and  $M_{+}$ from~\eqref{Mpm}   we have
\begin{displaymath}
\begin{split}
\frac{N^2}{c_N} \IE\left[\frac{ \svigi{ X_1^N}_2 \svigi{X_2^N}_2}{S_N^4} \one{S_N \geq N}
\right]
\geq &
\frac{N^2}{c_N} \IE\left[\frac{ \svigi{X_1^N}_2 \svigi{X_2^N}_2}{ \svigi{X_1^N + M_+ } ^4} \one{A_N}
\right]
\\
= &
\frac{N^2}{c_N} \IE\left[\frac{ \svigi{ X_1^N}_2 }{ \svigi{X_1^N + M_+ } ^4} \right] \IE\left[ \svigi{ X_2^N}_2\one{A_N}
\right]
\\
\geq &
\frac{N^2}{c_N} \IE\left[\frac{ \svigi{ X_1^N} _2 }{ \svigi{ X_1^N + M_+ } ^4} \one{X_1^N \leq M_+}\right]
\\
&
\times \IE\left[ \svigi{ X_2^N}_2\one{ \tolugildi{ \sum_{i=3}^N X_i^N - (N-2)m_N } \leq \frac{\beta_N}{2}}\one{ \tolugildi{ X_2^N - m_N} \leq \frac{\beta_N}{2}}
\right]
\\
\geq &
\frac{N^2}{c_N} \IE\left[\frac{ \svigi{ X_1^N}_2 }{16 M_+^4} \one{X_1^N \leq M_+}\right] 
\\
&
\times
\IE\left[ \svigi{ X_2^N}_2\one{ \tolugildi{ X_2^N - m_N} \leq \frac{\beta_N}{2}}
\right] \\
& \times  \IP\left[\left|\sum_{i=3}^N X_i^N - (N-2)m_N \right| \leq \frac{\beta_N}{2}\right]
\\
=  &
\cO\left(
\frac{N^2}{N^{1-\alpha}}
\frac{M^{2-\alpha}_+}{M_+^4}
\beta_N^{2-\alpha}
\right)
= 
\cO\left(
N^{1-\alpha}
\right)
,
\end{split}
\end{displaymath}
where we choose $\beta_N =N$ to optimise on the final line and
Lemma~\ref{lemma:largepsibound}
to deduce that the probability on the penultimate line
is $\mathcal{O}(1)$. {The proof of \eqref{eq:Lambdaeq1} is complete.}
\end{proof}

\begin{proof}[Proof of 
Proposition~\ref{pn:Lambda-convergence 1<a<2 full version}:
identifying the measure $\Lambda_{+}$]

We first note that \eqref{eq:Lambdaeq1} and Lemma~\ref{lemma:Schw6} 
reduce the proof of Proposition~\ref{pn:Lambda-convergence 1<a<2 full version}
to examining
\begin{displaymath}
\lim_{N \to \infty}\frac{N}{c_N} \IP[{\nu_{1}} > Nx] =
\lim_{N \to \infty}\frac{N}{c_N} \IP\left[\frac{X_{1}^N}{S_N} \one{S_N \geq N} > x\right]
\end{displaymath}
for $0 < x < 1$. We essentially follow~  \cite{schweinsberg03}, pp 129--131.
We start in a familiar way by conditioning on
$\set{\tilde{S}_N \geq (N-1)m_N - \beta_N} =\set{\tilde{S}_N \geq M_- }
$.
As before we assume $\beta_N/N$ is bounded (with $\beta_{N}$ for example as in ~\eqref{eq:6}), and recall the definition
of $A_N$ from Notation~\ref{defn:ANandM}, and of $M_{+}$ and $M_{-}$
from  \eqref{Mpm}. We see, using Lemma~\ref{lemma:largepsibound} for  $\prob{A_{N}^{c}}$, and that 
$1/c_N = \OO{N^{\alpha - 1}}$ from  Proposition \ref{pr:CNversion0},
\dsplit{
\frac{N}{c_N} \IP\left[\frac{X_{1}^N}{S_N} \one{S_N \geq N} > x\right]
\leq &
\frac{N}{c_N}
\IP\left[\frac{X_{1}^N}{X_1^N + M_-}  > x\right] +
\frac{N}{c_N}
\IP\left[ \tilde{S}_N < M_-  \right]\IP\left[ \frac{X_1^N}{N} \geq x \right]
\\
\leq &
\frac{N}{c_N}
\IP\left[X_{1}^N  > \frac{x M_-}{1-x} \right]
+
\frac{N}{c_N}
\IP\left[\comp{A}_N\right]\IP\left[ X_1^N \geq xN \right]
\\
\leq &
\frac{N}{c_N}\overline{f}\left( \frac{xM_-}{1-x} \right) \frac{1}{M_ -^\alpha}  \left( \left(\frac{1-x}{x} \right)^\alpha    -  \frac{M_ -^{\alpha}}{(\psi(N)+1)^\alpha}  \right)   
+
\cO\left(
\overline{f}(xN)\frac{N}{\beta_N^\alpha}
\right)
.
}
Considering a lower bound, we have 
\dsplit{
\frac{N}{c_N} \IP\left[\frac{X_{1}^N}{S_N} \one{S_N \geq N} > x\right]
\geq &
\frac{N}{c_N} \IP[A_N] \IP\left[\frac{X_1^N}{X_1^N + M_+} > x \right]
\\
= &
\frac{N}{c_N} \IP[A_N] \IP\left[ X_1^N > \frac{xM_+}{1-x} \right]
\\
\geq &
\frac{N}{c_N M_+^\alpha}\underline{g}\left( \frac{xM_+}{1-x }  \right) \left( \left( \frac{1-x}{x}\right)^\alpha    -  \frac{M_ +^{\alpha}}{ (\psi(N)+1)^\alpha}  \right)
\left( 1 - \OO{ \frac{ N  }{\beta_{N}^{\alpha}} } \right)
.
}
Combining these estimates with \eqref{star} we conclude from
Proposition \ref{prop:lambdaRule}, assuming $\beta_N/N$ is bounded,
convergence to the claimed coalescent; for any $A\ge 0$, $r > 0$, and
$0<x \le 1/(1 + A)$,
\begin{displaymath}
\braces{ \frac{1-x}{x}}^r - A^r =  r\int_{x}^{\tfrac{1}{1+A}} t^{-1-r}(1-t)^{r-1}\mathrm{d}t
\end{displaymath}
seen using the substitution  $u = (1-t)/t$.
Since  $\int y^{-1-\alpha}(1-y)^{\alpha -1}{\rm d}y  =  \int
 y^{-2} \left(y^{1-\alpha}(1-y)^{\alpha -1} \right){\rm d}y$ we have, as 
 required by Proposition~\ref{prop:lambdaRule}, identified a 
 measure $\Lambda_{+}$  so that the limit in ~\eqref{eq:20} holds.   
Using Lemma~\ref{lemma:Schw6}
to replace $c_N$ by $\mathcal{C}_N$ from 
Proposition~\ref{pr:CNversion0},
the proof of Proposition~\ref{pn:Lambda-convergence 1<a<2 full version} 
is complete.
\end{proof}


\textbf{Acknowledgements}

JACD supported by an Engineering and Physical Sciences Research
Council (EPSRC) grant number EP/L015811/1; BE supported in part by
EPSRC grant EP/G052026/1, by Deutsche Forschungsgemeinschaft
(DFG) grant BL 1105/3-1 to Jochen Blath through SPP Priority Programme
1590 Probabilistic Structures in Evolution, by DFG grant with
Projektnummer 273887127 through DFG SPP 1819: Rapid Evolutionary
Adaptation grant STE 325/17 to Wolfgang Stephan; AME and BE
acknowledge funding by Icelandic Centre of Research (Rann\'is) through
an Icelandic Research Fund (Ranns\'oknasj\'o{\dh}ur) Grant of
Excellence no.\ 185151-051 jointly with Einar \'Arnason, Katr\'in
Halld\'orsd\'ottir, Alison Etheridge, and Wolfgang Stephan; BE also
acknowledges SPP 1819 Start-up module grants jointly with Jere Koskela
and Maite Wilke Berenguer, and with Iulia Dahmer.  We thank Alison M.\
Etheridge for helpful comments and suggestions, and especially for
suggesting Lemma~\ref{sec:usefulapproxs}.

\textbf{Statements and Declarations}

The authors have no competing interests to declare that are relevant
to the content of this article. 

\textbf{MSC Classification}

60J10, 60J90, 60J95,  60-04

\textbf{Data availability}
The software developed for  this study is freely  available at \\  \url{https://github.org/eldonb/Beta\_coalescents\_when\_sample\_size\_is\_large}

\appendix 

\numberwithin{equation}{section}
\numberwithin{figure}{section}

\section{Comparing  Beta-coalescents}
\label{sec:furth-numer-exampl}

\begin{figure}[htp]
\centering
\caption{Comparing {$\overline \varrho_{i}(n)$}  between Beta-coalescents.
Estimates of  $\EE{R_{i}(n)}$ (recall
{\eqref{eq:8}}  in Section~\ref{numerics}) for sample size $n=100$
predicted by the Beta$(2-\alpha,\alpha)$-coalescent ($K=\infty$,  red line), the
incomplete Beta-coalescent for values of $K$ (recall \eqref{incbeta})
as shown and with $m_{\infty}$ approximated as in ~\eqref{eq:1}, and
the Kingman coalescent (black line). For all the Beta coalescents
$\alpha = 1.01$; results from $10^{5}$ experiments. }
\label{figK}%
\includegraphics[scale=1]{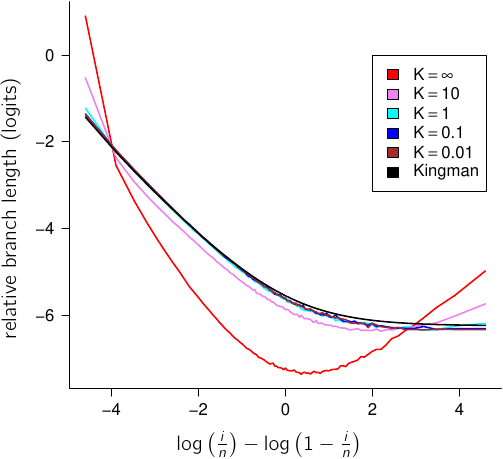}
\end{figure}

In this section we give a numerical   example  (see Figure~\ref{figK})
showing how estimates of   mean 
relative  branch lengths $\EE{R_{i}(n)}$, recall  ~\eqref{eq:8},  compare
between the   incomplete and the   complete Beta-coalescent.  
The aim is to understand how inference based on the site-frequency
spectrum might be influenced by the upper bound (recall~\eqref{eq:PXiJ}).  The main result is
that ignoring the upper bound  can
lead to wrong conclusions.  Recall the notation introduced in
Section~\ref{numerics}.

In Figure~\ref{figK} we compare relative branch lengths sampled under
different coalescents,  by iteratively sampling trees 
under a given coalescent, and recording the relative branch lengths
for each iteration, all for sample size $n= 100$.  The coalescents
considered are the Beta-coalescent of \cite{schweinsberg03} (see
Definition 
\ref{betacoal}; red line), the Kingman coalescent (black line), and
the incomplete Beta-coalescent (recall  \eqref{incbeta}) for
values of $K$ as shown.  For all the Beta-coalescents $\alpha = 1.01$.
Figure~\ref{figK} clearly shows that for small $K$, the site-frequency
spectrum predicted by the incomplete Beta-coalescent resembles that of
the Kingman coalescent more than that of the complete Beta-coalescent.
This should not be surprising given the form of the coalescent rate
for the incomplete Beta coalescent (recall  \eqref{incbeta}),
where we integrate over the interval $(0, K/(K+m_\infty))$, so that
smaller values of $K$ translate to smaller probability of each line
participating in a merger, reflecting the role of $K$ in the
prelimiting model (cf. Case~\ref{JCD1psi=N} in Proposition 
\ref{pr:CNversion0}). This conclusion holds for any
$\alpha \in (1,2)$, and applying the complete Beta coalescent in
inference, instead of the incomplete Beta coalescent, can lead to
overestimating $\alpha$. A `U-shaped' (an excess of singletons and
high-frequency variants relative to predictions of the
Kingman-coalescent) is sometimes seen as a characteristic of
$\Lambda$-coalescents \citep{BBE2013a,Freund2022.04.12.488084}.  Our
results show that, even though the offspring number distribution is
highly skewed and in the domain of attraction of a multiple-merger
coalescent, the predicted site-frequency spectrum is not necessarily
U-shaped.

Even though we have an explicit form of the cutoff point
($K/(K+m_{\infty})$) of the incomplete Beta-coalescent, in inference
one may view the stated coalescent as involving two parameters,
$\alpha \in (1,2)$, and (say) $\gamma \in (0,1]$, where $\gamma$
represents the cutoff point. One could  refer to the incomplete
Beta-coalescent as the Beta$(\gamma;2-\alpha, \alpha)$-coalescent (and
the Beta$(2-\alpha,\alpha)$-coalescent is the
Beta$(\gamma; 2-\alpha,\alpha)$-coalescent when $\gamma = 1$).  Given
population genetic  data one would then proceed to jointly estimate $\alpha$ and $\gamma$.

\clearpage
\pagebreak
\newpage

\section{Comparing     $\overline\varrho_{i}^{N}(n) $  and
$\overline \rho_{i}^{N}(n)$}
\label{sec:quenched}

\begin{figure}[H]
\centering
 \captionsetup[subfloat]{labelfont={scriptsize,sf,md,up},textfont={scriptsize,sf}}
\subfloat[$\psi(N) = \infty$, $\alpha = 1.01$]{\includegraphics[scale=0.7]{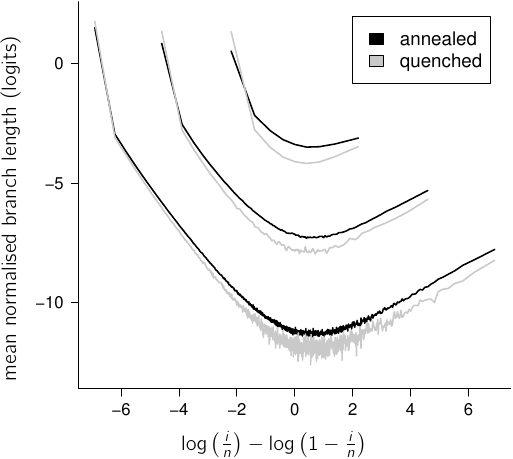}}
\subfloat[$\psi(N) = N$, $\alpha = 1.01$]{\includegraphics[scale=0.7]{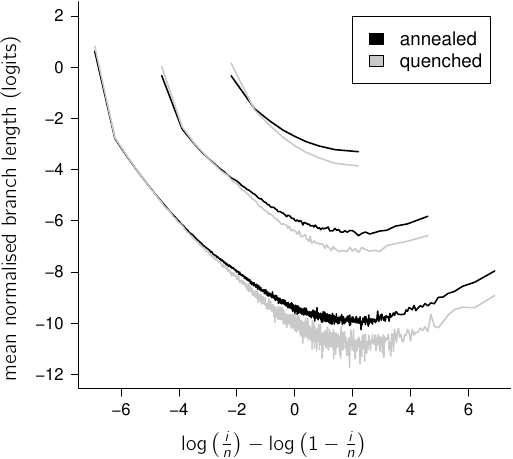}}\\
\subfloat[$\psi(N) = \infty$, $\alpha = 2$]{\includegraphics[scale=0.7]{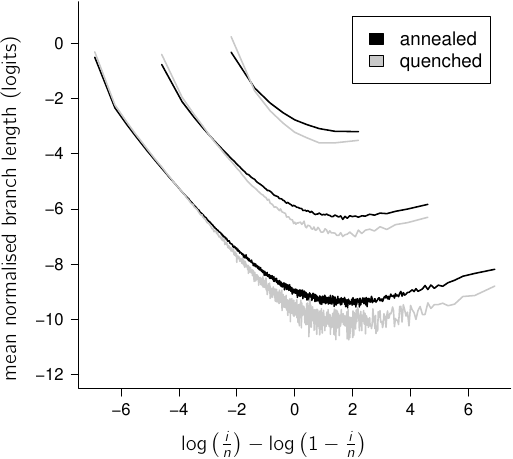}}
\subfloat[$\psi(N) = N/\log N$, $\alpha = 1.01$]{\includegraphics[scale=0.7]{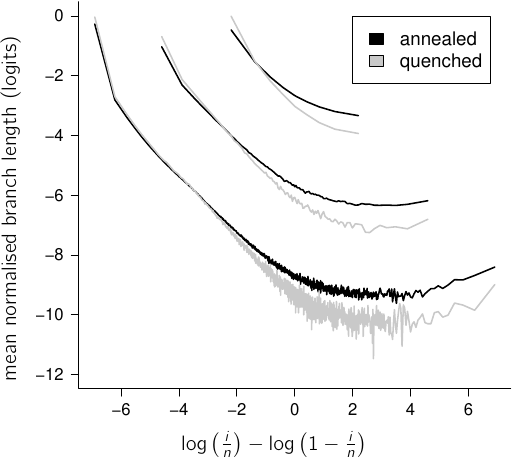}}\\
\caption{Comparing $\overline\varrho_{i}^{N}(n)$ and
$\overline \rho_{i}^{N}(n)$.   Estimates of mean
relative branch lengths for $N=10^{3}$,  $\alpha$  and
 $\psi(N)$ as  shown,  sample size $n=10^{1}, 10^{2}, 10^{3}$; graphing
$\log(h_{i}^{N}(n) ) - \log(1 -  h_{i}^{N}(n))$ as a function of
$\log(i/n) - \log(1 - i/n)$ for $i \in [n-1]$ where $h _{i}^{N}(n)$ is
$\overline \rho_{i}^{N}(n)$ 
an estimate of $\EE{\widetilde R_{i}^{N}(n)}$
(quenched; grey) and $\overline\varrho_{i}^{N}(n)$  (annealed; black); results
from $10^{4}$ resp.\ $10^{5}$  experiments. The limiting coalescent of
the unconditional  ancestral process (predicting  $\EE{R_{i}^{N}(n)}$)
would be  the complete Beta coalescent (a),  incomplete Beta
coalescent (b),  and the Kingman coalescent (c,d).   See
Appendix~\ref{sec:estimatequenched} 
for computing $\overline \rho_{i}^{N}(n)$, and Appendix~\ref{sec:code}
for $\overline\varrho_{i}^{N}(n)$. }
\label{fig:quenchedannealedsfsA}
\end{figure}

In this section we use simulations to investigate relative branch
lengths read off trees sampled from a  finite haploid panmictic  population
by conditioning on  the population ancestry.   This is done by
simulating the evolution of a finite haploid population of fixed size
$N$ evolving according to Definition~\ref{Schwm} and
~\eqref{eq:haploid_pxi} and keeping track of the ancestry of each
individual.  Every now and then we take a random sample of size $n$
and check if the sample coalesces, i.e.\ if a common ancestor of the
sampled leaves exists (regardless of the ancestry of the individuals
not sampled). If a (most recent) common ancestor of a given random
sample is not found the sample is discarded and the population evolves
further for some time until a new sample is drawn (independently of
the previous one).  If a common ancestor of a given sample is found,
the ancestry of the sample is fixed, and the branch lengths are read
off the fixed tree relating the sampled leaves. An estimate of the
mean relative branch lengths is then obtained by repeating this a
given number of times, each time starting from scratch with a new
population.  We denote the quantity thus approximated by
$\EE{ \widetilde R_{i}^{N}(n) }$, and the approximation as
$\overline\rho_{i}^{N}(n)$.    To our knowledge a coalescent that
would approximate the trees sampled in this way from a large
population has not been derived (and deriving such a coalescent is
outside the scope of the present work).  Therefore, we compare
$\overline \rho_{i}^{N}(n) $ to  $\overline \varrho_{i}^{N}(n)$
 and the results can be found in
Figure~\ref{fig:quenchedannealedsfsA}.  Even though the approximations
of $\EE{\widetilde R_{i}^{N}(n)}$ resp.\ $\EE{ R_{i}^{N}(n)}$ broadly
agree there is a noticeable difference.
Appendix~\ref{sec:estimatequenched} contains a brief description of an
algorithm for computing $\overline\rho_{i}^{N}(n)$  estimating  $\EE{ \widetilde R_{i}^{N}(n) }$, and
Appendix~\ref{sec:code} for computing  $\overline\varrho_{i}^{N}(n)$ estimating  $\EE{R_{i}^{N}(n) }$.


\section{Computing  $\overline \varrho_{i}^{N}(n) $}
\label{sec:code}
 In this section we briefly describe an
algorithm for  computing     $\overline \varrho_{i}^{N}(n)$, the
empirical mean  of  $R_{i}^{N}(n)$ \eqref{eq:8}  estimating   $\EE{R_{i}^{N}(n)}$ (recall
  Section~\ref{numerics}).

We describe an algorithm for sampling branch lengths of a gene
genealogy of a sample of size $n$ from a finite haploid population of
constant size $N$.  Suppose the population evolves according to
Definition~\ref{Schwm}.  Our algorithm returns a realisation
$\left(b_{1}^{N}(n), \ldots, b_{n-1}^{N}(n)\right)$ of
$\left( B_{i}^{N}(n), \ldots, B_{n-1}^{N}(n) \right)$ (recall
~\eqref{eq:32}--\eqref{eq:34} in Section~\ref{numerics}).  Since we are simulating a
Markov sequence (a Markov process with a countable state space and
evolving in discrete time) we only need to keep track of the current
block sizes $\set{ |\xi_{1}^{n}|, \ldots, |{\xi}_{b}^{n}| }$
given that there are $b$ blocks at the time where $|{\xi}_{i}^{n}|$
is the size (number of leaves the block is ancestral to) of block $i$,
so that $|{\xi}_{i}^{n}| \in \IN$ and
$|{\xi}_{1}^{n}| + \cdots + |{\xi}_{b}^{n}| = n$.  Let
$r_{i}^{N}(n)$ (initalised to zero) denote an estimate of
$\EE{R_{i}^{N}(n)}$ obtained in the following way:
\begin{enumerate}
\item Set each  $b_{i}^N(n)$ to zero 
\item set each block size $|{\xi}_{i}^{n}|$ to one 
\item while there are at least two blocks  repeat the following steps
in order
\begin{enumerate}
\item update the branch lengths of the current blocks, i.e.\
$b_{i}^{N}(n)
\leftarrow b_{i}^{N}(n) + 1$ for $i = |{\xi}_{1}^{n}|, \ldots,
|{\xi}_{b}^{n}|$ given that there are $b$ blocks at the time 
\item sample a realisation $x_{1}, \ldots, x_{N}$ of the juvenile
numbers  $X_{1}^{N},
\ldots, X_{N}^{N}$  (using that $X_{i}\sim \lfloor
U^{-1/\alpha}\rfloor$ when the law of $X_{i}$ is as in  \eqref{eq:26})
\item sample number of blocks per family according to a multivariate
hypergeometric with parameters $b$ (the current number of blocks)  and  $x_{1}, \ldots, x_{N}$
\item merge blocks at  random  according to the numbers
sampled in (c);  for example
suppose  blocks of  size $|{\xi}_{i_{1}}^{n}|,\ldots, |{\xi}_{i_{m}}^{n}|$
merge, then the continuing block has size $|{\xi}_{i_{1}}^{n}| +
\cdots +  |{\xi}_{i_{m}}^{n}| $
\end{enumerate}
\item update the estimate $r_{i}^{N}(n)$  of  $\EE{R_{i}^{N}(n)}$: $r_{i}^{N}(n) \leftarrow r_{i}^{N}(n) +  b_{i}^{N}(n)/\sum_{j}  b_{j}^{N}(n) $
\item after repeating steps (1) to (4) a given number ($\mathds{M}$)  of times
return  $ \mathds{M}^{-1} r_{i}^{N}(n)$ as  an approximation  of  $ \EE{R_{i}^{N}(n)}$ 
\end{enumerate}

\section{Computing  $\overline \rho_{i}^{N}(n) $}\label{sec:estimatequenched}


In this section we briefly describe an  algorithm for computing the
empirical mean    $\overline \rho_{i}^{N}(n) $ estimating 
$\EE{ \widetilde R_{i}^{N}(n) }$ (see
Figure~\ref{fig:quenchedannealedsfsA} in Appendix ~\ref{sec:quenched}).

\newcommand{\bcirc}{\scalebox{1.5}{$\circ$}}
\newcommand{\bbullet}{\scalebox{1.5}{$\bullet$}}
\newcommand{\x}{\scalebox{2}{{$\circ$}}}
\newcommand{\s}{\textcolor{lightgray}{\scalebox{2}{\ensuremath\bullet}}}
\begin{figure}[htp]
\centering
\captionsetup[subfloat]{labelfont={scriptsize,sf,md,up},textfont={scriptsize,sf}}
\subfloat[given the population ancestry the complete sample tree
(filled circles) is fixed]{ \label{fig:oneancestry}
\xymatrix @R=0pt @C=0.2pt @M=0pt  {%
& \text{population ancestry} & \phantom{\ldots} &  \text{random sample} & \phantom{\ldots} &  \text{ancestry
with complete sample tree}  & \\
\text{past}& \x\x\x\x\x\x\x\x\x && \x\x\x\x\x\x\x\x\x &&  \x\x\s\x\x\x\x\x\x  \\
& \x\x\x\x\x\x\x\x\x && \x\x\x\x\x\x\x\x\x &&  \x\x\x\x\s\x\x\x\x  \\
& \x\x\x\x\x\x\x\x\x && \x\x\x\x\x\x\x\x\x &&  \x\x\s\x\x\x\x\x\x  \\
& \x\x\x\x\x\x\x\x\x && \x\x\x\x\x\x\x\x\x &&  \x\x\s\x\x\x\s\x\x  \\
& \x\x\x\x\x\x\x\x\x && \x\x\x\x\x\x\x\x\x &&  \x\s\x\x\x\s\x\x\x  \\
& \x\x\x\x\x\x\x\x\x && \x\x\x\x\x\x\x\x\x &&  \s\x\s\x\x\x\s\x\x  \\
\text{present} \ar[uuuuuu] & \x\x\x\x\x\x\x\x\x && \x\s\x\x\s\x\s\x\x &&  \x\s\x\x\s\x\s\x\x \\
}
}  \\
\subfloat[several independent population ancestries with complete
fixed sample trees]{\label{fig:manyancestries}
\xymatrix @R=0pt @C=0.2pt @M=0pt  {%
& \text{ancestry 1} & {\ldots} &  \text{ancestry $M$}  & \\
\text{past}&\x\x\x\x\x\s\x\x\x\x\x\x\x\x  & &    \x\x\x\x\x\s\x\x\x\x\x\x\x\x &   \\
& \x\s\x\x\x\s\x\x\x\x\x\x\x\x  &&    \x\x\x\x\x\s\x\x\x\x\x\x\x\x &  \\
&\x\s\x\x\x\s\x\x\x\x\x\x\x\x    && \x\x\x\x\x\s\x\x\x\x\x\x\x\x & \\ 
& \x\s\x\x\x\x\x\x\x\x\s\x\x\x &&  \x\s\x\x\x\x\s\x\x\x\x\s\x\x & \\ 
& \x\x\s\x\x\x\s\x\x\x\s\x\x\x  &&   \x\x\s\x\x\x\s\x\s\x\x\x\x\x  & \\ 
&  \x\x\s\x\x\x\s\x\x\x\s\x\x\x  &&   \x\x\s\x\x\x\s\x\s\x\x\x\x\x & \\
\text{present} \ar[uuuuuu] &    \x\s\x\x\x\x\s\x\x\x\x\x\s\x  &&    \x\x\s\x\x\x\s\x\s\x\x\x\x\x   \\ 
}
}
\caption{Illustrating the algorithm for generating conditional
(quenched) gene genealogies. We take `ancestry' to mean that we know
the ancestral relations of the individuals (gene copies)  in the population at any
time.  The filled circles (\textcolor{lightgray}{\bbullet}) represent
sampled gene copies $(n=3)$ or gene copies ancestral to the sampled
ones.  The empty circles (\bcirc) are gene copies that are not
ancestors of the sampled ones. Figure~\ref{fig:quenchedannealedsfsA}
holds examples of approximations of $\EE{\widetilde R_{i}^{N}(n)}$
using the method illustrated here }
\label{fig:illustratingcondgg}
\end{figure}

 Let  $ \mathds{A}^{(N,n)} \equiv (A_{i}(g))_{g\in \IN\cup \{0\}, i\in [N]}$
denote the ancestry of the population.  In our construction the
population of $N$ haploid individuals is partitioned into $N$
levels. At any given time each level is occupied by one
individual. Let $A_{i}(g) \in [N]$ be the level of the immediate
ancestor of the individual occupying level $i$ at time $g$. Set
$A_{i}(0) = i$ for $i \in [N]$.  If $A_{i}(g) = A_{j}(g)$ for
$i \neq j$ the individuals occupying levels $i$ and $j$ at time $g$
derive from the same immediate ancestor. If the individual on level
$i$ at time $g$ produces $k$ surviving offspring then
$A_{j_{1}}(g+1) = \cdots = A_{j_{k}}(g+1) = i$.  Each individual
`points' to its' immediate ancestor.   
\begin{equation}
\label{eq:31}
\xymatrix @R=0.5pt @C=0.5pt @M=0.5pt {
& \text{levels} \\
\text{time}   & 1 & \ldots \quad &  \ell  & \quad \ldots  & \quad  N \\
\text{g}      &   &        &  i  \\
\text{g+1}    &   &  \ell & \ell & \ell  \\
}
\end{equation}
In \eqref{eq:31}  the individual on level $\ell$ at time $g$ has 3
offspring, and  his immediate ancestor occupies level $i$ at time
$g-1$.

A \emph{complete} sample tree
is one where the leaves have a common ancestor in the given population
ancestry (e.g.\ the `grey' sample in
Figure~\ref{fig:illustratingcondgg}).  Let
$r_{1}^{N}(n,\mathds{A}^{(N,n)}), \ldots,
r_{n-1}^{N}(n,\mathds{A}^{(N,n)})$ denote the realised relative branch
lengths of a complete sample tree whose ancestry is given by
$\mathds{A}^{(N,n)}$.  We approximate,  for $i \in \{1,2,\ldots, n-1\}$,
\begin{equation}
\label{eq:estimate}
\EE{ \widetilde R_{i}^{N}(n) } \approx  \frac 1M \sum_{j=1}^{M}   r_{i}^{N}\left(n, \mathds{A}_{j}^{(N,n)}  \right)
\end{equation}
where $M$ is the number of experiments, the number of realised ancestries $ \mathds{A}^{(N,n)}$.   

We  summarize the  algorithm for approximating   $\EE{ \widetilde
R_{i}^{N}(n) }$:
\begin{enumerate}
\item For each  experiment:
\begin{enumerate}
\item initialise the ancestry to $A_{i}(0) = i$ for $i\in [N]$
\item until a complete sample  tree is found:
\begin{enumerate}
\item draw a random sample, i.e.\ sample $n$ of $N$ levels at the most recent  time 
\item check if the tree of the given sample is complete,  if not discard the sample and  record the ancestry of  a new set of surviving offspring: 
\begin{enumerate}
\item   sample numbers of  potential offspring  $X_{1},\ldots, X_{N}$  
\item   given $X_{1},\ldots, X_{N}$
potential offspring sample the surviving offspring uniformly at random
without replacement and update the ancestry; if the individual on level $i$ at time
$g$ produces  $k$ surviving offspring then
$A_{j_{1}}(g+1) = \cdots = A_{j_{k}}(g+1) = i$.
\end{enumerate}
\end{enumerate}
\item given a complete tree  read  the branch lengths off the tree  and merge  blocks according to the
ancestry; suppose two  blocks have  ancestors
on  levels $i$ and $j$ at time $g$,  if  $A_{i}(g) = A_{j}(g)$ the blocks 
are merged;
\item given the branch lengths $\ell_{1}^{N}(n), \ldots, \ell_{n-1}^{N}(n)$  of a complete tree update the estimate $r_{i}^{N}(n,  \mathds{A}^{(N,n)}) $
of $\EE{ \widetilde R_{i}^{N}(n) }$ :  $r_{i}^{N}(n,  \mathds{A}^{(N,n)}) \leftarrow   r_{i}^{N}(n,  \mathds{A}^{(N,n)}) + \ell_{i}^{N}(n)/\sum_{i} \ell_{i}^{N}(n)  $
\end{enumerate}
\item From $M$ realised ancestries
$\mathds{A}_{1}^{(N,n)}, \ldots, \mathds{A}_{M}^{(N,n)}$ return 
   \eqref{eq:estimate} as an approximation of 
$\EE{\widetilde R_{i}^{N}(n) }$  for $i = 1,2, \ldots,
n-1$
\end{enumerate}
The idea is illustrated in
Figure~\ref{fig:illustratingcondgg}; (a) given a population ancestry
the complete  sample tree is  fixed as soon as the identity of the
sampled gene copies is known; (b) for each population ancestry we
read the branch lengths  off one  sample tree, and then average the
branch lengths over population ancestries.   



\end{document}